\pdfoutput=1
\documentclass[11pt,a4paper]{article}

\usepackage{fix-cm}
\usepackage{amssymb}
\usepackage{amsmath}
\usepackage{amsfonts}
\allowdisplaybreaks

\newcommand{\E}{\mathbb{E}}
\renewcommand{\P}{\mathbb{P}}
\newcommand{\I}{\mathbb{I}}

\newcommand{\asto}{\xrightarrow{\mathrm{a.s.}}}

\newcommand{\dto}{\xrightarrow{\mathrm{d}}}

\usepackage{float}
\usepackage{titlesec}
\usepackage{setspace}
\usepackage{hyperref}
\usepackage{geometry}
\usepackage{graphicx}

\usepackage{tcolorbox}

\usepackage{caption}
\hypersetup{
    colorlinks=true,
    linkcolor=blue,  
    citecolor=blue,  
    filecolor=black,
    urlcolor=cyan
}

\titleformat{\section}
   {\normalfont\fontsize{11}{11}\bfseries\centering} 
   {\thesection. }                       
   {0em}                                 
   {}                                    

 \titleformat{\subsection}
   {\normalfont\fontsize{11}{11}\bfseries} 
   {\thesubsection.$\;$}                       
   {0em}                                  
   {}                                     

\usepackage[style=authoryear,backend=biber, dashed=false, natbib=true, uniquename=false, maxbibnames=99, minbibnames=1, maxcitenames=2, mincitenames=1]{biblatex} 

\AtBeginBibliography{%

}

\newtheorem{theorem}{Theorem}[section]

\newtheorem{corollary}[theorem]{Corollary}

\newtheorem{lemma}[theorem]{Lemma}

\newtheorem{remark}[theorem]{Remark}

\numberwithin{equation}{section}
\begin{document}

\title{Two-Slope A-GARCH: Quasi Maximum Likelihood Estimation and Empirical Evidence%
\thanks{We thank Rasmus Søndergaard Pedersen, Heino Bohn Nielsen and Anders Rahbek for helpful comments. The authors are partly supported by the Independent Research Fund Denmark (DFF Grant 7015-00028).}}
\author{Jonas F. Frederiksen\footnote{University of Copenhagen; email: \tt{jff@econ.ku.dk}} \and Frederik Vilandt\footnote{University of Copenhagen; email: \tt{fvr@econ.ku.dk}}}
\date{}

\maketitle

\begin{abstract}
    The empirical news impact curve (NIC) of daily stock returns is typically higher at negative than at positive returns of the same size, has its minimum at a positive return, and is often flat above this point. The popular GJR and A-GARCH models each rule out one of these features. The Two-Slope A-GARCH (TS-AGARCH) model nests both by combining an estimated threshold, the point of asymmetry, with separate slopes on either side. Although the likelihood is not twice continuously differentiable in the threshold, the quasi maximum likelihood estimator is $\sqrt{n}$-consistent, and asymptotically normal unless the slope above the threshold is at its boundary value of zero. Either way, the GJR and A-GARCH models can be tested with standard normal critical values. For five major stock indices, we reject the GJR model for all five and the A-GARCH model for three. In a simple market model, assets inherit the asymmetry of the stock market through their comovement with it. Estimates for nearly 400 stocks, currencies, bonds, gold and hedging assets support this: the point of asymmetry is positive for assets that move with the market, near zero for neutral assets and negative for hedges, and the slopes follow a similar pattern.
    \end{abstract}


\newpage
\section{Introduction}
Generalized autoregressive conditional heteroskedasticity (GARCH) models describe a financial return $x_t$ and its volatility $\sigma_t^2$ over time $t$ by the multiplicative structure $x_t=\sigma_t\,z_t$, where
\begin{equation*}
  \sigma_t^2=g_\theta(x_{t-1},\sigma_{t-1}^2),
\end{equation*}
with $\{z_t\}$ an iid noise sequence, $\E [z_t]=0$ and $\E [z_t^2]=1$, and $g_\theta$ is a function known up to a finite-dimensional parameter $\theta$. Since the introduction of GARCH models by \citet{bollerslev1986generalized}, various specifications of $g_\theta$ that are asymmetric in $x_{t-1}$ have been proposed (see, e.g., \citealt{engle1990stock}, \citealt{glosten1993relation} and \citealt{zakoian1994threshold}), allowing volatility to respond differently to positive and negative returns. The distinctive features of a given specification are best described through its news impact curve (NIC), defined as the function $x\mapsto g_\theta(x,\sigma^2)$ for fixed $\theta$ and $\sigma^2$ (see \citealt{engle1993measuring}). Existing empirical studies of stock returns reveal a markedly asymmetric NIC. In particular, \textit{(i)} volatility reacts more strongly to negative returns than to positive returns of the same magnitude, known as the \emph{leverage effect} (see, e.g., \citealt{engle1993measuring}, \citealt{glosten1993relation} and \citealt{zakoian1994threshold}), \textit{(ii)} the point of asymmetry, where the NIC attains its minimum, is positive (see \citealt{engle1993measuring} and \citealt{hentschel1995all}), and \textit{(iii)} returns above this point often do not increase volatility at all (see, e.g., \citealt{glosten1993relation} and \citealt{patton2015good}). \smallskip

\begin{figure}[H]
\centering
\includegraphics[width=0.85\textwidth]{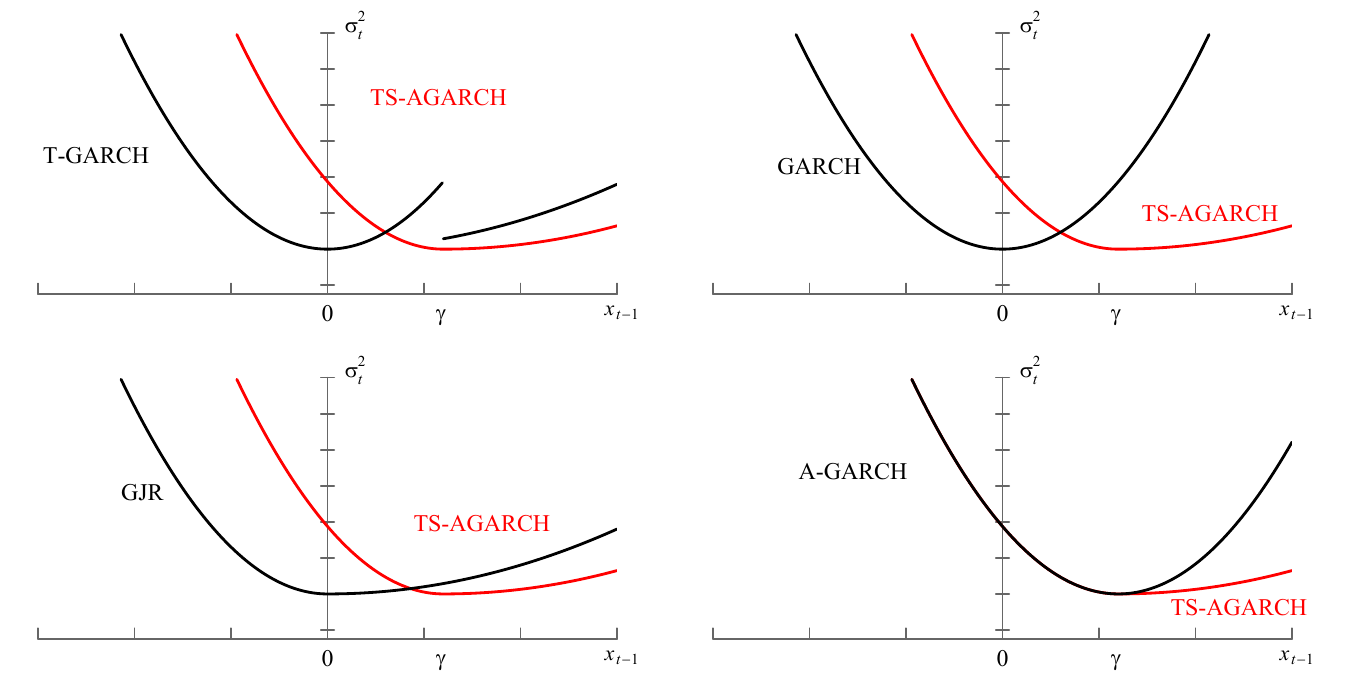}
\caption{NICs of the T-GARCH (top-left), GARCH (top-right), GJR (bottom-left) and A-GARCH (bottom-right). The red curve indicates the NIC of the TS-AGARCH. Here we have set $\omega+\beta\,\sigma_{t-1}^2=1, \ \alpha_1=0.2, \ \alpha_2=1.3$ and $\gamma=1.2$.}\label{fig: 4 NICs}
\end{figure}

In this paper, we consider the Two-Slope A-GARCH (TS-AGARCH) model, which specifies
\begin{equation}
  \sigma_t^2=\omega+\bigl(\alpha_1\I(x_{t-1}\geq\gamma)+\alpha_2\I(x_{t-1}<\gamma)\bigr)(x_{t-1}-\gamma)^2+\beta\sigma_{t-1}^2,\label{eq:stgarch}
\end{equation}
where the threshold $\gamma\in\mathbb{R}$ is the point of asymmetry, and $\alpha_1$ and $\alpha_2$ determine the slopes of the NIC to the right and left of $\gamma$, respectively. The TS-AGARCH captures all the above features: the leverage effect \textit{(i)} arises when $\alpha_2>\alpha_1$ or $\gamma>0$, while \textit{(ii)} and \textit{(iii)} correspond to $\gamma>0$ and $\alpha_1=0$, respectively. The two most widely used asymmetric GARCH specifications, the GJR model of \citet{glosten1993relation} and the A-GARCH of \citet{engle1990stock}, each rule out one of these features by construction. The GJR model, which is obtained from \eqref{eq:stgarch} by setting $\gamma=0$, pins the minimum of the NIC at zero and thereby rules out \textit{(ii)}, while the A-GARCH, obtained by setting $\alpha_1=\alpha_2$ in \eqref{eq:stgarch}, imposes equal slopes on either side of the minimum and thereby rules out \textit{(iii)} (see Figure \ref{fig: 4 NICs}).\smallskip

Our main contribution is providing the asymptotic likelihood theory needed to test these features within the TS-AGARCH. We show that the Gaussian quasi maximum likelihood estimator (QMLE) of $\theta=(\omega,\alpha_1,\alpha_2,\gamma,\beta)^\top$ is strongly consistent and asymptotically normal at the usual $\sqrt{n}$-rate under conditions standard in the GARCH literature, provided the true value of $\theta$ is an interior point of the parameter space. Consequently, the GJR hypothesis $\gamma=0$ and the A-GARCH hypothesis $\alpha_1=\alpha_2$ can be tested by standard $t$-tests. Rejecting at least one of these hypotheses in favor of $\gamma>0$ or $\alpha_2>\alpha_1$ supports \textit{(i)}, while rejecting $\gamma=0$ in favor of $\gamma>0$ supports \textit{(ii)}. Testing \textit{(iii)} is more delicate, as $\alpha_1 = 0$ lies on the boundary of the parameter space, where the QMLE is no longer asymptotically normal. We derive the limiting distribution in this case as well, and show that the $t$-statistic for $\alpha_1=0$ converges to the positive part of a standard normal variable, so that the hypothesis can be tested using one-sided standard normal critical values. Such boundary tests are available for asymmetric GARCH models whose threshold is fixed at zero (see \citealt{francq2019qml}), but, to the best of our knowledge, not for models in which it is estimated as in the TS-AGARCH. Hence, all three features \textit{(i)}--\textit{(iii)} can be assessed statistically within a single model. \smallskip

In an empirical application to daily returns on five major stock indices, we reject the GJR restriction $\gamma = 0$ in favor of $\gamma > 0$ for every index, in line with \textit{(i)} and \textit{(ii)}. The estimates of $\gamma$ lie between $0.32$ and $0.60$, so that markets are calmest after returns of about half a percent. The A-GARCH restriction $\alpha_1=\alpha_2$ is rejected for three of the indices, with $\alpha_2>\alpha_1$ throughout, and $\alpha_1=0$ in \textit{(iii)} is rejected for two, suggesting that the TS-AGARCH improves upon both the celebrated GJR and A-GARCH models. We then compare the NIC across nearly 400 S\&P 500 stocks, stock indices, currencies, Treasuries, gold and hedging assets. In the market model of \citet{sharpe1963simplified}, an asset inherits the asymmetry of the market through its CAPM beta, which suggests that its point of asymmetry, in units of its standard deviation, is that of the market times its correlation with the market, and that its slopes are interchanged when the correlation is negative. The estimates follow this pattern: the point of asymmetry is positive for assets that move with the market, close to zero for neutral assets and negative for hedges, and the ordering of the slopes follows the correlation in the same way. A fund that is short the S\&P 500, for instance, has the mirror image of the NIC of the index. The point of asymmetry, which the GJR rules out, is thus an economically meaningful feature of the NIC. \smallskip

Up to a reparametrization, the TS-AGARCH is already available in the econometric software PcGive (see Chapter 8 in \citealt{doornik2013pcgive}). However, to the best of our knowledge, it has not been studied in the academic literature, and neither conditions for stationarity nor an asymptotic theory for estimation and inference are available. In addition to the likelihood theory, we show that the TS-AGARCH is geometrically ergodic, and hence admits a strictly stationary solution, under conditions on $\theta$ that coincide with the well-known conditions for the GJR model and do not depend on the threshold parameter $\gamma$. The introduction of the threshold parameter thus costs nothing in terms of the stationarity region. \smallskip

For $\alpha_1\neq\alpha_2$, the TS-AGARCH is a threshold model in the tradition of the threshold heteroskedastic models of \citet{glosten1993relation} and \citet{zakoian1994threshold}, but, importantly, with the threshold parameter estimated rather than fixed at zero. A close relative is the threshold ARCH (T-ARCH) model of \citet{li2015asymptotic}, in which the threshold $\gamma$ is also freely varying. However, the T-ARCH, like the corresponding threshold GARCH (T-GARCH), replaces $(x_{t-1}-\gamma)^2$ in \eqref{eq:stgarch} with $x_{t-1}^2$, so that its NIC has a jump discontinuity at $\gamma$ (see Figure \ref{fig: 4 NICs}). We regard the continuous NIC of the TS-AGARCH as the more plausible description of financial returns, as it seems unrealistic that the effect of $x_{t-1}$ on $\sigma_t^2$ should vary drastically within an arbitrarily small neighborhood of $\gamma$. Moreover, in the T-ARCH, as in threshold autoregressive models, the threshold is only identified when the slopes differ, whereas $\gamma$ in the TS-AGARCH remains identified even when $\alpha_1=\alpha_2$, as it still determines the minimum point of the NIC, so that the A-GARCH hypothesis can be tested by a standard $t$-test. In addition, $\gamma$ can for the TS-AGARCH be estimated jointly with the other parameters by standard gradient-based optimization, whereas in the T-ARCH $\gamma$ must be estimated by grid search methods (see \citealt{li2015asymptotic}). \smallskip

In terms of proving asymptotic normality, the analysis of the TS-AGARCH differs from that of most GARCH-type models, since its likelihood function is not twice continuously differentiable. In fact, the second order directional derivatives of $\sigma_t^2(\theta)$ with respect to $\gamma$ have infinitely many jump discontinuities. This
rules out strategies used throughout the GARCH literature to establish asymptotic normality. Specifically, with $\mathcal{S}_n$ and $\mathcal{H}_n$ denoting the first and second order (directional) derivatives of the  log-likelihood, proofs typically rely on a mean
value expansion of $\mathcal{S}_n$ combined with a uniform law of large numbers (ULLN) for continuous functions applied to
$\mathcal{H}_n$ (see, e.g., \citealt{straumann2005estimation}, \citealt{meitz2011parameter} and \citealt{BerkesEtAl2003}). Since the log-likelihood for the TS-AGARCH is not twice continuously differentiable, the ULLN does not apply, and furthermore, measurability of the intermediate point in the mean value expansion is not automatic either (see \citealt{yang2026taylor} for a discussion). Moreover, this strategy relies on the first order condition $\mathcal{S}_n(\theta_n)=0$ for the QMLE $\theta_n$, which fails with positive probability when a parameter is on the boundary of the parameter space. Instead, we prove the validity of the following expansion for $\mathcal{S}_n$ around the true parameter $\theta_0$
\begin{equation*}
  n^{-1/2}\mathcal{S}_n(\theta_n)
  = n^{-1/2}\mathcal{S}_n(\theta_0)
  +\left(\int_0^1 n^{-1}\mathcal{H}_n\big(\theta_0+u(\theta_n-\theta_0)\big)\,du\right)n^{1/2}(\theta_n-\theta_0),
\end{equation*}
where the integral is a Lebesgue integral. Together with a direct argument showing that $n^{-1}\mathcal{H}_n$ converges uniformly over shrinking neighborhoods of $\theta_0$ despite its discontinuities, the expansion yields the asymptotic distribution of $\theta_n$. This holds both when $\theta_0$ is an interior point, where $\mathcal{S}_n(\theta_n)=0$, and when $\alpha_1$ is on the boundary, where $\mathcal{S}_n(\theta_n)$ is determined by the first order conditions for the constraint $\alpha_1\geq0$. The above strategy does not impose stronger moment conditions than $\mathbb{E}[z_0^4] < \infty$, which is standard in asymptotic theory for the QMLE in GARCH models. We therefore expect our approach to be of independent interest for other criterion functions whose curvature is only piecewise continuous. \smallskip

The $\sqrt{n}$-rate for the threshold estimator stands in contrast to the T-ARCH where $\sigma^2_t(\theta)$ is discontinuous. Specifically, \citet{li2015asymptotic} show that the QMLE of $\gamma$ in the T-ARCH is
$n$-consistent with a non-Gaussian limit distribution that is analytically intractable and must be simulated case by case. This contrast is analogous to threshold regression and threshold autoregression: in the continuous (kink) case, \citet{hansen2017regression} and \citet{Chan_Tsay1998} obtain joint $\sqrt{n}$-normality of the least squares slope and threshold estimates, whereas in the discontinuous case the threshold estimate is $n$-consistent with a non-standard limit (see, e.g., \citealt{chan1993consistency}). To establish asymptotic normality, \citet{hansen2017regression} and \citet{Chan_Tsay1998} rely on empirical process theory, and as a result require finite moments of higher order than in the corresponding non-threshold models. As mentioned above, our result only requires moment conditions standard for GARCH models, which is due to the fact that $\sigma^2_t(\theta)$ of the TS-AGARCH is smoother in $\theta$ than the conditional mean specification of these models. This additional smoothness allows us to avoid relying on empirical process theory arguments.

The remainder of the paper is organized as follows. Section \ref{sec: stochastic properties} gives the stationarity, ergodicity and moment results for the TS-AGARCH, Section \ref{sec: likelihood estimation} derives the score and Hessian of the quasi likelihood, paying particular attention to where differentiability fails, and Section \ref{sec: Asymptotic properties of the QMLE} contains the asymptotic theory for the QMLE, including the case where $\alpha_1$ is on the boundary. Section \ref{sec: Monte Carlo} reports Monte Carlo evidence on the finite sample accuracy of the asymptotic approximations, and Section \ref{sec: empirical section} contains the empirical applications. All proofs are given in the Appendix.

\section{The TS-AGARCH process} \label{sec: stochastic properties}
The Two-Slope A-GARCH (TS-AGARCH) process is given by
\begin{align}\label{eq:STGARCH_proc} 
    x_t = \sigma_t z_t, \quad \sigma_t^2 = 
    \begin{cases} 
    \omega + \alpha_1 (x_{t-1} - \gamma)^2 + \beta \sigma^2_{t-1},\quad \text{if} \quad x_{t-1} \geq \gamma, \\[0.05in] 
    \omega + \alpha_2 (x_{t-1} - \gamma)^2 + \beta \sigma^2_{t-1},\quad \text{if}  \quad x_{t-1} < \gamma, 
    \end{cases} \qquad t \in \mathbb{N},
\end{align}
for $(x_0,\sigma_0)$ given, parameters $\omega,\beta>0$, $\alpha_1,\alpha_2\geq0$ with $\max\{\alpha_1,\alpha_2\}>0$, and $\gamma\in\mathbb{R}$, and with $\{z_t:t\in\mathbb{Z}\}$ an iid sequence of random variables, where $z_0$ admits a continuous density which is strictly positive on $\mathbb{R}$, and satisfies $\E[z_0]=0$ and $\E[z_0^2]=1$. Here and throughout, we write $\{x_t\}$ as shorthand notation for a sequence $\{x_t:t\in\mathbb{N}\}$, unless otherwise stated.

The TS-AGARCH in \eqref{eq:STGARCH_proc} could be stated without assuming that $z_0$ has a continuous and strictly positive density. However, this condition plays a key role throughout the paper, so we present the model assuming it. 
Specifically, it ensures irreducibility of the Markov chain $\{\sigma_t^2\}$ in Theorem \ref{thm: geometric mixing and moments} below. It also implies that $\P(x_t=\gamma)=0$ for every $\gamma\in\mathbb{R}$, so that the second-order derivatives of the likelihood function are well defined almost surely (see Section \ref{sec: likelihood estimation}), and it is used to identify the true parameter $\theta_0$ (see Section \ref{sec: Asymptotic properties of the QMLE}).

In the following subsection, we provide conditions ensuring that the joint process $\{(x_t,\sigma_t^2)\}$ given in \eqref{eq:STGARCH_proc} is geometrically ergodic.
\subsection{Stochastic properties}
Most GARCH specifications of $\sigma_t^2$ arise as special cases of the stochastic recurrence equation $\sigma^2_t=A_t\,\sigma^2_{t-1}+B_t$ where $\{(A_t,B_t)\}$ is an iid sequence of $\mathbb{R}^2$-valued random vectors. This makes finding conditions for geometric ergodicity as well as existence of moments a straightforward task as stochastic recurrence equations (SREs) are well understood, see, e.g., \citet{buraczewski2016stochastic}. The TS-AGARCH in \eqref{eq:STGARCH_proc} follows an SRE only in the GJR-case where $\gamma=0$, for which
\begin{equation*}
    (A_t,B_t)=(\beta+(\alpha_1\I(z_{t-1}\ge0)+\alpha_2\I(z_{t-1}<0))z_{t-1}^2,\,\omega), \qquad t \in \mathbb{N},
\end{equation*}
and therefore well-known results yield in this case that $\{\sigma^2_t\}$ is geometrically ergodic, and hence admits a stationary representation, provided
\begin{equation}
    \phi(z_0)=\E[\log(\beta+(\alpha_1\I(z_{0}\geq 0)+\alpha_2\I(z_{0}<0))z_{0}^2)] <0. \label{eq: def phi}
\end{equation}
Furthermore, for $p>0$, the stationary distribution satisfies $\E[\sigma_0^{2p}]<\infty$ provided
\begin{equation}\label{eq: moment condition}
    \E[(\beta+(\alpha_1\I(z_{0}\geq 0)+\alpha_2\I(z_{0}<0))z_{0}^2)^{p}]<1.
\end{equation}
It can then be shown that the joint process $\{(x_t,\sigma^2_t)\}$ inherits the geometric ergodicity property of $\{\sigma_t^2\}$, and that $\E[x_0^{2p}]<\infty$ holds for the stationary distribution if the relation \eqref{eq: moment condition} holds.\medskip

When $\gamma\neq0$, no such results are directly available since $\sigma^2_t$ can no longer be represented as an SRE. However, $\{\sigma^2_t\}$ is still a Markov chain in this case. By relying on general Markov chain theory, we show in Theorem \ref{thm: geometric mixing and moments} that the exact same conditions as in the SRE case of $\gamma=0$, \eqref{eq: def phi}--\eqref{eq: moment condition}, give geometric ergodicity of $\{(x_t,\sigma^2_t)\}$ and existence of moments also when $\gamma\neq 0$.
\begin{theorem}\label{thm: geometric mixing and moments}
    Let $\{(x_t,\sigma_t^2)\}$ be given as in \eqref{eq:STGARCH_proc} and suppose that \eqref{eq: def phi} holds. Then the joint process $\{(x_t,\sigma_t^2)\}$ is geometrically ergodic. Furthermore, the stationary distribution satisfies $\E[x_0^{2p}]<\infty$ for $p>0$ if \eqref{eq: moment condition} holds and either $\min\{\alpha_1,\alpha_2\} > 0$ or $\E[z^{2p}_0] < \infty$.
\end{theorem}
\begin{remark} \label{remark: moments}
    Note that the stationarity condition in \eqref{eq: def phi} automatically yields that \eqref{eq: moment condition} is satisfied for some $0 < p < 1$. That is, for small $p > 0$, $\E[x^{2p}_0]$ and $\E[\sigma^{2p}_0]$ are finite for the stationary distribution.
\end{remark}
\section{Likelihood based estimation} \label{sec: likelihood estimation}
Given a sample $x_0,\dots,x_n$ generated according to \eqref{eq:STGARCH_proc} for some true parameter value $\theta_0=\left(\omega_0,\alpha_{1,0},\alpha_{2,0},\gamma_{0},\beta_{0}\right)^\top$, we consider quasi maximum likelihood estimation of $\theta_0$ based on the Gaussian likelihood function given by,
\begin{align}
    \hat{\mathcal{L}}_n(\theta)&=\sum^n_{t=1}\hat{\ell}_t(\theta), \quad \hat{\ell}_t(\theta) = -\frac{1}{2}\left(\log(\hat{\sigma}^2_t(\theta))+\frac{x^2_t}{\hat{\sigma}^2_t(\theta)}\right), \quad \theta \in \Theta, \label{eq: def Lhat(theta)}  \\[0.05in]
    \hat{\sigma}^2_t(\theta) &= \omega + (\alpha_1 \I(x_{t-1} \geq \gamma)+\alpha_2 \I(x_{t-1} < \gamma)) (x_{t-1}-\gamma)^2 + \beta  \hat{\sigma}^2_{t-1}(\theta) \label{eq: def sig2hat(theta)},
\end{align}
for an initial value $\hat{\sigma}^2_0(\theta) = \hat{\sigma}^2_0 > 0$ not depending on $\theta$ or $n$, but which may depend on the data.\footnote{In the simulation study in Section \ref{sec: Monte Carlo}, we use the backcast $\hat\sigma^2_0=\sum_{j=0}^{74}0.94^jx_j^2/\sum_{j=0}^{74}0.94^j$, i.e.\ an exponentially weighted average of the first squared observations, which is the default choice in the Python package \texttt{arch} of \citet{sheppard2025arch}. \label{fn: backcast}} The parameter space $\Theta$, $\theta_0 \in \Theta$, is given by the compact set
\begin{align}
    \Theta = [\omega_L,\omega_U]\times[0,\alpha_U]\times[\alpha_L,\alpha_U] \times  [\gamma_L,\gamma_U] \times [\beta_L,\beta_U], \label{eq: Theta}
\end{align}
where $0<\omega_L<\omega_U$, $0<\alpha_L<\alpha_U$, $\gamma_L<\gamma_U$ and $0<\beta_L<\beta_U<1$, and the QMLE is given by $\hat\theta_n=\arg\max_{\theta\in\Theta}\hat{\mathcal{L}}_n(\theta)$. Note that the existence of such a maximizer $\hat \theta_n$ is guaranteed since $\hat{\mathcal{L}}_n$ is continuous on the compact $\Theta$. The lower bound zero for $\alpha_1$ allows for the possibility that returns above the threshold do not affect the volatility. In contrast, $\alpha_2$ is bounded away from zero, which ensures that $\gamma_0$ and $\beta_0$ are identified also when $\alpha_{1,0}=0$ (see Lemma \ref{lem: cons 2} in the Appendix). \medskip

\begin{remark}
    The parameter space in \eqref{eq: Theta} is set up to allow for $\alpha_1 = 0$ while $\alpha_2 > 0$. The results proved in Section \ref{sec: Asymptotic properties of the QMLE} could equally well have been stated for the case where $\alpha_2 = 0$ is allowed while $\alpha_1 > 0$. We consider the parameter space in \eqref{eq: Theta} as it is the empirically relevant case (see Section \ref{sec: empirical section}). 
\end{remark}

As we justify below, maximization of the likelihood function $\hat{\mathcal{L}}_n$ can be accomplished using standard numerical optimization algorithms that rely on first and second order (partial) derivatives of $\hat{\mathcal{L}}_n$. In particular, estimating $\gamma$ does not require usage of grid search methods, which, in contrast, are necessary when estimating the threshold parameter of the T-ARCH model of \citet{li2015asymptotic}. The validity of such maximization procedures hinges on the fact that the first order partial derivative, $\hat{\mathcal{S}}_n$, of $\hat{\mathcal{L}}_n$ is well defined, and that the partial second order derivative, $\hat{\mathcal{H}}_n$, exists either as a left or right derivative with these being equal almost everywhere. 

More specifically, two complications arise when defining the partial derivatives of $\hat\sigma^2_t(\theta)$. First, since $\alpha_1 \geq 0$, as in \citet{francq2007quasi} for standard GARCH($p,q$) models, the partial derivatives with respect to $\alpha_1$ can only be defined as right derivatives at the boundary $\alpha_1 = 0$. The second complication is specific to the TS-AGARCH and has to do with the partial derivatives with respect to $\gamma$. In particular, as we show below, $\partial\hat\sigma_t^2(\theta)/\partial\theta$ is for any $\theta$ with $\alpha_1 >0$ a partial derivative in the usual sense, while $\partial^2\hat\sigma_t^2(\theta)/\partial\theta\partial\theta^\top$ exists only as a left or right derivative at some points. To handle both complications at once, we define all partial derivatives as right partial derivatives throughout. These coincide with the usual partial derivatives wherever the latter exist, and the convention has no influence on either the theoretical results or the maximization of $\hat{\mathcal{L}}_n$ in practice. The score and Hessian of $\hat{\mathcal{L}}_n$ are given by
\begin{align}
    \hat{\mathcal{S}}_n(\theta)&=\frac{\partial \hat{\mathcal{L}}_n(\theta)}{\partial\theta}=\sum_{t=1}^n\hat s_t(\theta), \quad \hat s_t(\theta) = \frac{1}{2}\left(\frac{x_t^2}{\hat\sigma_t^2(\theta)}-1\right)\frac{\partial\hat\sigma_t^2(\theta)/\partial\theta}{\hat\sigma_t^2(\theta)} ,\label{eq: def shat}\\
    \hat{\mathcal{H}}_n(\theta)&=\frac{\partial^2\hat{\mathcal{L}}_n(\theta)}{\partial\theta\partial\theta^\top}=\sum_{t=1}^n \hat h_t(\theta), \label{eq: def hhat} \\ \hat h_t(\theta)&=\frac{1}{2}\left(\frac{x_t^2}{\hat\sigma_t^2(\theta)}-1\right)\frac{\partial^2\hat\sigma_t^2(\theta)/\partial\theta\partial\theta^\top}{\hat\sigma_t^2(\theta)}-\frac{1}{2}\left(2\frac{x_t^2}{\hat\sigma_t^2(\theta)}-1\right)\frac{\partial\hat\sigma_t^2(\theta)/\partial\theta}{\hat\sigma_t^2(\theta)}\frac{\partial\hat\sigma_t^2(\theta)/\partial\theta^\top}{\hat\sigma_t^2(\theta)} \notag,
\end{align}
for any $\theta \in \Theta$.

The left and right partial derivatives of $\hat{\sigma}_t^2(\theta)$ with respect to $\gamma$ are equal everywhere. In particular, the directional derivatives are both zero at the set of potential discontinuities,
\begin{equation*}
    D_t:=\{\theta\in\Theta:\gamma = x_{t-i} \text{ for some } i \in\{1,\dots,t\}\}.
\end{equation*}
The partial derivative $\partial\hat\sigma_t^2(\theta)/\partial\theta$ is therefore well defined in the usual sense and given by
\begin{align}
    \frac{\partial\hat\sigma_t^2(\theta)}{\partial\theta}=\begin{pmatrix}
        \frac{\partial\hat\sigma_t^2(\theta)}{\partial\omega} \\[0.02in]
        \frac{\partial\hat\sigma_t^2(\theta)}{\partial\alpha_1} \\[0.02in]
        \frac{\partial\hat\sigma_t^2(\theta)}{\partial\alpha_2} \\[0.02in]
        \frac{\partial\hat\sigma_t^2(\theta)}{\partial\gamma} \\[0.02in]
        \frac{\partial\hat\sigma_t^2(\theta)}{\partial\beta} 
    \end{pmatrix}=\sum_{i=1}^t \beta^{i-1}\begin{pmatrix}
        1 \\
        \I(x_{t-i}\geq \gamma)(x_{t-i}-\gamma)^2 \\
        \I(x_{t-i}< \gamma)(x_{t-i}-\gamma)^2 \\
        2(\gamma-x_{t-i})(\alpha_{1}\I(x_{t-i}\geq \gamma)+\alpha_{2}\I(x_{t-i}<\gamma)) \\
        \hat\sigma_{t-i}^2(\theta)
    \end{pmatrix} \label{eq: def sighat2 deriv}
\end{align}
for all $\theta \in \Theta$ and $t\geq1$.

Next, all the second order derivatives of $\hat\sigma^2_t(\theta)$ except $\partial^2\hat\sigma^2_t(\theta)/\partial \gamma^2$ are well defined in the usual sense on $\Theta$. For the latter, the left and right partial derivatives of $\partial\hat\sigma^2_t(\theta)/\partial\gamma$ with respect to $\gamma$ exist for every $\theta \in \Theta$ but differ at the points in $D_t$ when $\alpha_1\neq\alpha_2$. To illustrate this, let $\partial^2_+\hat\sigma^2_t(\theta)/\partial\gamma^2$ and $\partial^2_-\hat\sigma^2_t(\theta)/\partial\gamma^2$ denote the right and left second order partial derivatives at a point $\theta \in \Theta$ defined as the limits
\begin{align*}
    \frac{\partial^2_+\hat\sigma^2_t(\theta)}{\partial\gamma^2} &= \lim_{\bar\gamma \downarrow \gamma} \frac{\partial\hat\sigma^2_t(\theta(\bar\gamma))/\partial\gamma-\partial\hat\sigma^2_t(\theta)/\partial\gamma}{\bar{\gamma}-\gamma},\quad \frac{\partial^2_-\hat\sigma^2_t(\theta)}{\partial\gamma^2} = \lim_{\bar\gamma \uparrow \gamma}\frac{\partial\hat\sigma^2_t(\theta(\bar\gamma))/\partial\gamma-\partial\hat\sigma^2_t(\theta)/\partial\gamma}{\bar{\gamma}-\gamma},
\end{align*}
where $\theta(\bar\gamma)=(\omega,\alpha_1,\alpha_2,\bar\gamma,\beta)^\top$.
Using that
\begin{align*}
    \lim_{\bar\gamma \downarrow \gamma}\I(x \geq \bar\gamma)= \I(x > \gamma), \quad \text{and} \quad \lim_{\bar\gamma \uparrow \gamma}\I(x \geq \bar\gamma)= \I(x \geq \gamma),
\end{align*}
it follows by lengthy calculations (see Section \ref{sec: def score and info} in the Appendix for details) that the left and right derivatives are given by
\begin{align}
\frac{\partial^2_+\hat\sigma^2_t(\theta)}{\partial\gamma^2} &= \sum_{i=1}^t \beta^{i-1}2(\alpha_1\I(x_{t-i}> \gamma)+\alpha_2\I(x_{t-i}\leq \gamma)), \label{eq: right deriv sig2hat} \\
\frac{\partial^2_-\hat\sigma^2_t(\theta)}{\partial\gamma^2}&=\sum_{i=1}^t \beta^{i-1}2(\alpha_1\I(x_{t-i}\geq \gamma)+\alpha_2\I(x_{t-i}< \gamma)). \notag
\end{align}
Thus, at any $\theta \in D_t$ with $\alpha_1\neq\alpha_2$, the left and right derivatives differ:
\begin{align*}
    \frac{\partial^2_+\hat\sigma^2_t(\theta)}{\partial\gamma^2} - \frac{\partial^2_-\hat\sigma^2_t(\theta)}{\partial\gamma^2} = 2(\alpha_2-\alpha_1)\sum_{\{i \leq t \ : \ x_{t-i}=\gamma\}}\beta^{i-1}. 
\end{align*}
Hence, the derivative is not well defined in the usual sense. However, since $\P(\cup_{i\geq1}(x_{t-i}=\gamma))\leq\sum_{i=1}^\infty\P(x_{t-i}=\gamma)=0$ for all\footnote{Since $\sigma_t>0$ is independent of $z_t$, and $z_t$ is assumed to have a Lebesgue density, it holds that $\P(x_t=\gamma)=\E[\P(z_t=\gamma/\sigma_t\,|\,\sigma_t)]=0$ for any $\gamma\in\mathbb{R}$.\label{fn: null set}} $\gamma$, it holds that the left and right derivatives are equal almost surely. In accordance with our convention, $\partial^2\hat\sigma^2_t(\theta)/\partial\gamma^2$ denotes the right derivative in \eqref{eq: right deriv sig2hat}.

These considerations show in particular that $\hat{\mathcal{H}}_n$ has discontinuities on the set $D_n$. This is in contrast to the matrix of second derivatives of other GARCH-type models, which is typically a smooth function of $\theta$ (see, e.g., \citealt{francq2019garch} and \citealt{straumann2005estimation}). The remaining non-zero entries of $\partial^2\hat\sigma_t^2(\theta)/\partial\theta\partial\theta^\top$ are given by \eqref{eq: expression 2nd partials} in the Appendix, with the sums $\sum_{i=1}^\infty$ replaced by $\sum_{i=1}^t$ and $\sigma^2_{t-i}(\theta)$ replaced by $\hat\sigma^2_{t-i}(\theta)$.
\section{Asymptotic properties of the QMLE}\label{sec: Asymptotic properties of the QMLE}
In this section, we establish under standard assumptions that the QMLE, $\hat\theta_n$, is strongly consistent (Theorem \ref{thm: consistency}), and we derive its asymptotic distribution (Theorem \ref{thm: asym normality}). The latter covers both the case where $\theta_0$ is an interior point of $\Theta$, in which the limit distribution of $\hat\theta_n$ is Gaussian, and the case where $\alpha_{1,0}=0$ is a boundary point, where the limit distribution is non-Gaussian. In both cases, the rate of convergence is the standard $\sqrt{n}$. We use these results to test the GJR hypothesis $\mathrm{H}_0: \gamma = 0$ and the hypothesis $\mathrm{H}_0:\alpha_1=0$ that returns above the threshold do not affect the volatility (Corollary \ref{corr: t-test}).
\medskip

As is standard in the GARCH literature (see, e.g., \citealt{FrancqZakoian2004} and \citealt{straumann2005estimation}), we assume that the observed sample $x_0,\dots,x_n$ is generated by a realization of a two-sided strictly stationary and ergodic TS-AGARCH process $\{x_t:t\in\mathbb{Z}\}$ with $\theta_0$ satisfying the stationarity condition in \eqref{eq: def phi}. That a stationary and ergodic one-sided process $\{x_t\}$ can be extended to a two-sided process $\{x_t:t\in\mathbb{Z}\}$ whose finite-dimensional distributions coincide is a well-known fact, see, e.g., Chapter 25 in \citet{kallenberg1997foundations}. 

To prove Theorems \ref{thm: consistency}--\ref{thm: asym normality}, we approximate the feasible likelihood $\hat{\mathcal{L}}_n$ by the unobserved quasi log-likelihood function $\mathcal{L}_n$ given by,
\begin{align}
    \mathcal{L}_n(\theta)&=\sum^n_{t=1}\ell_t(\theta), \quad  \ell_t(\theta)= -\frac{1}{2}\left(\log\left(\sigma^2_t(\theta)\right)+\frac{x^2_t}{\sigma^2_t(\theta)}\right), \qquad \theta \in \Theta, \label{eq: loglik} \\
    \sigma^2_t(\theta) &= \frac{\omega}{1-\beta} + \sum_{i=1}^\infty\beta^{i-1}(\alpha_1 \I(x_{t-i} \geq \gamma)+\alpha_2 \I(x_{t-i} < \gamma)) (x_{t-i}-\gamma)^2, \label{eq: sig2(theta)}
\end{align}
and analyze first the infeasible estimator $\theta_n=\arg\max_{\theta\in\Theta}\mathcal{L}_n(\theta)$. Similar to \eqref{eq: def shat}--\eqref{eq: def hhat}, we denote the first and second (right) derivatives of $\mathcal{L}_n$ as $\mathcal{S}_n$ and $\mathcal{H}_n$, and we let $\mathcal{I}_n=-\mathcal{H}_n$ denote the observed information. Note that $\{\ell_t(\theta):t\in\mathbb{N}\}$ is a stationary and ergodic sequence for any $\theta\in\Theta$ by stationarity and ergodicity of $\{x_t : t \in \mathbb{Z}\}$, which simplifies the analysis compared to the feasible likelihood $\hat{\mathcal{L}}_n$. 

Strong consistency of $\hat\theta_n$ follows by establishing this property for $\theta_n$ and next showing that $\hat{\mathcal{L}}_n/n$ and $\mathcal{L}_n/n$ are uniformly close on $\Theta$, which proves $\theta_n-\hat\theta_n \asto 0$. Similarly, the asymptotic distribution is first derived for the infeasible estimator $\theta_n$, and then transferred to $\hat\theta_n$ by proving that $\sqrt{n}(\theta_n-\hat\theta_n)\asto0$. The first main result on $\hat\theta_n$ is the following theorem.
\begin{theorem} \label{thm: consistency}
    Suppose that $\theta_0\in\Theta$ and that condition \eqref{eq: def phi} of Theorem \ref{thm: geometric mixing and moments} holds. Then the QMLE is strongly consistent, i.e.\ $\hat \theta_n \asto \theta_0$ as $n \to \infty$.
\end{theorem}
By construction of $\sigma^2_t(\theta)$, note that the threshold parameter, $\gamma_0$, is identified even in the case where $\alpha_{1,0}=\alpha_{2,0}$ as it dictates the minimum point of the NIC, see also Figure \ref{fig: 4 NICs}. This is in contrast to the T-ARCH model, where identification of $\gamma_0$ requires $\alpha_{1,0}\neq \alpha_{2,0}$ (see, e.g., Theorem 3.1 of \citealt{li2015asymptotic}), and also the CTAR of \citet{Chan_Tsay1998} that requires the autoregressive coefficients to differ. In particular, the A-GARCH hypothesis $\mathrm{H}_0:\alpha_1=\alpha_2$ can be tested by standard tests, see Section \ref{sec: empirical section}. Moreover, $\theta_0$ remains identified when $\alpha_{1,0}=0$, i.e.\ when returns above $\gamma_0$ do not affect the volatility. In this case, $\gamma_0$ is identified from the left branch of the NIC, whose slope is positive since $\alpha_{2,0} >0$. \medskip

Our second main result given in Theorem \ref{thm: asym normality} establishes the asymptotic distribution of $\hat\theta_n=(\hat\omega_n,\hat\alpha_{1,n},\hat\alpha_{2,n},\hat\gamma_n,\hat\beta_n)^\top$ at the usual $\sqrt{n}$-rate. In the GARCH literature, asymptotic normality is typically established by relying on a mean value expansion of the score function and the first order condition $\mathcal{S}_n(\theta_n)=0$ for the maximizer $\theta_n$ of $\mathcal{L}_n$, which holds provided $\theta_0$ is an interior point of $\Theta$, to obtain
\begin{equation*}
    n^{1/2}(\theta_n-\theta_0)=-[n^{-1}\mathcal{H}_n(\theta_n^*)]^{-1}\,n^{-1/2}\mathcal{S}_n(\theta_0),
\end{equation*}
where $\theta_n^*$ is a point between $\theta_0$ and $\theta_n$. Then asymptotic normality would follow by showing that $n^{-1/2}\mathcal{S}_n(\theta_0)$ obeys a central limit theorem and that a uniform law of large numbers applies to $n^{-1}\mathcal{H}_n(\theta)$ (see, for instance, \citealt{straumann2005estimation} and \citealt{meitz2011parameter}). However, to justify the latter, one typically requires that $\mathcal{H}_n$ is a continuous function in $\theta$, which as outlined in Section \ref{sec: likelihood estimation} does not hold for the TS-AGARCH model. Furthermore, as $\mathcal{H}_n$ is not continuous, i.e.\ $\mathcal{L}_n$ is not twice continuously differentiable, it is not obvious that $\theta_n^\ast$ is measurable (see, e.g., \citealt{yang2026taylor} for sufficient conditions, which are not fulfilled in our case). Finally, when $\alpha_{1,0}=0$ is on the boundary of $\Theta$, the first order condition fails with positive probability. We avoid these complications by instead proving the validity of another expansion of the score. We utilize that the score contributions are Lipschitz continuous along line segments in a neighborhood of $\theta_0$ (see in particular Lemma \ref{lem: FTC} in the Appendix), and the second fundamental theorem of calculus for Lipschitz functions then yields, with $g_n(u)=\theta_0+u(\theta_n-\theta_0)$, $u\in[0,1]$,
\begin{equation}
    n^{-1/2}\mathcal{S}_n(\theta_n)=n^{-1/2}\mathcal{S}_n(\theta_0)-\left(\int_0^1n^{-1}\mathcal{I}_n(g_n(u))du\right)n^{1/2}(\theta_n-\theta_0). \label{eq:expan-score}
\end{equation}
Despite the discontinuities of $\mathcal{I}_n$, we prove uniform convergence of $n^{-1}\mathcal{I}_n(\theta)$ over shrinking neighborhoods of $\theta_0$, such that the integral term converges a.s.\ to the positive definite matrix $\Omega$ given in Theorem \ref{thm: asym normality}. If $\theta_0$ is an interior point of $\Theta$, then $\mathcal{S}_n(\theta_n)=0$ for $n$ sufficiently large, such that
\begin{equation*}
    n^{1/2}(\theta_n-\theta_0)=\left(\int_0^1n^{-1}\mathcal{I}_n(g_n(u))du\right)^{-1}\,n^{-1/2}\mathcal{S}_n(\theta_0),
\end{equation*}
and asymptotic normality follows as $n^{-1/2}\mathcal{S}_n(\theta_0)$ obeys a CLT. If instead $\alpha_{1,0}=0$, then $\theta_n=(\omega_n,\alpha_{1,n},\alpha_{2,n},\gamma_n,\beta_n)^\top$ may have $\alpha_{1,n}=0$ while the partial derivative of $\mathcal{L}_n$ with respect to $\alpha_1$ is negative. However, as $\theta_0$ is an interior point in the remaining coordinates, all entries of $\mathcal{S}_n(\theta_n)$ except the second one, $\mathcal{S}^{(2)}_n(\theta_n)=\partial\mathcal{L}_n(\theta_n)/\partial\alpha_1$, are zero for $n$ sufficiently large, while, as $\theta_n$ maximizes $\mathcal{L}_n$, it holds that
\begin{equation*}
    \mathcal{S}^{(2)}_n(\theta_n)\leq0\quad\text{if }\alpha_{1,n}=0,\qquad\text{and}\qquad\mathcal{S}^{(2)}_n(\theta_n)=0\quad\text{if }\alpha_{1,n}>0.
\end{equation*}
Inserting these relations in the expansion \eqref{eq:expan-score} determines $\mathcal{S}^{(2)}_n(\theta_n)$, and we solve explicitly for $n^{1/2}(\theta_n-\theta_0)$, which gives the limit in Theorem \ref{thm: asym normality}(b). 
\begin{theorem} \label{thm: asym normality}
    Suppose that the condition \eqref{eq: def phi} of Theorem \ref{thm: geometric mixing and moments} is satisfied, and that $\E[z^4_0]<\infty$. Then, as $n \to \infty$, the following holds for the QMLE.
    \begin{itemize}
        \item[(a)] If $\theta_0$ is an interior point of $\Theta$, then
        \begin{align}
            \sqrt{n}(\hat\theta_n-\theta_0)\dto \mathrm{N}(0,\Sigma),\qquad \Sigma = \frac{\mathbb{V}[z^2_0]}{2}\Omega^{-1}. \label{eq: def Sigma}
        \end{align}
        \item[(b)] If $\alpha_{1,0}=0$, and $(\omega_0,\alpha_{2,0},\gamma_0,\beta_0)$ is an interior point of $[\omega_L,\omega_U]\times[\alpha_L,\alpha_U]\times[\gamma_L,\gamma_U]\times[\beta_L,\beta_U]$, then
        \begin{align}
            \sqrt{n}(\hat\theta_n-\theta_0)\dto\hat\lambda:=Z-\frac{\min\{Z_2,0\}}{\Sigma_{22}}\,\Sigma \, e_2,\qquad Z\sim\mathrm{N}(0,\Sigma), \label{eq: boundary limit}
        \end{align}
        with $\Sigma$ as in \eqref{eq: def Sigma}, $e_2=(0,1,0,0,0)^\top$, $Z_2=e_2^\top Z$ and $\Sigma_{22}=e_2^\top\Sigma\, e_2$.
    \end{itemize}
     In both cases, the positive definite $\Omega$ is given by
     \begin{align*}
         \Omega=\frac{1}{2}\E\left[\frac{\partial \sigma^2_t(\theta_0)/\partial\theta}{\sigma^2_t}\left(\frac{{\partial \sigma^2_t(\theta_0)/\partial\theta}}{\sigma^2_t}\right)^\top\right],
     \end{align*}
     where $\sigma^2_t:=\sigma^2_t(\theta_0)$, and $\partial\sigma^2_t(\theta_0)/\partial\theta$ is obtained from \eqref{eq: def sighat2 deriv} by replacing $t$ by $\infty$ in the summation and $\hat\sigma^2_{t-i}(\theta)$ by $\sigma^2_{t-i}(\theta)$, and evaluating at $\theta=\theta_0$.
\end{theorem}
We have the following remarks.
\begin{remark}
    In case (a) of Theorem \ref{thm: asym normality}, the threshold estimator $\hat\gamma_n$ is asymptotically normal at the standard $\sqrt{n}$-rate with an asymptotic variance that can be estimated directly from the data (see Corollary \ref{corr: t-test}). This is different from the asymptotic behavior of the threshold estimator for the T-ARCH model in \citet{li2015asymptotic}. In particular, their Theorem 3.3 shows that the threshold estimator is $n$-consistent with a non-Gaussian asymptotic distribution, which is in general analytically intractable and must be simulated on a case-by-case basis (see the comments below their Theorem 3.3).
\end{remark}

\begin{remark} \label{remark: boundary}
    Case (b) of Theorem \ref{thm: asym normality} is an instance of the general result of \citet{andrews1999estimation}, where the limit is the projection of $Z$ onto the cone $\mathbb{R}\times[0,\infty)\times\mathbb{R}^3$ with respect to the norm $(\lambda^\top\Omega\lambda)^{1/2}$, see also \citet{francq2007quasi} for GARCH models with coefficients equal to zero. It implies that $\sqrt{n}\,\hat\alpha_{1,n}\dto\max\{Z_2,0\}$, which is a censored normal distribution with an atom of mass $1/2$ at zero. Moreover, for each $j\neq2$,
    \begin{equation*}
        \frac{\sqrt{n}(\hat\theta_{j,n}-\theta_{j,0})}{\Sigma_{jj}^{1/2}}\dto(1-\eta_j^2)^{1/2}\xi_j+\eta_j\max\{\xi,0\},\qquad\eta_j=\frac{\Sigma_{j2}}{(\Sigma_{jj}\Sigma_{22})^{1/2}},
    \end{equation*}
    where $\xi$ and $\xi_j$ are independent $\mathrm{N}(0,1)$-distributed random variables, and $\eta_j$ is the correlation between the $j$th and the second coordinate of $Z$. Hence, when $\alpha_{1,0}=0$, the estimator $\hat\theta_{j,n}$ of each of the remaining parameters is not asymptotically normal, unless $\eta_j=0$.
\end{remark}

An interesting application of the result in Theorem \ref{thm: asym normality} is to test hypotheses about the NIC by $t$-statistics. We state the following corollary whose proof is immediate from Slutsky's theorem, Theorem \ref{thm: asym normality} and Remark \ref{remark: boundary}.
\begin{corollary} \label{corr: t-test}
    Let $\hat{\Sigma}_n$ be a consistent estimator of $\Sigma$ defined in \eqref{eq: def Sigma}, and for $j=1,\dots,5$, let
    \begin{equation*}
        T_{j,n}=n^{1/2}(\hat{\theta}_{j,n}-\theta_{j,0})/(\hat{\Sigma}_n^{(j,j)})^{1/2}
    \end{equation*}
    denote the $t$-statistic for $\mathrm{H}_0:\theta_j=\theta_{j,0}$, where $\hat{\Sigma}_n^{(j,j)}$ is the $j$th diagonal entry of $\hat{\Sigma}_n$. Then, as $n\to\infty$, the following holds.
    \begin{itemize}
        \item[(a)] Under the assumptions of Theorem \ref{thm: asym normality}(a), $T_{j,n}\dto\mathrm{N}(0,1)$ for $j=1,\dots,5$.
        \item[(b)] Under the assumptions of Theorem \ref{thm: asym normality}(b), $T_{2,n}\dto\max\{\xi,0\}$, and $T_{j,n}\dto{(1-\eta_j^2)^{1/2}}\xi_j+\eta_j\max\{\xi,0\}$ for $j\neq2$, with $\xi$, $\xi_j$ and $\eta_j$ as in Remark \ref{remark: boundary}.
    \end{itemize}
\end{corollary}
Throughout the remainder of the paper, $\tau_n$ and $t_n$ denote the $t$-statistics for $\gamma$ and $\alpha_1$, respectively, i.e.
\begin{equation*}
    \tau_n=T_{4,n}=n^{1/2}(\hat{\gamma}_n-\gamma_0)/(\hat{\Sigma}_n^{(4,4)})^{1/2}\qquad\text{and}\qquad t_n=T_{2,n}=n^{1/2}(\hat{\alpha}_{1,n}-\alpha_{1,0})/(\hat{\Sigma}_n^{(2,2)})^{1/2}.
\end{equation*}
Since $\gamma_0=0$ is not a boundary point of the parameter space when $\gamma_L<0<\gamma_U$, the GJR hypothesis $\mathrm{H}_0:\gamma=0$ can be tested by $\tau_n$ with $\mathrm{N}(0,1)$ critical values when $\alpha_{1,0}>0$ by Corollary \ref{corr: t-test}(a), and Remark \ref{remark: GJR at boundary} below shows that this test remains valid when $\alpha_{1,0}=0$. Corollary \ref{corr: t-test}(b) can be used to test whether returns above the threshold affect the volatility, i.e.\ to test $\mathrm{H}_0:\alpha_1=0$ against $\alpha_1>0$. For $a\in(0,1/2)$, the test that rejects $\mathrm{H}_0$ when $t_n>z_{1-a}$, the $(1-a)$-quantile of the $\mathrm{N}(0,1)$-distribution, has asymptotic size $a$, since $t_n\dto\max\{\xi,0\}$ under $\mathrm{H}_0$. Hence, the test of $\mathrm{H}_0:\alpha_1=0$ uses the usual one-sided critical values, e.g.\ $1.64$ at the $5\%$ level.

The following remark gives a standard choice of $\hat{\Sigma}_n$. The proof of its consistency is standard, but lengthy, and is therefore omitted.
\begin{remark} \label{remark: covar estimators}
The covariance $\Sigma$ defined in \eqref{eq: def Sigma} can be consistently estimated by the well-known sandwich formula
    \begin{align}
         \hat{\Sigma}_n = [n^{-1}\hat{\mathcal{H}}_n(\hat\theta_n)]^{-1} n^{-1}\sum^n_{t=1}\hat s_t(\hat\theta_n) \hat s_t(\hat\theta_n)^\top[n^{-1}\hat{\mathcal{H}}_n(\hat\theta_n)]^{-1}, \label{eq: def covar estimator 2}
    \end{align}
    with $\hat{\mathcal{H}}_n$ defined in \eqref{eq: def hhat} and $\hat s_t$ defined in \eqref{eq: def shat}. Consistency holds under the assumptions of Theorem \ref{thm: asym normality}, including the case $\alpha_{1,0}=0$.
\end{remark}
\begin{remark} \label{remark: GJR at boundary}
    If $\alpha_{1,0}=0$, Corollary \ref{corr: t-test}(b) gives $\tau_n\dto\tau:=(1-\eta^2)^{1/2}\xi_\gamma+\eta\max\{\xi,0\}$, with $\xi_\gamma:=\xi_4$ and $\eta:=\eta_4$, which is not $\mathrm{N}(0,1)$ unless $\eta=0$. Nevertheless, the two-sided test that rejects the GJR hypothesis when $|\tau_n|>z_{1-a/2}$ still controls the size asymptotically. To see this, write $\tau=X\I(\xi<0)+Y\I(\xi\geq0)$, with $X:=(1-\eta^2)^{1/2}\xi_\gamma\sim\mathrm{N}(0,1-\eta^2)$ independent of $\xi$, and $Y:=X+\eta\xi\sim\mathrm{N}(0,1)$. As $(Y,\xi)=_{\text{d}}(-Y,-\xi)$, $\P(|Y|>y,\,\xi\geq0)=\P(|Y|>y)/2$ for any $y$, and as $\tau$ has a continuous distribution,
    \begin{align*}
        \P(|\tau_n|>z_{1-a/2})&\to\tfrac12\P(|X|>z_{1-a/2})+\tfrac12\P(|Y|>z_{1-a/2})\\
        &=1-\Phi\big(z_{1-a/2}/(1-\eta^2)^{1/2}\big)+a/2\in(a/2,a],
    \end{align*}
    with equality to $a$ if and only if $\eta=0$. Hence, the test can be applied without knowing whether the nuisance parameter $\alpha_1$ is on the boundary, which otherwise complicates inference in this type of problem (see \citealt{andrews2001testing} and \citealt{cavaliere2022bootstrap}). It is conservative when $\alpha_{1,0}=0$, but its asymptotic size is close to $a$ when $\eta$ is small (see Section \ref{sec: Monte Carlo}).
\end{remark}
\section{Monte Carlo simulations} \label{sec: Monte Carlo}
In this section, we investigate by Monte Carlo (MC) simulations how well the asymptotic results in Theorem \ref{thm: asym normality} and Corollary \ref{corr: t-test} approximate the finite sample distributions, both when $\theta_0$ is an interior point of $\Theta$ and when $\alpha_{1,0}=0$ is on the boundary of $\Theta$. We focus on the estimator of the threshold parameter $\gamma$, the test of the GJR hypothesis $\mathrm{H}_0:\gamma=0$, and the test of $\mathrm{H}_0:\alpha_1=0$. The setup is as follows.

The data generating process (DGP) uses a burn-in sample of $b=1000$ observations. That is, the $x_t$'s are generated recursively as
\begin{align}
    x_t &= \sigma_t z_t, \qquad t = -(b-1),\dots,-1,0,1,\dots,n, \notag
    \\ \sigma^2_t&=  \begin{cases}
    \omega_0 + \alpha_{1,0} (x_{t-1} - \gamma_0)^2 + \beta_0 \sigma^2_{t-1}, \quad \text{if}\quad x_{t-1} \geq \gamma_0, \\[0.05in]
    \omega_0 + \alpha_{2,0} (x_{t-1} - \gamma_0)^2 + \beta_0 \sigma^2_{t-1},  \quad \text{if}\quad x_{t-1} < \gamma_0,
    \end{cases}
\end{align}
with $\{z_t:t=-b,\dots,n\}$ an iid sequence of mean zero random variables with $\mathbb{V}[z_0]=1$, and initial values chosen as $\sigma_{-b}=\sqrt{\omega_0/(1-\beta_0)}$ and $x_{-b}=\sigma_{-b}z_{-b}$. The true parameter values are $\omega_0=1$, $\alpha_{2,0}=0.4$, $\gamma_0 = 0$ and $\beta_0 = 0.8$, and either $\alpha_{1,0}=0.05$, such that $\theta_0$ is an interior point of $\Theta$, or $\alpha_{1,0}=0$, such that $\alpha_{1,0}$ is on the boundary of $\Theta$. In particular, the GJR hypothesis $\gamma_0=0$ holds in all designs. The innovations are either $\mathrm{N}(0,1)$-distributed (such that $\hat\theta_n$ is the MLE) or scaled $t_5$-distributed, i.e.\ Student's $t$-distribution with $5$ degrees of freedom scaled to have unit variance (such that $\hat\theta_n$ is a QMLE) with $\E[z_0^4]=9<\infty$. This gives four designs: (i) $z_t\sim\mathrm{N}(0,1)$ and $\alpha_{1,0}=0.05$, (ii) scaled $t_5$ and $\alpha_{1,0}=0.05$, (iii) $z_t\sim\mathrm{N}(0,1)$ and $\alpha_{1,0}=0$, and (iv) scaled $t_5$ and $\alpha_{1,0}=0$.

The true parameter values ensure that the stationarity condition \eqref{eq: def phi} is satisfied in all four designs. In terms of finite moments, let $h(p)$ denote the left-hand side of the moment condition \eqref{eq: moment condition} evaluated at $\theta_0$, i.e.
\begin{align*}
    h(p):=\E[(\beta_0+(\alpha_{1,0} \I(z_0 \geq 0)+\alpha_{2,0}\I(z_0<0))z^2_0)^p],\qquad p >0.
\end{align*}
As $\gamma_0=0$, $\sigma^2_t$ follows the SRE $\sigma^2_t=A_t\sigma^2_{t-1}+\omega_0$ of Section \ref{sec: stochastic properties}, where it is well known that $\E[\sigma_0^{2p}]<\infty$ if and only if $h(p)<1$. For both distributions of $z_0$, it holds that $h(1)=(\alpha_{1,0}+\alpha_{2,0})/2+\beta_0$, which equals $1.025$ when $\alpha_{1,0}=0.05$ and $1$ when $\alpha_{1,0}=0$, while $h(1/2)<1$ in all four designs. Hence, $\E[|x_0|]<\infty$, while $\E[x_0^2]=\infty$. This is allowed by Theorem \ref{thm: asym normality}, which imposes no moment conditions on $x_t$, only that $\E[z_0^4]<\infty$.

The estimate $\hat{\theta}_n$ is obtained by maximizing the feasible log-likelihood \eqref{eq: def Lhat(theta)} with the backcast initial value $\hat\sigma^2_0$ given in Footnote \ref{fn: backcast}, using the L-BFGS-B algorithm with the analytical score in \eqref{eq: def shat}, over the parameter space $\Theta=[0.001,20]\times[0,2]\times[0.001,2]\times[-5,5]\times[0.001,0.999]$. The asymptotic covariance $\Sigma$ is estimated by the sandwich estimator in \eqref{eq: def covar estimator 2}. The sample size is $n=5000$, which is close to the sample size of the empirical application in Section \ref{sec: empirical section}, and the number of MC replications is $M=10^4$. Results for $n=20000$, $n=50000$ and $n=200000$ are given in Table \ref{tab: mc rejection} in Appendix \ref{sec: supp mc}.

\begin{figure}[H]
    \centering
    \includegraphics[width=\linewidth]{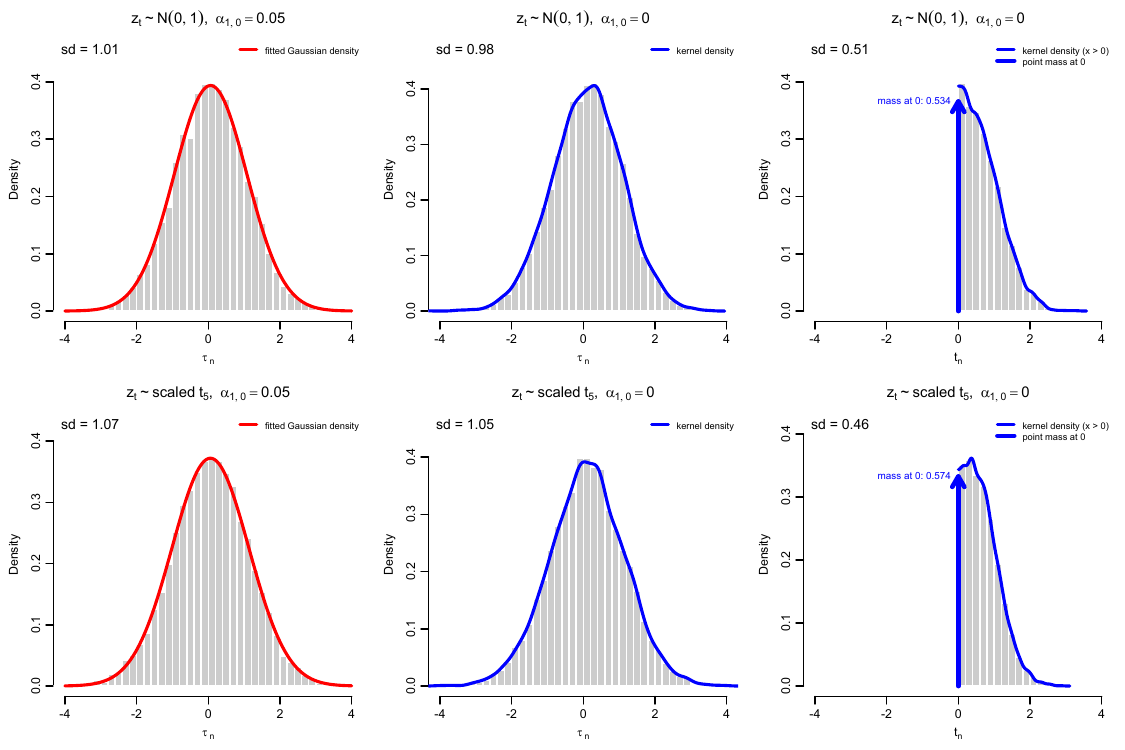}
    \caption{MC distributions for $n=5000$ of $\tau_n$ for $\alpha_{1,0}=0.05$ (left) and $\alpha_{1,0}=0$ (middle), and of $t_n$ for $\alpha_{1,0}=0$ (right). Rows: $z_t\sim\mathrm{N}(0,1)$ (top) and scaled $t_5$ (bottom). The MC standard deviation is reported as \textsf{sd}. The red curves are fitted Gaussian densities, and the blue curves are Gaussian kernel density estimates with Silverman's rule-of-thumb bandwidth. In the right column, the arrow indicates the point mass at zero, whose size, i.e.\ the share of replications with $\hat\alpha_{1,n}=0$, is reported in the panel, while the histogram and the kernel density estimate (reflected at zero) describe the distribution on $(0,\infty)$.}
    \label{fig: mc}
\end{figure}

For each design, we consider the $t$-statistic $\tau_n$ for the GJR hypothesis $\mathrm{H}_0:\gamma=0$ and, in the boundary designs, the $t$-statistic $t_n$ for $\mathrm{H}_0:\alpha_1=0$, both with $\hat\Sigma_n$ given by \eqref{eq: def covar estimator 2}. Figure \ref{fig: mc} shows their MC distributions. The MC distributions of $\sqrt{n}(\hat\gamma_n-\gamma_0)$ (Figure \ref{fig: mc gamma} in the Appendix) are close to Gaussian in all four designs, and their standard deviations are nearly unchanged for larger $n$, which indicates that the $\sqrt{n}$-rate is correct.

\noindent In the interior designs, the $\mathrm{N}(0,1)$-approximation of $\tau_n$ is accurate (left column of Figure \ref{fig: mc}). The two-sided test at the $5\%$ level rejects in $5.4\%$ of the replications for $\mathrm{N}(0,1)$ and in $6.8\%$ for scaled $t_5$ innovations, and the slight over-rejection for scaled $t_5$ innovations vanishes as $n$ increases. In the boundary designs, the limit $\tau$ of $\tau_n$ in Remark \ref{remark: GJR at boundary} is close to $\mathrm{N}(0,1)$, as $\eta\approx0.17$ for $\mathrm{N}(0,1)$ and $\eta\approx0.12$ for scaled $t_5$ innovations,\footnote{We compute $\eta=\Sigma_{42}/(\Sigma_{44}\Sigma_{22})^{1/2}$ with $\Sigma$ proportional to $\Omega^{-1}$, where $\Omega$ is approximated by the average of $\frac12(\partial\sigma^2_t(\theta_0)/\sigma^2_t)(\partial\sigma^2_t(\theta_0)/\sigma^2_t)^\top$ over a simulated path of length $10^6$.} and the asymptotic rejection probability of the test is $4.8\%$ and $4.9\%$, respectively. Accordingly, the MC distributions of $\tau_n$ in the middle column are close to $\mathrm{N}(0,1)$, and the test rejects in $4.8\%$ and $6.3\%$ of the replications, where the latter falls to $5.0\%$ for $n=50000$. Hence, the test of the GJR hypothesis based on $\mathrm{N}(0,1)$ critical values is valid also when $\alpha_{1,0}=0$.

In the boundary designs, the limit $\max\{\xi,0\}$ of $t_n$ has an atom of mass $1/2$ at zero. The share of replications with $\hat\alpha_{1,n}=0$ is $53.4\%$ for $\mathrm{N}(0,1)$ and $57.4\%$ for scaled $t_5$ innovations, and the right tail of $t_n$ is thinner than that of the limit (right column of Figure \ref{fig: mc}). Consequently, the one-sided test of $\mathrm{H}_0:\alpha_1=0$, which rejects when $t_n>1.64$, rejects in only $3.0\%$ and $1.7\%$ of the replications. Thus, the test controls the size, but it is conservative, in particular for heavy-tailed innovations, and the right tail of $t_n$ approaches that of the limit only slowly as $n$ increases (see the Appendix).

\section{Empirical evidence} \label{sec: empirical section}
In this section, we use the TS-AGARCH and the asymptotic theory of the QMLE to study the asymmetric response of volatility to returns. In Section \ref{sec: emp indices}, we test the GJR hypothesis $\gamma=0$, the A-GARCH hypothesis $\alpha_1=\alpha_2$ and the hypothesis $\alpha_1=0$ for five stock indices, and show by simulation that, when the data are generated by the TS-AGARCH, the GJR and the A-GARCH overstate their single source of asymmetry, as we also find for the indices. In Section \ref{sec: emp assets}, we show that, across a wide range of assets, the location of the minimum of the NIC and the ordering of the slopes follow the correlation of the asset with the stock market, as the market model suggests. The estimated NIC of an asset, which uses only its own return, thus reveals how the asset moves with the market. To simplify notation, we drop the subscript $n$ for the parameter estimates, but not for the test statistics.

\subsection{Stock indices}\label{sec: emp indices}
We consider daily log returns (in percent) of the SP500, FTSE 100, EURO STOXX 50 (ESTX 50), Nikkei 225 and Swiss Market Index (Swiss MI) stock indices from April 10th 2007 to August 14th 2026, which gives between $4728$ and $4885$ observations per index. The data were obtained from \textsf{finance.yahoo.com} with the \texttt{R}-package \texttt{quantmod}. For each index, we estimate the TS-AGARCH in \eqref{eq:STGARCH_proc}.\footnote{The returns are slightly autocorrelated. With a constant, a first-order autoregression or both in the conditional mean, which our theory does not cover, the estimates of $\gamma$ change by at most $0.02$, and each of the tests below leads to the same conclusion for every index. For the TS-AGARCH in \eqref{eq:STGARCH_proc}, the Ljung--Box tests with 20 lags do not reject at the $5\%$ level for $\hat z_t=x_t/\sigma_t(\hat\theta)$ or $\hat z_t^2$ for any index. Figure \ref{fig: zhat for all series} in the Appendix shows the autocorrelations of $\hat z_t$ and $\hat z_t^2$ and a QQ-plot of $\hat z_t$.} We also estimate the GJR ($\gamma=0$) and the A-GARCH ($\alpha_1=\alpha_2$). All three models are estimated by maximizing the feasible quasi log-likelihood \eqref{eq: def Lhat(theta)} with the backcast initial value $\hat\sigma^2_0$ of Footnote \ref{fn: backcast}, using the L-BFGS-B algorithm with the lower bound zero for $\alpha_1$, as in Section \ref{sec: Monte Carlo}. The optimizer for the TS-AGARCH is also started at the GJR and A-GARCH estimates, so that its log-likelihood is at least as large as theirs. Standard errors are computed with the sandwich formula \eqref{eq: def covar estimator 2}, where $\hat{\mathcal{H}}$ is obtained by numerical differentiation of the analytical score.

Note that our asymptotic results also cover the QMLEs of the GJR and the A-GARCH, as they are both obtained by imposing linear restrictions on $\theta$. Specifically, with $\theta=A\varphi$, where $\varphi=(\omega,\alpha_1,\alpha_2,\beta)^\top$ for the GJR and $\varphi=(\omega,\alpha,\gamma,\beta)^\top$ for the A-GARCH, and $A$ is the $5\times4$ matrix that sets $\gamma=0$ and $\alpha_1=\alpha_2=\alpha$, respectively, Theorem \ref{thm: asym normality}(a) yields that the QMLE of $\varphi$ is strongly consistent and asymptotically normal with covariance $(\mathbb{V}[z_0^2]/2)(A^\top\Omega A)^{-1}$, provided $\theta$ is not on the boundary of $\Theta$. Similarly, Theorem \ref{thm: asym normality}(b) covers the GJR with $\alpha_{1,0}=0$, in line with the results of \citet{francq2019qml} for asymmetric GARCH models with the threshold fixed at zero. 

Within the TS-AGARCH, we test the GJR hypothesis $\mathrm{H}_0:\gamma=0$ by $\tau_n$, the hypothesis $\mathrm{H}_0:\alpha_1=0$ against $\alpha_1>0$ by $t_n$ with the one-sided critical value $1.64$ (see Corollary \ref{corr: t-test}), and the A-GARCH hypothesis $\mathrm{H}_0:\alpha_1=\alpha_2$ by
\begin{align*}
    \kappa_n=\frac{n^{1/2}v^\top\hat{\theta}}{(v^\top\hat{\Sigma}v)^{1/2}},\qquad v=(0,-1,1,0,0)^\top,
\end{align*}
which is asymptotically $\mathrm{N}(0,1)$ under $\mathrm{H}_0$ by Theorem \ref{thm: asym normality}(a). By Remark \ref{remark: GJR at boundary}, the test based on $\tau_n$ is valid whether or not $\alpha_{1,0}=0$, and $\kappa_n$ is not affected by the boundary, since $\alpha_1=\alpha_2$ implies $\alpha_1\geq\alpha_L>0$. For the same reason, the hypotheses $\alpha_1=0$ and $\alpha_1=\alpha_2$ cannot both hold. Within the GJR and the A-GARCH, we compute the corresponding $t$-statistics for the parameters that each model has. Panel A of Table \ref{tab: indices} reports the estimates of the three models, together with the test statistics that each model allows.
\begin{itemize}
  \item \textit{TS-AGARCH.} The estimate $\hat\gamma$ is between $0.32$ and $0.60$, and the GJR hypothesis is rejected in favor of $\gamma>0$ for all five indices. The markets are thus calmest after returns of about half a percent. The hypothesis $\alpha_1=0$ is rejected for the SP500 and the Nikkei 225, and the A-GARCH hypothesis is rejected for the FTSE 100, the ESTX 50 and the Nikkei 225, in all cases with $\hat\alpha_2>\hat\alpha_1$. Hence, for the FTSE 100 and the ESTX 50 the NIC is nearly flat to the right of $\hat\gamma$ and steep to its left, for the Nikkei 225, $0<\alpha_1<\alpha_2$, and for the SP500, $\alpha_1>0$, while equal slopes are not rejected. For the Swiss MI, neither hypothesis about the slopes is rejected, which is logically inconsistent, and can possibly be explained by an imprecise estimate of $\alpha_1$.
  \item \textit{GJR.} The estimate of $\alpha_1$ is close to zero, and $\alpha_1=0$ is not rejected for four of the five indices, the exception being the Nikkei 225. Equal slopes are rejected for all five indices, with differences between the slopes of $0.16$ to $0.22$, against $0.06$ to $0.13$ in the TS-AGARCH. Taken at face value, the GJR suggests that volatility does not respond to returns above zero and that the NIC is much steeper to the left. For the SP500, the GJR and the TS-AGARCH reach opposite conclusions on both hypotheses. That is, $\alpha_1 = \alpha_2$ is rejected with the GJR while $\alpha_1 = 0$ is not in this model, and the converse is true for the TS-AGARCH. We give a possible explanation of this below.
  \item \textit{A-GARCH.} The minimum of the NIC is at $0.60$ to $0.81$, to the right of the TS-AGARCH estimate for all five indices. Again, we provide a possible explanation below.
\end{itemize}

\begin{table}[H]
\centering
\resizebox{\textwidth}{!}{\begin{tabular}{l|cc|c|ccc}
\hline\hline
 & \multicolumn{2}{c|}{GJR} & A-GARCH & \multicolumn{3}{c}{TS-AGARCH}\\
 & $\hat\alpha_{1}$ & $\hat\alpha_{2}-\hat\alpha_{1}$ & $\hat\gamma$ & $\hat\alpha_{1}$ & $\hat\alpha_{2}-\hat\alpha_{1}$ & $\hat\gamma$\\
\hline
\multicolumn{7}{l}{\textit{A. Data: estimates ($t$-statistics $t_n$, $\kappa_n$ and $\tau_n$ in parentheses)}}\\
SP500 & 0.005 (0.28) & 0.206 (6.19) & 0.60 (6.49) & 0.073 (1.69) & 0.076 (1.53) & 0.51 (4.06)\\
FTSE 100 & 0.001 (0.13) & 0.206 (7.36) & 0.64 (6.16) & 0.013 (1.02) & 0.134 (4.59) & 0.32 (2.87)\\
ESTX 50 & 0.000 (0.00) & 0.211 (6.40) & 0.81 (9.52) & 0.015 (0.73) & 0.126 (3.30) & 0.60 (4.09)\\
Nikkei 225 & 0.042 (2.74) & 0.164 (5.17) & 0.71 (5.68) & 0.072 (3.22) & 0.075 (2.36) & 0.47 (3.92)\\
Swiss MI & 0.013 (0.57) & 0.219 (7.04) & 0.67 (7.74) & 0.061 (1.00) & 0.061 (0.72) & 0.58 (2.71)\\
\hline
\multicolumn{7}{l}{\textit{B. Simulated from the TS-AGARCH estimates in A: average estimates [rejection frequency in \%]}}\\
SP500 & 0.026 [55] & 0.225 [100] & 0.60 [100] & 0.071 [82] & 0.084 [53] & 0.49 [98]\\
FTSE 100 & 0.006 [7] & 0.204 [100] & 0.62 [100] & 0.015 [15] & 0.132 [97] & 0.33 [80]\\
ESTX 50 & 0.004 [3] & 0.232 [100] & 0.78 [100] & 0.016 [11] & 0.130 [97] & 0.56 [96]\\
Nikkei 225 & 0.042 [95] & 0.171 [100] & 0.73 [100] & 0.075 [97] & 0.072 [48] & 0.49 [78]\\
Swiss MI & 0.015 [28] & 0.212 [100] & 0.66 [100] & 0.059 [62] & 0.068 [43] & 0.55 [95]\\
\hline\hline
\end{tabular}}
\caption{GJR, A-GARCH and TS-AGARCH, all with a zero conditional mean. Panel A: estimates for the indices, with the $t$-statistics $t_n$ for $\mathrm{H}_0:\alpha_1=0$ (one-sided), $\kappa_n$ for $\mathrm{H}_0:\alpha_1=\alpha_2$ and $\tau_n$ for $\mathrm{H}_0:\gamma=0$ in parentheses, each computed within the model in the column. Panel B: means of the estimates over $M=1000$ samples simulated from the TS-AGARCH estimates in Panel A, with innovations resampled from $\hat z_t$, and, in brackets, rejection frequencies (in \%) of the tests at the $5\%$ level.}
\label{tab: indices}
\end{table}

The GJR and the A-GARCH both have only one source of asymmetry, whereas the TS-AGARCH allows both sources. When the true model is TS-AGARCH, this induces overestimation of the asymmetric effects in the restricted models. The GJR, whose NIC has its minimum at zero, fits a DGP with $\gamma >0$ by flattening its right branch, i.e.\ by lowering $\alpha_1$, possibly to zero, and by steepening its left branch. The A-GARCH, whose slopes are equal, can make its NIC steeper for negative than for positive returns only by moving its minimum to the right in order to fit a DGP with $\alpha_1 < \alpha_2$.

To see whether this explains the estimates in Panel A, we simulate, for each index, $M=1000$ samples from the TS-AGARCH with the estimated parameters and the sample size of that index. The innovations are drawn with replacement from the standardized residuals $\hat z_t$ of the index, and a burn-in sample of $1000$ observations is discarded. In each sample, we estimate the three models as above. Panel B of Table \ref{tab: indices} reports the results.
\begin{itemize}
  \item \textit{GJR.} On average, the GJR estimates $\alpha_1$ at between a quarter and three fifths of its true value, e.g.\ at $0.026$ for the SP500, where the true value is $0.073$. Within the GJR, $\alpha_1=0$ is not rejected in $45\%$ of the simulated samples for the SP500, in $72\%$ for the Swiss MI and in more than $90\%$ for the FTSE 100 and the ESTX 50, although $\alpha_{1,0}>0$ in all designs. The GJR estimates the difference between the slopes at $0.17$ to $0.23$, between $1.5$ and $3.5$ times the true difference.
  \item \textit{A-GARCH.} The A-GARCH places the minimum at $0.60$ to $0.78$ on average, $0.08$ to $0.30$ to the right of the true location.
\end{itemize}
The GJR and A-GARCH estimates on the actual series are thus what one would expect if the TS-AGARCH were the true model: the GJR estimates $\alpha_1$ close to zero for all five indices, and for the FTSE 100, the ESTX 50 and the Nikkei 225, where the A-GARCH hypothesis is rejected, the A-GARCH places the minimum $0.21$ to $0.32$ to the right of the TS-AGARCH estimate, against $0.09$ for the SP500 and the Swiss MI. Hence, the conclusion of the GJR that volatility does not respond to returns above zero reflects that it cannot capture a minimum of the NIC away from zero, and part of the location estimated by the A-GARCH reflects that it cannot capture different slopes. Only the TS-AGARCH separates the two sources of asymmetry.

\subsection{The location and slopes across assets}\label{sec: emp assets}
The location of the minimum of the NIC is positive for all five stock indices: volatility is lowest after a moderately positive return. In this section, we investigate whether this is a general phenomenon and, in particular, whether the asymmetry of an individual asset is inherited from the market or from news specific to the asset.
To motivate the analysis, we consider the market model of \citet{sharpe1963simplified}, which describes the return of an individual asset $x_{i,t}$, at time $t$, as
\begin{align}
    x_{i,t} = b_i x_{\text{m},t} + u_{i,t}, \label{eq:market-model-CAPM}  
\end{align}
where $b_i$ is the market beta of the asset in the CAPM, with $x_{\text{m},t}$ representing the market return, and $u_{i,t}$ is idiosyncratic news, independent of the market. Suppose that the market $x_{\text{m},t}$ follows the TS-AGARCH in \eqref{eq:STGARCH_proc} with parameters $\omega_\text{m}$, $\alpha_{1,\text{m}}$, $\alpha_{2,\text{m}}$, $\gamma_\text{m}$ and $\beta_\text{m}$. The individual asset in \eqref{eq:market-model-CAPM} inherits market-level asymmetry through $b_i$, while the idiosyncratic part may carry asymmetry of its own. An asset's comovement with the market therefore separates the two sources of asymmetry. 

Consider initially the market part alone, $u_{i,t}=0$, and assume $b_i \neq 0$, such that the individual asset return in \eqref{eq:market-model-CAPM} reduces to a TS-AGARCH with $\gamma_i$ and $(\alpha_{1,i},\alpha_{2,i})$ in \eqref{eq:STGARCH_proc} given by
\begin{align}
    \gamma_i = b_i \gamma_\text{m}, \qquad (\alpha_{1,i},\alpha_{2,i}) = \begin{cases}
        (\alpha_{1,\text{m}},\alpha_{2,\text{m}}) \quad &\text{if} \quad b_i > 0, \\
        (\alpha_{2,\text{m}},\alpha_{1,\text{m}}) \quad &\text{otherwise}.
    \end{cases} \label{eq:parameter-mapping-CAPM-implied}
\end{align}
The asset is calmest when the market is calmest, since $x_{\text{m},t-1}=\gamma_{\text{m}}$ yields by \eqref{eq:market-model-CAPM} that $x_{i,t-1}=b_i \gamma_{\text{m}}=\gamma_i$. Note that if $b_i<0$, the NIC of the asset is, up to a shift, the mirror image of that of the market: the location changes sign and the slopes are interchanged. Thus, one would expect hedges of the market, such as the short S\&P 500 fund below, for which $b_i \approx - 1$, to have a $\gamma_i \approx - \gamma_\text{m}$ and opposite sign of the tilt of the market defined as
\begin{align*}
    \lambda_{i} := \frac{\alpha_{2,i}-\alpha_{1,i}}{\alpha_{2,i}+\alpha_{1,i}} = \text{sgn}(b_i)\lambda_{\text{m}}, \label{eq:tilt-CAPM}
\end{align*}
where $\lambda_{\text{m}}$ denotes the market tilt. Rewriting $\gamma_i$ and $\lambda_i$ in terms of their correlation with the market $\rho_{i,\text{m}}$, the following formulas that link the parameters to the market parameters are obtained  
\begin{equation}
  \frac{\gamma_i}{\sigma_i}=\frac{\gamma_{\text{m}}}{\sigma_{\text{m}}} \, \rho_{i,\text{m}}, \qquad \lambda_{i} =\lambda_{\text{m}} \, \rho_{i,\text{m}}, \label{eq: location benchmark}
\end{equation}
where $\rho_{i,\text{m}}\in\{-1,1\}$ for $u_{i,t}=0$, and $\gamma_{\text{m}}>0$ and $\lambda_{\text{m}}>0$ by Section \ref{sec: emp indices}. In general, $u_{i,t}\neq0$ and $\rho_{i,\text{m}}\in[-1,1]$, but if the market independent part of \eqref{eq:market-model-CAPM} is small or symmetric, one would expect the estimates of a TS-AGARCH fitted to the asset return to follow \eqref{eq: location benchmark} with the estimated correlation $\hat\rho_{i,\text{m}}$. We investigate this empirically below.

\begin{figure}[H]
\centering
\includegraphics[width=\textwidth]{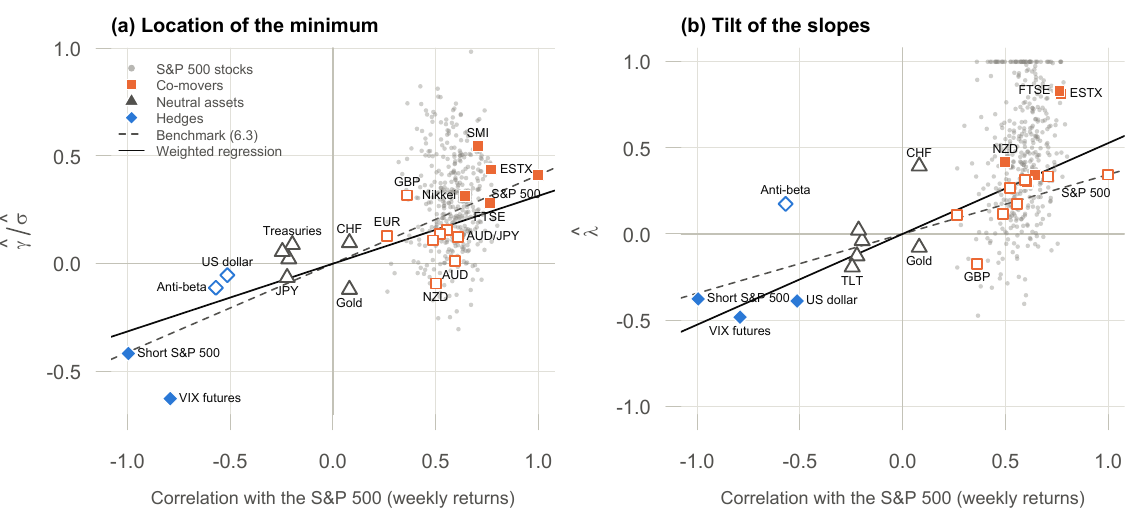}
\caption{Location of the minimum of the NIC, $\hat\gamma/\hat\sigma$ (a), and tilt of the slopes, $\hat\lambda$ (b), against the correlation $\hat\rho_m$ of weekly returns with the S\&P 500. Filled symbols: significant at the $5\%$ level ($|\tau_n|>1.96$ in (a), $|\kappa_n|>1.96$ in (b)). Gray dots: S\&P 500 stocks. Dashed lines: the benchmarks \eqref{eq: location benchmark}, with $\hat\gamma_m/\hat\sigma_m=0.41$ and $\hat\lambda_m=0.34$ for the S\&P 500. Solid lines: regressions through the origin across the 21 series other than the stocks, the S\&P 500 and the short fund, weighted by the inverse squared standard errors, with slopes $0.31$ ($0.06$) in (a) and $0.53$ ($0.10$) in (b).}
\label{fig: cross asset}
\end{figure}

We estimate the same model as in Section \ref{sec: emp indices}, the TS-AGARCH with a zero conditional mean, with the same QMLE, using daily log returns in percent over the same period, April 10th 2007, or the first trading day of the series, to August 14th 2026.\footnote{Stocks and exchange-traded funds (ETFs): \textsf{finance.yahoo.com} via \texttt{quantmod}, adjusted closing prices. Currencies and the dollar index: Federal Reserve H.10 noon buying rates and the nominal broad dollar index, from FRED.} The S\&P 500 plays the role of the market, and we group the other assets by their role relative to it, according to the estimated market correlation, $\hat\rho_{i,\text{m}}$. Although the TS-AGARCH parameters are estimated using daily returns, the correlation is based on weekly returns, as this avoids accounting for the different closing times of the stock and currency markets. We define the following categories of assets based on their comovement with the market:
\begin{itemize}
  \item \textit{co-movers} ($0.25<\hat\rho_{i,\text{m}}$): the current constituents of the S\&P 500 index with a complete history of daily returns over the period (398 stocks),\footnote{The constituents are those as of November 2026. We exclude 105 constituents without a complete history over the period, mostly firms that were listed later.} and the four other stock indices of Section \ref{sec: emp indices}. The category also includes currencies that tend to rise against the US dollar with the stock market.
  \item \textit{neutral assets} ($-0.25\leq\hat\rho_{i,\text{m}}\leq0.25$): ETFs (Exchange Traded Funds) of US Treasuries with maturities of 1--3, 7--10 and more than 20 years (SHY, IEF, TLT), the Japanese yen and the Swiss franc, and gold (GLD).
  \item \textit{hedges} ($\hat\rho_{i,\text{m}}<-0.25$): an ETF that is short the S\&P 500 (SH), an ETF of short-term VIX futures (VIXY, from 2011), a market-neutral ``anti-beta'' equity fund that is long low-beta and short high-beta stocks (BTAL, from 2011), and the trade-weighted US dollar.
\end{itemize}
Figure \ref{fig: cross asset} illustrates the proposed linear relationships in \eqref{eq: location benchmark}. Specifically, Figure \ref{fig: cross asset}(a) plots $\hat\gamma_i/\hat\sigma_i$ against $\hat\rho_{i,\text{m}}$, where $\hat \gamma_i$ and $\hat\sigma_i$ denote, respectively, the estimated $\gamma$ parameter and standard deviation of the returns of asset $i$, while Figure \ref{fig: cross asset}(b) shows $\hat \lambda_i$ plotted against $\hat\rho_{i,\text{m}}$. Furthermore, Table \ref{tab: assets} in the Appendix provides the exact parameter estimates and corresponding test statistics.

The main takeaways are the following. Figure \ref{fig: cross asset} is clearly indicative of the linear relationships in \eqref{eq: location benchmark}, with positive slopes as $\gamma_\text{m}$ and $\lambda_\text{m}$ are positive: the location of the minimum of the NIC is positive for assets that move with the stock market, close to zero and insignificant for neutral assets, and negative for hedges, most clearly for VIX futures, and the ordering of the slopes follows the correlation in the same way. An asset thus inherits its location and the ordering of its slopes from the stock market through its correlation with the market. The location, which the GJR rules out, is therefore an economically meaningful feature of the NIC, and the TS-AGARCH makes it possible to estimate and test it separately from the difference between the slopes.

\section*{Disclosure statement}
The authors report there are no competing interests to declare.

\section*{Declaration of generative AI use}
During the preparation of this manuscript, the authors used Claude Opus 5.5 (Anthropic) to generate the code used in the simulations and empirical analysis, and as a language-editing tool to improve clarity and readability. All content was subsequently reviewed and edited by the authors, who take full responsibility for the manuscript.

\section*{Data availability statement}
The data and the code that reproduce all tables and figures of the paper are provided in a replication package. The data are publicly available: daily prices of the stock indices, the stocks and the exchange traded funds from Yahoo Finance (\url{https://finance.yahoo.com}), and the exchange rates and the US dollar index from Federal Reserve Economic Data (FRED) of the Federal Reserve Bank of St.\ Louis (\url{https://fred.stlouisfed.org}).

\appendix
\newpage

\printbibliography[heading=bibintoc, title={References}]
\newpage
\section{Appendix}
The Appendix contains additional Monte Carlo results (Section \ref{sec: supp mc}) and empirical results (Section \ref{sec: additional figures}), and the proofs of the results in the paper (Sections \ref{sec: proofs stoch} and \ref{sec: proofs qmle}).

Throughout the Appendix, for any $X \in \mathbb{R}^k$, $k \in \mathbb{N}$, we let $|X|=(X^\top X)^{1/2}$, and for a matrix $X$, we let $|X|=(\mathrm{tr}(X^\top X))^{1/2}$. For any $p > 0$, we write $\lVert X\rVert_p = (\E[|X|^p])^{1/p}$.
\subsection{Additional Monte Carlo results} \label{sec: supp mc}
\begin{figure}[H]
    \centering
    \includegraphics[width=0.8\linewidth]{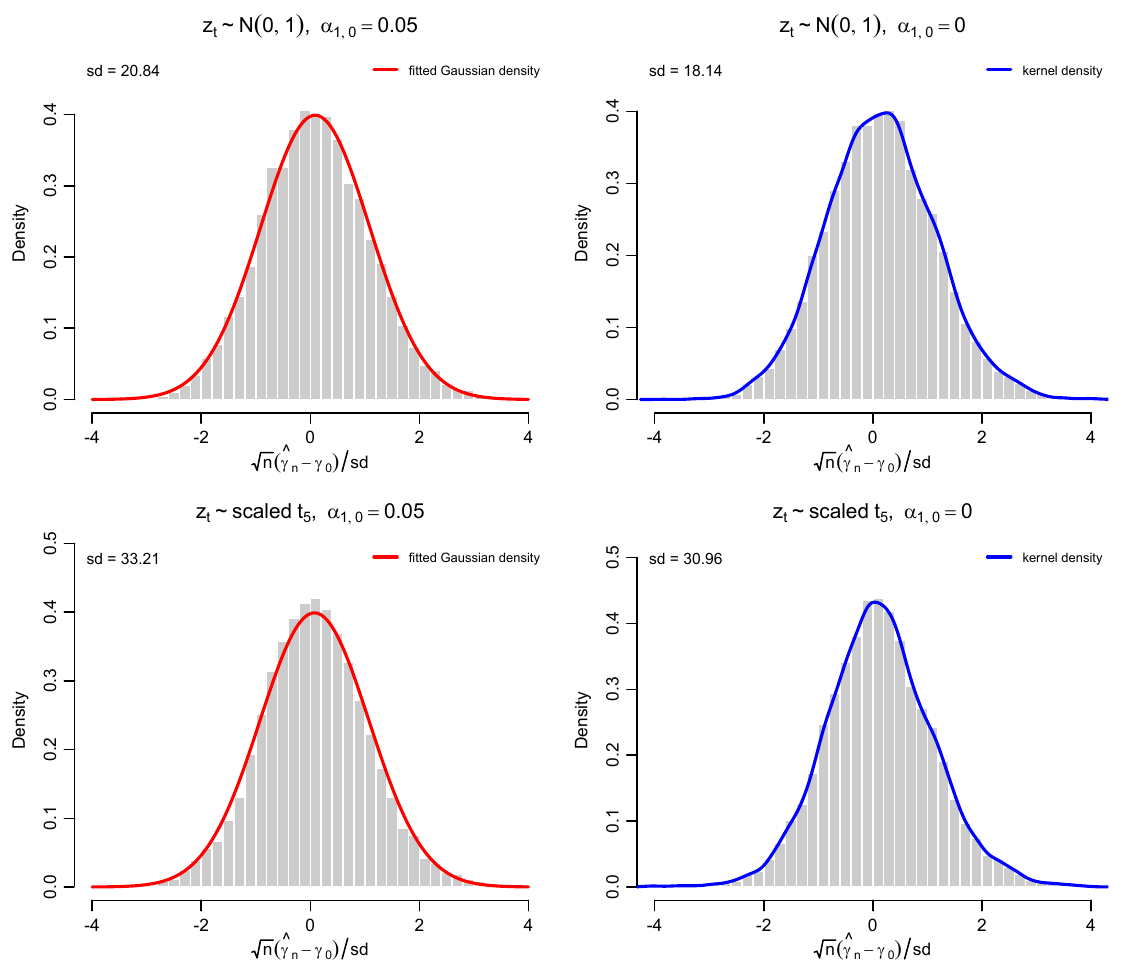}
    \caption{MC distributions of $\sqrt{n}(\hat\gamma_n-\gamma_0)$ for $n=5000$, divided by the MC standard deviation, which is reported as \textsf{sd}. Rows: $z_t\sim\mathrm{N}(0,1)$ (top) and scaled $t_5$ (bottom). Columns: $\alpha_{1,0}=0.05$ (left) and $\alpha_{1,0}=0$ (right). The red curves are fitted Gaussian densities, and the blue curves are Gaussian kernel density estimates with Silverman's rule-of-thumb bandwidth.}
    \label{fig: mc gamma}
\end{figure}

This section complements the Monte Carlo study in Section \ref{sec: Monte Carlo}, whose designs (i)--(iv) and notation we use. Figure \ref{fig: mc gamma} shows the MC distributions of $\sqrt{n}(\hat\gamma_n-\gamma_0)$ for $n=5000$, and Table \ref{tab: mc rejection} reports the rejection frequencies of the tests based on $\tau_n$ and $t_n$ for $n=5000$, $20000$, $50000$ and $200000$.

\noindent In the interior designs, the distributions of $\sqrt{n}(\hat\gamma_n-\gamma_0)$ in Figure \ref{fig: mc gamma} are close to Gaussian, in line with Theorem \ref{thm: asym normality}(a), although the tails are slightly heavier than Gaussian for scaled $t_5$ innovations. The standard deviation of $\sqrt{n}(\hat\gamma_n-\gamma_0)$ is $20.8$ for $\mathrm{N}(0,1)$ and $33.2$ for scaled $t_5$ innovations, and it is nearly unchanged for larger $n$, e.g.\ $20.1$ and $33.9$ for $n=50000$, which indicates that the $\sqrt{n}$-rate is correct. The larger standard deviation for scaled $t_5$ innovations reflects the factor $\mathbb{V}[z_0^2]/2$ in $\Sigma$, see \eqref{eq: def Sigma}, where $\mathbb{V}[z_0^2]=2$ for $z_0\sim\mathrm{N}(0,1)$, while $\mathbb{V}[z_0^2]=2(\nu-1)/(\nu-4)=8$ for a scaled $t_\nu$-distribution with $\nu=5$, although $\Omega$ also depends on the distribution of $z_t$. In the boundary designs, the limit of $\sqrt{n}(\hat\gamma_n-\gamma_0)/\Sigma_{44}^{1/2}$ is $\tau = (1-\eta^2)^{1/2}\xi_\gamma+\eta\max\{\xi,0\}$ by Remark \ref{remark: boundary}, where $\xi_\gamma$ and $\xi$ are independent $\mathrm{N}(0,1)$-distributed, and $\eta$ is the correlation between the fourth and the second coordinate of $Z\sim\mathrm{N}(0,\Sigma)$ in Theorem \ref{thm: asym normality}(b). This limit is not Gaussian, but in our designs $\eta$ is small, $\eta\approx0.17$ for $\mathrm{N}(0,1)$ and $\eta\approx0.12$ for scaled $t_5$ innovations (see Section \ref{sec: Monte Carlo}), such that the limit is close to Gaussian. Accordingly, the kernel density estimates in the right column of Figure \ref{fig: mc gamma} have nearly Gaussian shapes.

\begin{table}[H]
\centering
\begin{tabular}{llcccc}
\hline
\hline
 & & (i) & (ii) & (iii) & (iv) \\
 & & $\mathrm{N}(0,1)$ & scaled $t_5$ & $\mathrm{N}(0,1)$ & scaled $t_5$ \\
 & $n$ & $\alpha_{1,0}=0.05$ & $\alpha_{1,0}=0.05$ & $\alpha_{1,0}=0$ & $\alpha_{1,0}=0$ \\
\hline
$\tau_n$ & 5000 & 5.4 & 6.8 & 4.8 & 6.3 \\
 & 20000 & 5.3 & 5.6 & 4.7 & 5.4 \\
 & 50000 & 4.9 & 5.2 & 4.6 & 5.0 \\
 & 200000 & 5.0 & 4.9 & 4.8 & 5.1 \\
\hline
$t_n$ & 5000 & 5.4 & 8.2 & 3.0 & 1.7 \\
 & 20000 & 5.0 & 6.2 & 3.6 & 2.2 \\
 & 50000 & 4.7 & 5.3 & 4.1 & 2.2 \\
 & 200000 & 5.1 & 4.9 & 4.6 & 2.7 \\
\hline
$\hat\alpha_{1,n}=0$ & 5000 & 0.0 & 0.0 & 53.4 & 57.4 \\
 & 20000 & 0.0 & 0.0 & 52.8 & 54.2 \\
 & 50000 & 0.0 & 0.0 & 51.1 & 53.9 \\
 & 200000 & 0.0 & 0.0 & 50.6 & 53.7 \\
\hline
\hline
\end{tabular}
\caption{Rejection frequencies (in percent) of the tests based on $\tau_n$ and $t_n$ at the $5\%$ level, and the share (in percent) of replications with $\hat\alpha_{1,n}=0$. The test based on $\tau_n$ is the two-sided test of $\mathrm{H}_0:\gamma=0$ with critical value $1.96$. The test based on $t_n$ is the two-sided test of $\mathrm{H}_0:\alpha_1=0.05$ with critical value $1.96$ in designs (i)--(ii), and the one-sided test of $\mathrm{H}_0:\alpha_1=0$ against $\alpha_1>0$ with critical value $1.64$ in designs (iii)--(iv). Both $t$-statistics use the sandwich estimator in \eqref{eq: def covar estimator 2}. With $M=10^4$, the MC standard error of a rejection frequency of $5\%$ is about $0.2$ percentage points.}
\label{tab: mc rejection}
\end{table}
\noindent For $\tau_n$, Table \ref{tab: mc rejection} shows that the over-rejection for scaled $t_5$ innovations at $n=5000$, in designs (ii) and (iv), vanishes as $n$ increases.

In the interior designs, the table also reports the two-sided test of $\mathrm{H}_0:\alpha_1=0.05$ based on $t_n$. For $\mathrm{N}(0,1)$ innovations, the $\mathrm{N}(0,1)$-approximation of $t_n$ is accurate, and the test rejects in $5.4\%$ of the replications for $n=5000$. For scaled $t_5$ innovations, the distribution of $t_n$ is skewed to the left (skewness $-0.69$),\footnote{Here, the skewness refers to the sample skewness of $t_n$ across the MC replications.} and the two-sided test rejects in $8.2\%$ of the replications. This finite sample distortion diminishes as $n$ increases: for $n=20000$, $n=50000$ and $n=200000$, the skewness is $-0.39$, $-0.29$ and $-0.22$, and the rejection frequency is $6.2\%$, $5.3\%$ and $4.9\%$, respectively. The slower convergence for scaled $t_5$ innovations is likely due to their heavy tails, since $\E[|z_0|^q]=\infty$ for $q\geq5$, the score contributions $s_t(\theta_0)$ have infinite third order moments.

In the boundary designs, the share of replications with $\hat\alpha_{1,n}=0$ decreases slowly toward $50\%$ as $n$ increases, to $50.6\%$ and $53.7\%$ for $n=200000$. Moreover, the right tail of $t_n$ is thinner than that of the limit $\max\{\xi,0\}$, e.g.\ for $n=5000$, the $97.5\%$ quantile of $t_n$ is $1.71$ and $1.51$, respectively, compared to $1.96$ for $\max\{\xi,0\}$. As $n$ increases, the right tail of $t_n$ approaches that of the limit, but slowly. In particular, for $n=200000$, the $97.5\%$ quantile of $t_n$ is $1.89$ and $1.69$, and the test rejects in $4.6\%$ and $2.7\%$ of the replications, respectively. Thus, the test controls the size, but it is conservative, in particular for heavy-tailed innovations, where the rejection frequency is only about half the nominal level even for $n=200000$.
\subsection{Additional empirical results} \label{sec: additional figures}
\begin{figure}[H]
    \centering
    \includegraphics[width=0.7\textwidth]{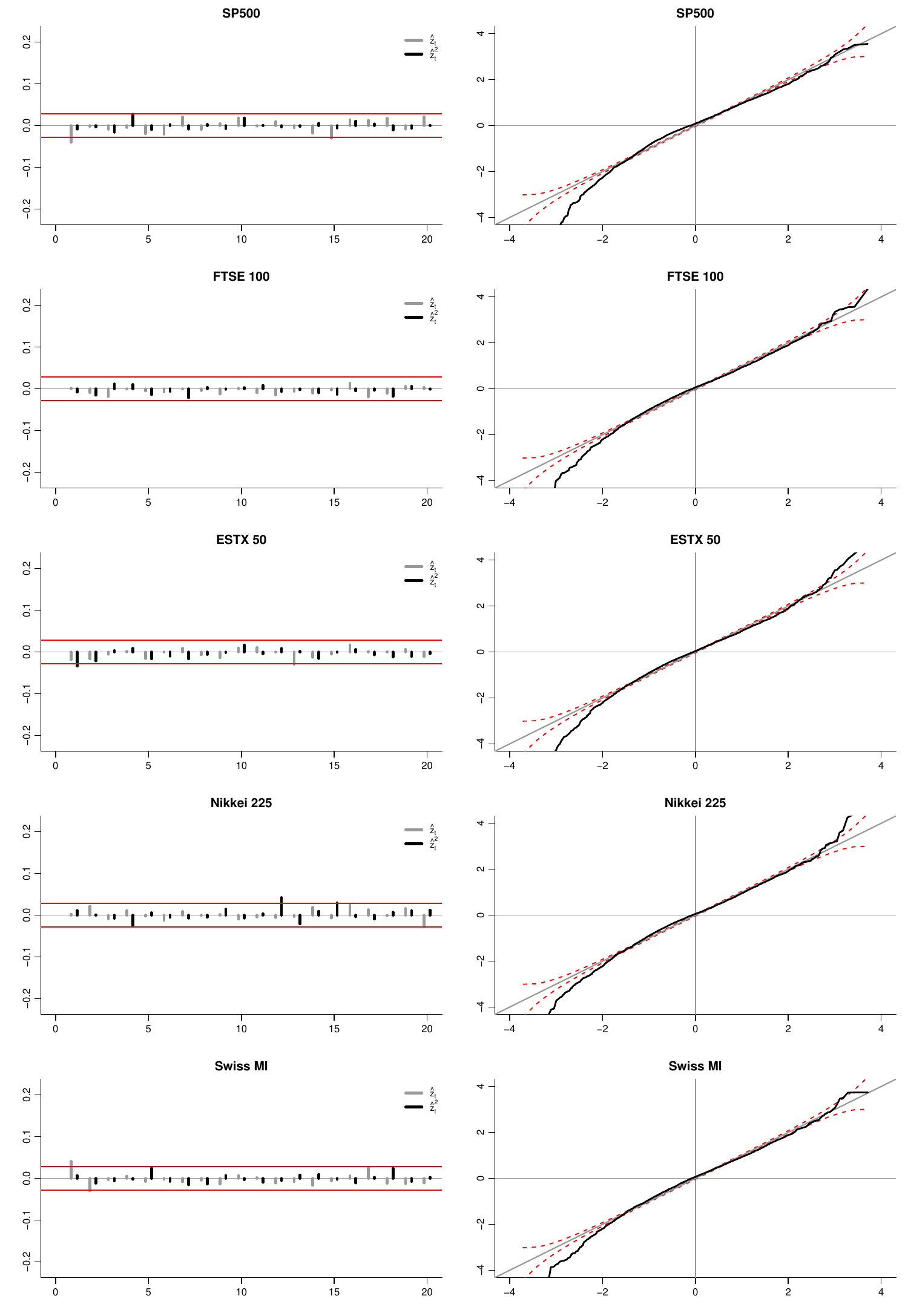}
    \caption{Misspecification plots for the TS-AGARCH fitted to the index returns in Section \ref{sec: emp indices}. Left: ACF of $\hat z_t$ (gray) and $\hat z_t^2$ (black) with the red lines indicating a 0.95 confidence interval. Right: QQ-plot of quantiles of $\hat z_t$ plotted against the quantiles of the $\mathrm{N}(0,1)$-distribution (black line). The gray line indicates the 45-degree line, while the red lines indicate a 0.95 confidence interval.} \label{fig: zhat for all series}
\end{figure}
\begin{table}[H]
\centering
\resizebox{\textwidth}{!}{\begin{tabular}{l r | r r r r | r r r}
\hline\hline
Series & $\hat\rho_m$ & $\hat\gamma$ & $\tau_n$ & $\hat\gamma/\hat\sigma$ & $\hat\rho_m\hat\gamma_m/\hat\sigma_m$ & $\hat\alpha_1$ & $\hat\alpha_2$ & $\kappa_n$\\
\hline
S\&P 500 stocks (median of 398) & 0.59 & 0.571 & 1.51 & 0.29 & 0.24 & 0.031 & 0.077 & 1.81\\
\hline
S\&P 500 & 1.00 & 0.513 & 4.06 & 0.41 & 0.41 & 0.073 & 0.149 & 1.53\\
ESTX 50 & 0.77 & 0.601 & 4.09 & 0.44 & 0.32 & 0.015 & 0.140 & 3.30\\
FTSE 100 & 0.76 & 0.315 & 2.87 & 0.28 & 0.31 & 0.013 & 0.148 & 4.59\\
Swiss MI & 0.71 & 0.583 & 2.71 & 0.54 & 0.29 & 0.061 & 0.121 & 0.72\\
Nikkei 225 & 0.64 & 0.474 & 3.92 & 0.31 & 0.26 & 0.072 & 0.147 & 2.36\\
AUD/JPY & 0.61 & 0.123 & 1.40 & 0.12 & 0.25 & 0.066 & 0.123 & 1.96\\
Australian dollar & 0.60 & 0.009 & 0.10 & 0.01 & 0.24 & 0.036 & 0.068 & 1.81\\
NZD/JPY & 0.56 & 0.152 & 1.40 & 0.16 & 0.23 & 0.050 & 0.072 & 1.03\\
Canadian dollar & 0.52 & 0.076 & 1.07 & 0.14 & 0.21 & 0.029 & 0.050 & 1.46\\
New Zealand dollar & 0.50 & -0.074 & -0.71 & -0.09 & 0.21 & 0.023 & 0.056 & 2.33\\
Norwegian krone & 0.49 & 0.087 & 0.97 & 0.11 & 0.20 & 0.039 & 0.049 & 0.68\\
British pound & 0.36 & 0.192 & 1.46 & 0.32 & 0.15 & 0.063 & 0.045 & -0.36\\
Euro & 0.26 & 0.072 & 0.71 & 0.13 & 0.11 & 0.029 & 0.036 & 0.54\\
Swiss franc & 0.08 & 0.061 & 0.68 & 0.10 & 0.03 & 0.029 & 0.067 & 1.28\\
Gold (GLD) & 0.08 & -0.137 & -0.51 & -0.12 & 0.03 & 0.065 & 0.055 & -0.24\\
\hline
Treasury 1--3y (SHY) & -0.20 & 0.008 & 1.75 & 0.09 & -0.08 & 0.084 & 0.078 & -0.44\\
Treasury 7--10y (IEF) & -0.21 & 0.010 & 0.15 & 0.02 & -0.09 & 0.057 & 0.060 & 0.12\\
Japanese yen & -0.22 & -0.041 & -0.38 & -0.06 & -0.09 & 0.051 & 0.040 & -0.70\\
Treasury 20y+ (TLT) & -0.25 & 0.053 & 0.32 & 0.06 & -0.10 & 0.074 & 0.050 & -0.88\\
US dollar index & -0.51 & -0.018 & -0.48 & -0.05 & -0.21 & 0.053 & 0.024 & -2.46\\
Anti-beta (BTAL) & -0.57 & -0.119 & -1.16 & -0.11 & -0.23 & 0.067 & 0.096 & 1.28\\
VIX futures (VIXY) & -0.79 & -2.674 & -3.87 & -0.63 & -0.33 & 0.161 & 0.056 & -2.04\\
Short S\&P 500 (SH) & -1.00 & -0.512 & -4.38 & -0.42 & -0.41 & 0.159 & 0.072 & -1.99\\
\hline\hline
\end{tabular}
}
\caption{TS-AGARCH estimates (zero conditional mean) for individual stocks, stock indices and other assets, sorted by the correlation $\hat\rho_m$ of weekly returns with the S\&P 500. First row: medians across the S\&P 500 stocks of each column. $\tau_n$: $t$-statistic for $\mathrm{H}_0:\gamma=0$; $\kappa_n$: $t$-statistic for $\mathrm{H}_0:\alpha_1=\alpha_2$. $\hat\sigma$: standard deviation of the returns. $\hat\rho_m\hat\gamma_m/\hat\sigma_m$: the benchmark \eqref{eq: location benchmark}, with $\hat\gamma_m/\hat\sigma_m=0.41$ for the SP500. Sample: April 10th 2007 to August 14th 2026, except for VIX futures (from January 2011) and the anti-beta fund (from September 2011).}
\label{tab: assets}
\end{table}
\subsection{Proof of stochastic properties} \label{sec: proofs stoch}
\noindent\textbf{Proof of Theorem \ref{thm: geometric mixing and moments}:} Observe that $\{\sigma_t^2\}$ is a Markov chain, which from \eqref{eq:STGARCH_proc} can be represented as $\sigma_t^2=\Psi(\sigma_{t-1}^2,z_{t-1})$, with the map $\Psi : (0,\infty)\times \mathbb{R}\to (0,\infty)$ given by
\begin{equation}
    \Psi(x,u)=B(u)\,x+C(x,u), \label{eq: Psi decomp}
\end{equation}
where
\begin{align*}
    B(u)&=\big(\alpha_1\,\mathbb{I}(u\geq 0)+\alpha_2\,\mathbb{I}(u<0)\big)\,u^2+\beta, \\ C(x,u)&=C_{(1)}(x,u)+C_{(2)}(x,u)+C_{(3)}(x,u),
\end{align*}
with
\begin{align*}
    &C_{(1)}(x,u)=\omega+(\alpha_1\mathbb{I}(u\geq0)+\alpha_2\,\mathbb{I}(u<0))(\gamma^2-2\gamma\,\sqrt{x}\,u), \\[0.03in]
    &C_{(2)}(x,u)=(\alpha_2-\alpha_1)\,\mathbb{I}(0\leq\sqrt{x}\,u< \gamma)(\sqrt{x}\,u-\gamma)^2, \\[0.03in]
    &C_{(3)}(x,u)=(\alpha_1-\alpha_2)\,\mathbb{I}(\gamma\leq\sqrt{x}\,u< 0)(\sqrt{x}\,u-\gamma)^2.
\end{align*}
Note furthermore that $|C(x,u)|\leq c(1 + \sqrt{x}|u|)$ for some constant $c > 0$. 
\smallskip

We first show that $\{\sigma^2_t\}$ is geometrically ergodic if \eqref{eq: def phi} holds by applying Theorem 15.2.4 in \citet{douc2018markov}. For this, note initially by Lemma \ref{lem: chotard and auger} that $\{\sigma^2_t\}$ is an aperiodic, $\nu$-irreducible Markov chain on $(0,\infty)$,  with bounded subsets of $(0,\infty)$ being small and therefore petite. It remains to verify the drift criterion for a drift function $V: (0,\infty) \to [1,\infty)$ given in Definition 14.1.5 of \citet{douc2018markov}. That is, for $E$ a bounded (hence petite) subset of $(0,\infty)$, we show the existence of $\lambda \in (0,1)$ and $b \geq 0$, such that
\begin{align}
    \E[V(\sigma^2_t) | \sigma^2_{t-1} = x] \leq \lambda V(x) + b \I(x \in E), \quad x > 0. \label{eq:driftcond}
\end{align}
In our case, the drift function has the form $V(x)=1+x^s$ for some $s\in(0,1)$ satisfying $\E[B(z_0)^s] < 1$. Such an $s$ exists by dominated convergence from the assumption that $\E[\log(B(z_0))]<0$ in \eqref{eq: def phi} and the fact that $z_0$ has finite moments of some order.

To this end, fix $\varepsilon >0$ such that $\lambda := \E[B(z_0)^s] + \varepsilon <1$. Using that $x^{s/2}=o(V(x))$ as $x \to \infty$, it holds for some $m >0$ that
\begin{align}
    \E[V(\sigma^2_t) | \sigma^2_{t-1} = x]  &\leq \left(\E[B(z_0)^s] + \frac{1+\E[|C(x,z_0)|^s]}{V(x)}\right) V(x) \notag \\
    &\leq \begin{cases}
        \lambda V(x),  &x \in (m,\infty)\\
        \lambda V(x) + 1 + c^s(1+m^{s/2}\E[|z_0|^s]), &x \in (0,m].
    \end{cases} \label{eq:driftcondbound}
\end{align}
Setting $E = (0,m]$ and $b = 1 + c^s(1+m^{s/2}\E[|z_0|^s])$ thus verifies the condition in \eqref{eq:driftcond}.

That the joint process $\{(x_t,\sigma^2_t)\}$ is geometrically ergodic follows from Proposition 4 in \citet{carrasco2002mixing}, which applies since conditions (i)--(v) in \citet{carrasco2005erratum} (the latter being an erratum to the first) are satisfied. 

\medskip
Next, we show that for the stationary distribution $\E[\sigma^{2p}_0]<\infty$ when \eqref{eq: moment condition} holds, i.e.\ $\E[B(z_0)^p]<1$ for some $p > 0$, and either $\E[z_0^{2p}]<\infty$ or $\min\{\alpha_1,\alpha_2\}>0$. In the latter case, $\E[z_0^{2p}]<\infty$ is automatically satisfied since
\begin{align*}
    \E[z^{2p}_0] \leq \frac{\E[B(z_0)^p]}{(\min\{\alpha_1,\alpha_2\})^p}.
\end{align*}
To establish $\E[\sigma^{2p}_0]<\infty$, we apply Lemma 14.1.10 of \citet{douc2018markov}. That is, we show that \eqref{eq:driftcond} holds with $E=(0,\infty)$ and $V(x)=1+x^p$. 
The conditions of the lemma are satisfied for $p < 1$ by the arguments above. 

Suppose on the other hand that $p \geq 1$. Define $\kappa = (1-\delta)^{1-p}$ and $\eta = \delta^{1-p}$, where $\delta >0$ is chosen sufficiently small such that $\kappa \E[B(z_0)^p]<1$. It then follows by convexity of $y \mapsto y^p$ that  
\begin{align*}
    &|B(u)x+C(x,u)|^p=|(1-\delta)B(u)x/(1-\delta)+\delta C(x,u)/\delta|^p \leq \kappa B(u)^px^p+\eta|C(x,u)|^p.  
\end{align*}
Next, using that $|C(x,u)|^p \leq c^p(1+\sqrt{x}|u|)^p$, and the fact that $\E[|z_0|^p]<\infty$ by assumption, it follows by repeating the arguments used to establish \eqref{eq:driftcondbound} that \eqref{eq:driftcond} holds with $E=(0,\infty)$ and $V(x)=1+|x|^p$.

Finally, since $z_0$ and $\sigma^2_0$ are independent, it holds for the stationary distribution that $\E[x^{2p}_0]=\E[z_0^{2p} \sigma_0^{2p}]=\E[z_0^{2p}]\E[ \sigma_0^{2p}]<\infty$.\hfill $\square$
\medskip
  
\begin{lemma} \label{lem: chotard and auger}
    Under the assumptions of Theorem \ref{thm: geometric mixing and moments}, it holds that $\{\sigma_t^2\}$ is an aperiodic, $\nu$-irreducible Markov chain on $(0,\infty)$. Furthermore, bounded subsets of $(0,\infty)$ are small for $\{\sigma_t^2\}$.
\end{lemma}

\noindent\textbf{Proof of Lemma \ref{lem: chotard and auger}:} We apply Theorem 4.4 of \citet{chotard2019verifiable} to $\{\sigma_t^2\}$. Observe
\begin{align*}
    \sigma_{t+1}^2=F(\sigma_{t}^2,\,z_t)=\omega+\beta\sigma_t^2+\psi(\sigma_t^2,z_t),
\end{align*}
where $F(x,\,u)=\omega+\beta x+\psi(x,u)$ and
\begin{align}
    \psi(x,u)=h(\sqrt{x}\,u),\qquad h(v)=\big(\alpha_1\I(v\geq\gamma)+\alpha_2\I(v<\gamma)\big)(v-\gamma)^2. \label{eq: def psi Chotard Auger}
\end{align}
Since $z_0$ has the continuous density $f_z$, which does not depend on $x$, Assumptions A1--A4 of \citet{chotard2019verifiable} are satisfied. Moreover, $h$ is a second order polynomial on $(-\infty,\gamma)$ and on $(\gamma,\infty)$ and hence differentiable on these intervals. Furthermore, $0\leq h(v)\leq\max\{\alpha_1,\alpha_2\}(v-\gamma)^2$ implies that $h'(\gamma)=0$. Therefore,
\begin{equation}
    h'(v)=2\big(\alpha_1\I(v\geq\gamma)+\alpha_2\I(v<\gamma)\big)(v-\gamma),\qquad v\in\mathbb{R}, \label{eq: h prime}
\end{equation}
which is continuous, such that $F$ is $C^1$ on $(0,\infty)\times\mathbb{R}$, and Assumption A5 is satisfied as well.

It remains to show the existence of a so-called steadily attracting state $x^\ast$. A point $x^\ast$ is a steadily attracting state if for any $x_0 >0$ and open set $U$ with $x^\ast\in U$, there exists a $T\in \mathbb{N}$ such that for all $k\geq T$, there is a $k$-step path from $x_0$ to $U$. Since $f_z$ is strictly positive on $\mathbb{R}$, there exists a $k$-step path from $x_0$ to $U$ provided one can choose $u_1,\dots,u_k\in\mathbb{R}$ that ensure $\sigma_k^2\in U$, with $\sigma_k^2=F(\sigma_{k-1}^2,u_k)$ and $\sigma_0^2=x_0$. Observe that for any $x > 0$ and $y > 0$, there exists $u \in \mathbb{R}$ such that $y = \psi(x,u)$. This follows by noting that if $\alpha_2>0$, then $h$ maps $(-\infty,\gamma)$ onto $(0,\infty)$, and if $\alpha_1>0$, then $h$ maps $(\gamma,\infty)$ onto $(0,\infty)$. As $\max\{\alpha_1,\alpha_2\}>0$, it follows that for any $x>0$ and $y>0$ there exists $u\in\mathbb{R}$ with $\psi(x,u)=y$, namely $u=(\gamma-\sqrt{y/\alpha_2})/\sqrt{x}$ if $\alpha_2>0$, and $u=(\gamma+\sqrt{y/\alpha_1})/\sqrt{x}$ if $\alpha_2=0$. It therefore suffices to choose $\psi_1,\dots,\psi_k$ with $\psi_i > 0$ that ensure $\sigma^2_k \in U$ with $\sigma^2_k$ given by the recursion
\begin{equation*}
\sigma_{k}^2=\omega + \beta \sigma^2_{k-1}+\psi_k,\qquad \sigma^2_0=x_0.
\end{equation*}
\textit{Existence of steadily attracting state:} Let $x^\ast = \omega/(1-\beta)+a$ for some $a >0$. By substitution
\begin{align*}
    \sigma_{k}^2 = \beta^kx_0+\omega\sum_{i=0}^{k-1}\beta^{i}+\sum_{i=0}^{k-1}\beta^{i}\psi_{k-i}.
\end{align*}
Since $\beta<1$, it holds $\omega\sum_{i=0}^{k-1}\beta^{i}\to\omega/(1-\beta)$ and $\beta^k\,x_0\to0$ as $k\to\infty$ for any given $x_0\in(0,\infty)$. By choosing
\begin{equation*}
    \psi_{k-i}=\frac{a}{k\,\beta^{i}}>0,\qquad 0\leq i\leq k-1,
\end{equation*}
such that $\sum_{i=0}^{k-1}\beta^{i}\psi_{k-i}=a$, it thus follows that $\sigma^2_k \to x^\ast$ as $k \to \infty$. This convergence implies that for any open $U$ containing $x^\ast$, there exists a $T\in\mathbb{N}$ such that $\sigma_k^2\in U$ for all $k\geq T$. Therefore, we conclude that $x^\ast$ is steadily attracting.\medskip

Finally, since $f_z>0$ on $\mathbb{R}$, the control set $O^1_{x^\ast}=\{u\in\mathbb{R}:f_z(u)>0\}$ of \citet{chotard2019verifiable} equals $\mathbb{R}$, so it suffices to find one $u^\ast\in\mathbb{R}$ with $\partial F(x^\ast,u^\ast)/\partial u\neq0$. By \eqref{eq: h prime}, any $u^\ast<\gamma/\sqrt{x^\ast}$ works if $\alpha_2>0$, and any $u^\ast>\gamma/\sqrt{x^\ast}$ if $\alpha_1>0$. For instance, let $u^\ast=(\gamma-1)/\sqrt{x^\ast}$ if $\alpha_2>0$, and $u^\ast=(\gamma+1)/\sqrt{x^\ast}$ if $\alpha_2=0$. Then the controllability matrix of the one-step map at $x^\ast$ is given by
\begin{align}\label{eq: controllability matrix}
    \frac{\partial F(x^\ast,u^\ast)}{\partial u}=\sqrt{x^\ast}\,h'\big(\sqrt{x^\ast}\,u^\ast\big)=
    \begin{cases}
        -2\alpha_2\sqrt{x^\ast}, & \alpha_2>0,\\[0.03in]
        2\alpha_1\sqrt{x^\ast}, & \alpha_2=0.
    \end{cases}
\end{align}
As it is non-zero, it has full rank (equal to 1), and it follows that all the conditions of Theorem 4.4 are satisfied.\medskip

Therefore, by Theorem 4.4 of \citet{chotard2019verifiable}, $\{\sigma_t^2\}$ is an aperiodic, $\nu$-irreducible $T$-chain, and every \textit{compact} subset of $(0,\infty)$ is small for $\{\sigma_t^2\}$. It remains to show that every \textit{bounded} subset $C\subset(0,\infty)$ is also small, i.e.\ that for every bounded subset $C$, there exist $k\in\mathbb{N}$ and a non-zero measure $\mu$ such that for all $x\in C$ and Borel $A\subset(0,\infty)$,
\begin{align*}
    P^k(x,A)\geq \mu(A),
\end{align*}
where $P^k$ is the $k$-step transition kernel of  $\{\sigma_t^2\}$, see Definition 9.1.1 in \citet{douc2018markov}.

For any bounded set $C\subset(0,\infty)$, there exists $M>0$ such that $C\subset(0,M]$. Consequently, with $\sigma_1^2=\omega+\beta\,x+\psi(x,z_0)$, where $\psi$ is given by \eqref{eq: def psi Chotard Auger}, any initial $x\in C$ satisfies
\begin{equation*}
    \omega\leq\sigma_1^2\leq \omega+\beta\,M+\max\{\alpha_1,\alpha_2\}(\sqrt{M}|z_0|+|\gamma|)^2,
\end{equation*}
such that for any $b > \omega+\beta M$
\begin{align*}
    \mathbb{P}\left(|z_0|\leq \sqrt\frac{b-\omega-\beta M}{M\max\{\alpha_1,\alpha_2\}}-|\gamma|/\sqrt M\right) \leq \P(\sigma^2_1 \in [\omega,b] \  | \  \sigma^2_0=x)=P(x,[\omega,b]).
\end{align*}
Choosing $b$ sufficiently large, it follows as $z_0$ is supported on all of $\mathbb{R}$ that for some $\eta>0$
\begin{align}
    \eta < \inf_{x\in C}P(x,K) \label{eq: inf ineq 1}
\end{align}
for $K = [\omega,b]$. That is, there is positive probability for $\sigma^2_1=\omega + \beta x + \psi(x,z_0)$ to reach the compact $K$ for any starting point $x \in C$. Since $K$ is compact and hence small, any Borel $A$ can be reached from $K$ in a finite number of steps, i.e.\ there exists $m\in\mathbb{N}$ such that
\begin{equation}
    \inf_{x\in K}P^m(x,A)\geq \bar\mu(A), \label{eq: inf ineq 2}
\end{equation}
for some non-zero measure $\bar\mu$.

To conclude the proof for $\{\sigma^2_t\}$, using the Chapman--Kolmogorov equations and the established \eqref{eq: inf ineq 1}--\eqref{eq: inf ineq 2}, by choosing $k = m+1$, it holds for all $x\in C$ and Borel $A\subset(0,\infty)$ that
\begin{align*}
    P^k(x,A)&= \int_{(0,\infty)} P(x,dy)P^{m}(y,A)\geq \int_KP(x,dy)P^{m}(y,A) \\
    &\geq \bar\mu(A)\int_K P(x,dy)=\bar\mu(A)P(x,K)\geq \bar\mu(A)\,\eta=:\mu(A).
\end{align*}
The right-hand side is a non-zero measure establishing that any bounded set $C \subset (0,\infty)$ is small for $\{\sigma_t^2\}$.
 \hfill $\square$

\subsection{Proofs of asymptotic properties of the QMLE} \label{sec: proofs qmle}
Throughout this section we assume the conditions of Theorem \ref{thm: consistency}. For $\theta\in\Theta$ and $x\in\mathbb{R}$, we let $a_\theta(x)=\alpha_1\I(x\geq\gamma)+\alpha_2\I(x<\gamma)$ and $a_0(x)=a_{\theta_0}(x)$, such that $\sigma^2_t(\theta)=\omega+a_\theta(x_{t-1})(x_{t-1}-\gamma)^2+\beta\sigma^2_{t-1}(\theta)$ by \eqref{eq: sig2(theta)}. Moreover, $\mathcal{F}_t=\sigma(z_t,z_{t-1},\dots)$.
\subsubsection{Consistency}
Let $\mathcal{L}(\theta) = \mathbb{E}[\ell_0(\theta)], \, \theta \in \Theta$. As is noted in \citet{straumann2005estimation}, since
\begin{align*}
    \ell_0(\theta)=-\frac{1}{2}\left(\log(\sigma^2_0(\theta))+\frac{x^2_0}{\sigma^2_0(\theta)}\right), \qquad \theta \in \Theta,
\end{align*}
with $\log(\sigma^2_0(\theta))\geq \log(\omega_L)>-\infty$, then $\ell_0(\theta)$ is not necessarily integrable if $\E[x^2_0]=\infty$, but the expectation of the positive part is finite. Using this and Tonelli's theorem, it follows that we may write $\mathcal{L}(\theta)$ as
\begin{align*}
    \mathcal{L}(\theta) = -\frac{1}{2}\E\left[\log(\sigma^2_0(\theta)) + \frac{\sigma^2_0}{\sigma^2_0(\theta)}\right], \qquad \theta \in \Theta.
\end{align*}
Although $\mathcal{L}(\theta)$ can possibly have value $-\infty$, it does hold that $\mathcal{L}(\theta_0) \in \mathbb{R}$ since
\begin{align*}
    \E[|\ell_0(\theta_0)|]\leq \E[|\log(\sigma^2_0)|]+\E[z^2_0]<\infty,
\end{align*}
where $\E[|\log(\sigma^2_0)|]<\infty$ holds as $\sigma^2_0$ is bounded below and $\E[\sigma^{2s}_0]<\infty$ for some $0 < s < 1$. It thus follows that the difference $\mathcal{L}(\theta_0)-\mathcal{L}(\theta)$ is well defined and given by
\begin{align}
     \mathcal{L}(\theta_0)-\mathcal{L}(\theta)
    &= \E[\ell_0(\theta_0)-\ell_0(\theta)]=\frac{1}{2} \E\left[\log\left(\frac{\sigma^2_0(\theta)}{\sigma^2_0}\right) +  \frac{\sigma^2_0}{\sigma^2_0(\theta)} -1\right],\qquad \theta \in \Theta. \label{eq: loglik difference}
\end{align}
Finally since $f(x):=\log(x)+1/x-1, \, x > 0$, satisfies $f(x)\geq0$, it follows that $\mathcal{L}$ attains its maximum value on $\Theta$ at $\theta_0$. The following identification lemma shows that this maximum is unique.
\begin{lemma} \label{lem: cons 2}
    The true parameter $\theta_0$ is the unique maximizer of $\mathcal{L}(\theta)$ on $\Theta$.
\end{lemma}
\noindent\textbf{Proof:} Assume $\mathcal{L}$ attains its maximum value at some $\theta\in\Theta$, such that by \eqref{eq: loglik difference} it holds $\mathcal{L}(\theta_0)=\mathcal{L}(\theta)$. We now show that $\theta=\theta_0$.

The restriction $\mathcal{L}(\theta_0)=\mathcal{L}(\theta)$ is equivalent to $\E[f(\sigma^2_t(\theta)/\sigma^2_t)]=0$. Since $f$ is non-negative with $f(x)=0$ iff $x=1$, this implies that $\sigma^2_t = \sigma^2_t(\theta)$ almost surely for every $t$. Imposing $\sigma^2_t=\sigma^2_t(\theta)$ and $\sigma^2_{t-1}=\sigma^2_{t-1}(\theta)$ in the recursions for $\sigma^2_t$ and $\sigma^2_t(\theta)$ yields that, almost surely,
\begin{equation}
    \omega_0+a_0(x_{t-1})(x_{t-1}-\gamma_0)^2+\beta_0\sigma^2_{t-1}=\omega+a_\theta(x_{t-1})(x_{t-1}-\gamma)^2+\beta\sigma^2_{t-1}. \label{eq: identification identity}
\end{equation}
Conditionally on $\mathcal{F}_{t-2}$, $\sigma_{t-1}$ is fixed, and $x_{t-1}=\sigma_{t-1}z_{t-1}$ has a density which is strictly positive on $\mathbb{R}$, as $z_{t-1}$ is independent of $\mathcal{F}_{t-2}$. As both sides of \eqref{eq: identification identity} are continuous functions of $x_{t-1}$, it follows that, almost surely, \eqref{eq: identification identity} holds for all values $x\in\mathbb{R}$ of $x_{t-1}$, with $\sigma_{t-1}$ fixed.

If $x\geq\max\{\gamma,\gamma_0\}$, then both sides of \eqref{eq: identification identity} are second order polynomials in $x$,
\begin{equation*}
    \omega_0+\alpha_{1,0}(x-\gamma_0)^2+\beta_0\sigma^2_{t-1}=\omega+\alpha_1(x-\gamma)^2+\beta\sigma^2_{t-1},
\end{equation*}
and comparing coefficients yields $\alpha_1=\alpha_{1,0}$ and $\alpha_1\gamma=\alpha_{1,0}\gamma_0$, as well as $\omega_0-\omega+\alpha_{1,0}(\gamma_0^2-\gamma^2)+(\beta_0-\beta)\sigma^2_{t-1}=0$ almost surely. The random variable $\sigma^2_{t-1}$ is non-degenerate, since, conditionally on $\mathcal{F}_{t-3}$, it is a non-constant continuous function of $x_{t-2}$ (as $\alpha_{2,0}>0$), and $x_{t-2}$ has a density which is strictly positive on $\mathbb{R}$. Hence $\beta=\beta_0$.

If $x<\min\{\gamma,\gamma_0\}$, then \eqref{eq: identification identity} reads, using $\beta=\beta_0$,
\begin{equation*}
    \omega_0+\alpha_{2,0}(x-\gamma_0)^2=\omega+\alpha_2(x-\gamma)^2,
\end{equation*}
which yields $\alpha_2=\alpha_{2,0}$ and $\alpha_2\gamma=\alpha_{2,0}\gamma_0$. As $\alpha_{2,0}>0$, it follows that $\gamma=\gamma_0$, and then $\omega=\omega_0$. This proves $\theta=\theta_0$.

Note that the argument does not require $\alpha_{1,0}>0$. If $\alpha_{1,0}=0$, the first half-line yields $\alpha_1=0$ and $\beta=\beta_0$, while $\gamma_0$ is identified from the left branch of the NIC on the second half-line. \hfill $\square$
\bigskip

\noindent\textbf{Proof of Theorem \ref{thm: consistency}:} Following the arguments in Theorem 5.3.1 of \citet{straumann2005estimation}, we show that
\begin{enumerate}
    \item $\sup_{\theta\in\Theta}|\mathcal{L}_n(\theta)-\hat{\mathcal{L}}_n(\theta)|/n\asto 0$.
    \item $\limsup_{n\to \infty}\sup_{\theta \in \Theta \cap K} \mathcal{L}_n(\theta)/n \leq \sup_{\theta \in \Theta \cap K}\mathcal{L}(\theta)$ for any compact $K$, with probability 1.
    \item $\mathcal{L}(\theta)$ is upper semicontinuous on $\Theta$.
\end{enumerate}
\smallskip

\noindent \emph{Proof of 1.} For $t\geq1$, $\sigma^2_t(\theta)$ and $\hat\sigma^2_t(\theta)$ satisfy the same recursion, see \eqref{eq: def sig2hat(theta)} and \eqref{eq: sig2(theta)}, with initial values $\sigma^2_0(\theta)$ and $\hat\sigma^2_0$, respectively, such that $|\sigma^2_t(\theta)-\hat{\sigma}^2_t(\theta)| \leq \beta_U^t|\sigma^2_0(\theta)-\hat{\sigma}^2_0|\leq\beta_U^tV_0$ with $V_0=\sup_{\theta\in\Theta}\sigma^2_0(\theta)+\hat\sigma^2_0$. As $\hat{\sigma}^2_t(\theta) \wedge \sigma^2_t(\theta)\geq \omega_L$, the mean value theorem gives $|\log\sigma^2_t(\theta)-\log\hat\sigma^2_t(\theta)|\leq|\sigma^2_t(\theta)-\hat\sigma^2_t(\theta)|/\omega_L$ and $|1/\sigma^2_t(\theta)-1/\hat\sigma^2_t(\theta)|\leq|\sigma^2_t(\theta)-\hat\sigma^2_t(\theta)|/\omega_L^2$, and hence
\begin{equation*}
    \sup_{\theta\in\Theta}|\mathcal{L}_n(\theta)-\hat{\mathcal{L}}_n(\theta)|\leq\sum_{t=1}^\infty\sup_{\theta\in\Theta}|\ell_t(\theta)-\hat\ell_t(\theta)|\leq c\,V_0\sum_{t=1}^\infty\beta_U^t(1+x_t^2),
\end{equation*}
where $c>0$ is a constant. By Remark \ref{remark: moments}, $\E[|x_0|^{2s}]<\infty$ for some $s\in(0,1)$, and, using $(\sum_ia_i)^s\leq\sum_ia_i^s$ for $a_i\geq0$, $\E[(\sum_{t=1}^\infty\beta_U^t(1+x_t^2))^s]\leq\sum_{t=1}^\infty\beta_U^{ts}(1+\E[|x_0|^{2s}])<\infty$. In the same way, $V_0<\infty$ almost surely, as $\sup_{\theta\in\Theta}\sigma^2_0(\theta)\leq c(1+\sum_{i=1}^\infty\beta_U^{i-1}x_{-i}^2)$ by \eqref{eq: sig2(theta)}. Hence the right-hand side is finite almost surely, which proves 1.\medskip

\noindent \emph{Proof of 2. and 3.} Let $Q(\theta) := -\mathcal{L}(\theta)$ and similarly $q_t(\theta) := -\ell_t(\theta)$, with $\ell_t(\theta)$ the log-likelihood contribution defined in \eqref{eq: loglik}. The statement in 2. is equivalent to showing that for any compact set $K$,
\begin{align*}
    \inf_{\theta \in K \cap \Theta}Q(\theta) \leq \liminf_{n\to \infty} \inf_{\theta \in K \cap \Theta}\frac{1}{n}\sum_{t=1}^nq_t(\theta), \qquad \text{a.s.},
\end{align*}
and 3. is equivalent to showing that $Q(\theta)$ is lower semicontinuous on $\Theta$. Both statements follow by verifying the conditions (1)--(3) in Lemma 3.11 of \citet{pfanzagl1969measurability}. The lemma is stated for iid observations, which its proof uses only through the strong law of large numbers for $n^{-1}\sum_{t=1}^n\inf_{\theta\in U}q_t(\theta)$, $U\subseteq\Theta$ open. As $\{\inf_{\theta\in U}q_t(\theta)\}$ is stationary and ergodic, and bounded from below by $\log(\omega_L)/2$ as $\sigma^2_t(\theta)\geq\omega_L$, the ergodic theorem can be used instead, as in the proof of Theorem 5.3.1 in \citet{straumann2005estimation}.

Condition (1) holds with probability 1: each term of the series in \eqref{eq: sig2(theta)} is continuous in $\theta$, and the series converges uniformly on $\Theta$ when $\sum_{i=1}^\infty\beta_U^{i-1}x_{t-i}^2<\infty$, which holds for all $t$ outside a null set by the proof of 1. Redefining $q_t(\theta):=\log(\omega_L)/2$ on this null set does not affect 2. and 3. For condition (2), note that for any compact $C\subset \Theta$, there exists a subset $D_C \subset C$ which is countable and dense in $C$. Hence as $q_0(\theta)$ is continuous, it holds $\inf_{\theta \in C} q_0(\theta) = \inf_{\theta \in D_C} q_0(\theta)$. As the infimum on the right-hand side is taken over a countable set, $\inf_{\theta \in C} q_0(\theta)$ is measurable since $q_0(\theta)$ is a measurable function of $\{x_{t} : t \leq 0\}$ by \eqref{eq: sig2(theta)}. Finally, condition (3) is satisfied since $q_0(\theta)$ is bounded below as stated above and hence $\E[\inf_{\theta \in \Theta}q_0(\theta)] > -\infty$. This proves 2. and 3.
\medskip

Combining 1. and 2., we obtain for any compact $K$
\begin{align}
\limsup_{n\to \infty}\sup_{\theta \in \Theta \cap K} \hat{\mathcal{L}}_n(\theta)/n \leq \sup_{\theta \in \Theta \cap K}\mathcal{L}(\theta) \label{eq: pfanzagl},   
\end{align}
with probability 1.
We will now use this property along with 1. to prove the theorem.

Fix any $\varepsilon > 0$ and define $\Theta_\varepsilon = \{\theta \in \Theta : |\theta - \theta_0|\geq \varepsilon\}$. As $\mathcal{L}$ is upper semicontinuous on $\Theta$, it attains its supremum on the compact set $\Theta_\varepsilon$. Since $\theta_0$ is the unique maximizer of $\mathcal{L}(\theta)$ on $\Theta$ by Lemma \ref{lem: cons 2}, it therefore holds $ \sup_{\theta \in \Theta_\varepsilon} \mathcal{L}(\theta)<\mathcal{L}(\theta_0)$, which combined with \eqref{eq: pfanzagl} yields
\begin{align*}
    \limsup_{n\to \infty}\sup_{\theta \in \Theta_\varepsilon} \hat{\mathcal{L}}_n(\theta)/n < \mathcal{L}(\theta_0)
\end{align*}
with probability 1.
Next, recalling that $\ell_t(\theta_0)$ is integrable, by the ergodic theorem $\mathcal{L}_n(\theta_0)/n=n^{-1}\sum_{t=1}^n\ell_t(\theta_0)\asto \mathcal{L}(\theta_0)$, and hence by 1. $\hat{\mathcal{L}}_n(\theta_0)/n\asto \mathcal{L}(\theta_0)$ yielding
\begin{align*}
     \limsup_{n\to \infty}\sup_{\theta \in \Theta_\varepsilon} \hat{\mathcal{L}}_n(\theta)/n < \lim_{n\to \infty} \hat{\mathcal{L}}_n(\theta_0)/n.
\end{align*}
Finally, since $\hat{\mathcal{L}}_n(\hat{\theta}_n)\geq \hat{\mathcal{L}}_n(\theta_0)$ for all $n$, it follows with probability 1 that
\begin{align*}
     \limsup_{n\to \infty}\sup_{\theta \in \Theta_\varepsilon} \hat{\mathcal{L}}_n(\theta)/n < \liminf_{n\to \infty} \hat{\mathcal{L}}_n(\hat{\theta}_n)/n.
\end{align*}
From the above inequality, we conclude that, with probability 1, $\hat{\theta}_n\notin \Theta_\varepsilon$ for all sufficiently large $n$. Since this holds for each $\varepsilon=1/m$, $m\in\mathbb{N}$, and a countable union of null sets is a null set, it follows that $\hat{\theta}_n\asto \theta_0$. \hfill $\square$
\subsubsection{Defining the score and information} \label{sec: def score and info}
Below we introduce the derivatives for the infeasible likelihood. These are used repeatedly throughout the proofs of the asymptotic distribution of the QMLE given in Section \ref{sec: proofs asymp normality} below.

Define the first order derivatives of $\sigma^2_t(\theta)$,
\begin{align}
    \partial\sigma^2_t(\theta):=\frac{\partial\sigma^2_t(\theta)}{\partial\theta}=\sum^\infty_{i=1}\beta^{i-1}\begin{pmatrix}
        1 \\
        \I(x_{t-i}\geq \gamma)(x_{t-i}-\gamma)^2 \\
        \I(x_{t-i}< \gamma)(x_{t-i}-\gamma)^2 \\
        2(\gamma-x_{t-i})(\alpha_{1}\I(x_{t-i}\geq \gamma)+\alpha_{2}\I(x_{t-i}<\gamma)) \\
        \sigma_{t-i}^2(\theta)
    \end{pmatrix}, \label{eq: partial sig2(theta)}
\end{align}
where, as in Section \ref{sec: likelihood estimation}, all partial derivatives are right partial derivatives. Next, we define the matrix $\partial^2\sigma^2_t(\theta)$ of second order partial derivatives of $\sigma^2_t(\theta)$. All the second order derivatives of $\sigma^2_t(\theta)$ except $\partial^2\sigma^2_t(\theta)/\partial \gamma^2$ are well defined in the usual sense on $\Theta$. However, the left and right second order partial derivatives of $\sigma^2_t(\theta)$ with respect to $\gamma$ exist for every $\theta \in \Theta$ but differ, when $\alpha_1\neq\alpha_2$, at the countable set of points
\begin{align}
    D=\big\{\gamma\in[\gamma_L,\gamma_U]: \gamma=x_{t-i} \text{ for some } i \geq 1\big\} \label{eq: countable set}.
\end{align}
To illustrate, let $\partial^2_+\sigma^2_t(\theta)/\partial\gamma^2$ and $\partial^2_-\sigma^2_t(\theta)/\partial\gamma^2$ denote the right and left second order partial derivatives at a point $\theta \in \Theta$ defined as the limits
\begin{align*}
    \frac{\partial^2_+\sigma^2_t(\theta)}{\partial\gamma^2} &= \lim_{\bar\gamma \downarrow \gamma} \frac{\partial\sigma^2_t(\theta(\bar\gamma))/\partial\gamma-\partial\sigma^2_t(\theta)/\partial\gamma}{\bar{\gamma}-\gamma},\quad \frac{\partial^2_-\sigma^2_t(\theta)}{\partial\gamma^2} = \lim_{\bar\gamma \uparrow \gamma}\frac{\partial\sigma^2_t(\theta(\bar\gamma))/\partial\gamma-\partial\sigma^2_t(\theta)/\partial\gamma}{\bar{\gamma}-\gamma},
\end{align*}
where $\theta(\bar\gamma)=(\omega,\alpha_1,\alpha_2,\bar\gamma,\beta)^\top$.
Considering the right derivative, it holds for $\bar\gamma \geq \gamma$
\begin{align*}
   &\frac{\partial\sigma^2_t(\theta(\bar\gamma))/\partial\gamma-\partial\sigma^2_t(\theta)/\partial\gamma}{\bar{\gamma}-\gamma} \\
   &=\sum_{i=1}^\infty \beta^{i-1}2(\alpha_1\I(x_{t-i}\geq \bar\gamma)+\alpha_2\I(x_{t-i}< \bar\gamma))+2(\alpha_2-\alpha_1)\sum_{i=1}^\infty \beta^{i-1}\frac{\gamma-x_{t-i}}{\bar\gamma-\gamma}\I(\gamma < x_{t-i} < \bar\gamma)
\end{align*}
where we have used that for any $t$
\begin{align*}
    &(\bar\gamma-x_{t})(\alpha_1\I(x_{t}\geq \bar\gamma)+\alpha_2\I(x_{t}< \bar\gamma))-(\gamma-x_{t})(\alpha_1\I(x_{t}\geq \gamma)+\alpha_2\I(x_{t}< \gamma)) \\&= \alpha_1 (\bar\gamma-\gamma)\I(x_{t}\geq \bar\gamma)+\alpha_2 (\bar\gamma-\gamma)\I(x_{t}< \bar\gamma) + (\alpha_2-\alpha_1)(\gamma - x_{t})\I(\gamma < x_{t} < \bar\gamma).
\end{align*}
Considering the first sum, using that $\I(x \geq \bar\gamma) \to \I(x > \gamma)$ and that $\I(x < \bar\gamma) \to \I(x \leq \gamma)$ as $\bar\gamma \downarrow \gamma$, it follows by dominated convergence that as $\bar\gamma \downarrow \gamma$
\begin{align*}
    \sum_{i=1}^\infty \beta^{i-1}2(\alpha_1\I(x_{t-i}\geq \bar\gamma)+\alpha_2\I(x_{t-i}< \bar\gamma)) \asto \sum_{i=1}^\infty \beta^{i-1}2(\alpha_1\I(x_{t-i}> \gamma)+\alpha_2\I(x_{t-i}\leq \gamma)).
\end{align*}
As for the second sum, fix any $\varepsilon >0$ and choose $k \in \mathbb{N}$ such that $\beta^k/(1-\beta)<\varepsilon$. Observe
\begin{align*}
    &\left|\sum_{i=1}^\infty \beta^{i-1}\frac{\gamma-x_{t-i}}{\bar\gamma-\gamma}\I(\gamma < x_{t-i} < \bar\gamma) \right| \leq \left|\sum_{i=1}^k \I(\gamma < x_{t-i} < \bar\gamma) \right| + \beta^k/(1-\beta),
\end{align*}
which follows since $|\gamma-x_{t-i}|/(\bar\gamma-\gamma) \I(\gamma < x_{t-i} < \bar\gamma) \leq 1$ for all $i$. As there are only a finite number of terms in the first sum, there exists $\gamma^\ast$ satisfying $\gamma<\gamma^*<\min\{x_{t-i}:i\leq k,x_{t-i}> \gamma\}$, such that the $k$ indicators are equal to zero. Hence we conclude
\begin{align*}
    \frac{\partial^2_+\sigma^2_t(\theta)}{\partial\gamma^2} = \sum_{i=1}^\infty \beta^{i-1}2(\alpha_1\I(x_{t-i}> \gamma)+\alpha_2\I(x_{t-i}\leq \gamma)).
\end{align*}
By similar considerations, it follows that the left derivative is given by
\begin{align*}
\frac{\partial^2_-\sigma^2_t(\theta)}{\partial\gamma^2}=\sum_{i=1}^\infty \beta^{i-1}2(\alpha_1\I(x_{t-i}\geq \gamma)+\alpha_2\I(x_{t-i}< \gamma)).
\end{align*}
Thus, at any $\theta$ with $\gamma\in D$ and $\alpha_1\neq\alpha_2$, the left and right derivatives differ:
\begin{align*}
    \frac{\partial^2_+\sigma^2_t(\theta)}{\partial\gamma^2} - \frac{\partial^2_-\sigma^2_t(\theta)}{\partial\gamma^2} = 2(\alpha_2-\alpha_1)\sum_{\{i \ : \ x_{t-i}=\gamma\}}\beta^{i-1}. 
\end{align*}
However, since $\P(\cup_{i\geq1}(x_{t-i}=\gamma))\leq\sum_{i=1}^\infty\P(x_{t-i}=\gamma)=0$ for all $\gamma$ (see Footnote \ref{fn: null set}), the left and right derivatives are equal almost surely. In accordance with the convention of Section \ref{sec: likelihood estimation}, the entry $\partial^2\sigma^2_t(\theta)/\partial \gamma^2$ of the matrix of second order partial derivatives of $\sigma^2_t(\theta)$ is the right derivative
\begin{align}
    \frac{\partial^2\sigma^2_t(\theta)}{\partial\gamma^2} = \sum_{i=1}^\infty \beta^{i-1}2(\alpha_1\I(x_{t-i}> \gamma)+\alpha_2\I(x_{t-i}\leq \gamma)). \label{eq: second order deriv sig2(theta) wrt gamma}
\end{align}
That is, we let
\begin{align}
&\partial^2\sigma^2_t(\theta)=\frac{\partial^2\sigma^2_t(\theta)}{\partial \theta \partial \theta^\top}, \label{eq: 2nd partial deriv}
\end{align}
where the remaining non-zero entries of the partial derivative are given by
\begin{align}
    &\frac{\partial^2\sigma_t^2(\theta)}{\partial\alpha_1\partial\gamma}=-2\sum_{i=1}^\infty\beta^{i-1}\I(x_{t-i}\geq\gamma)(x_{t-i}-\gamma), \notag\\[0.05in]
    &\frac{\partial^2\sigma_t^2(\theta)}{\partial\alpha_2\partial\gamma}=-2\sum_{i=1}^\infty\beta^{i-1}\I(x_{t-i}<\gamma)(x_{t-i}-\gamma), \notag\\[0.05in]
    &\frac{\partial^2\sigma_t^2(\theta)}{\partial\alpha_1\partial\beta}=\sum_{i=1}^\infty\beta^{i-2}(i-1)\I(x_{t-i}\geq\gamma)(x_{t-i}-\gamma)^2, \notag\\[0.05in]
    &\frac{\partial^2\sigma_t^2(\theta)}{\partial\omega\partial\beta}=\sum_{i=1}^\infty\beta^{i-2}(i-1), \notag \\[0.05in]
    &\frac{\partial^2\sigma_t^2(\theta)}{\partial\alpha_2\partial\beta}=\sum_{i=1}^\infty\beta^{i-2}(i-1)\I(x_{t-i}<\gamma)(x_{t-i}-\gamma)^2, \notag\\[0.05in]
    &\frac{\partial^2\sigma_t^2(\theta)}{\partial\beta^2}=2\sum_{i=1}^\infty\beta^{i-2}(i-1)\sigma_{t-i}^2(\theta),  \notag \\[0.05in]
    &\frac{\partial^2\sigma_t^2(\theta)}{\partial\gamma\partial\beta}=2\sum_{i=1}^\infty\beta^{i-2}(i-1)(\gamma-x_{t-i})(\alpha_1\I(x_{t-i}\geq\gamma)+\alpha_2\I(x_{t-i}<\gamma)). \label{eq: expression 2nd partials} 
\end{align}
Define the score $\mathcal{S}_n$ as a function of $\theta \in \Theta$ as
\begin{align}
    \mathcal{S}_n(\theta)=\sum^n_{t=1}s_t(\theta),\quad s_t(\theta) = \frac{\partial\ell_t(\theta)}{\partial\theta} = -\frac{1}{2}\left(1-\frac{x^2_t}{\sigma^2_t(\theta)}\right)\frac{\partial \sigma^2_t(\theta)}{\sigma^2_t(\theta)} \label{eq: def of score},
\end{align}
and let $\mathcal{S}(\theta)=\E[s_t(\theta)]$. Similarly, let $\mathcal{I}_n$ denote the observed information
\begin{align}
    &\mathcal{I}_n(\theta)=\sum^n_{t=1}\zeta_t(\theta), \notag \\ &\zeta_t(\theta) = -\frac{\partial s_t(\theta)}{\partial \theta^\top} = -\frac{1}{2}\left(\frac{x^2_t}{\sigma^2_t(\theta)}-1\right)\frac{\partial^2\sigma^2_t(\theta)}{\sigma^2_t(\theta)} + \frac{1}{2}\left(2\frac{x^2_t}{\sigma^2_t(\theta)} -1\right)\frac{\partial \sigma^2_t(\theta)}{\sigma^2_t(\theta)}\frac{\partial \sigma^2_t(\theta)^\top}{\sigma^2_t(\theta)}, \label{eq: def information}
\end{align}
and let $\zeta_t=\zeta_t(\theta_0)$. Furthermore, $\partial \sigma^2_t=\partial \sigma^2_t(\theta_0)$.\bigskip

\subsubsection{Asymptotic distribution} \label{sec: proofs asymp normality}
In this section and all sections below, we assume the conditions of Theorem \ref{thm: asym normality}, in case (a) or in case (b). Since $\hat\theta_n$ is strongly consistent by Theorem \ref{thm: consistency}, and the infeasible estimator $\theta_n$ is as well by the proof of Theorem \ref{thm: consistency}, we may, for the asymptotic theory, assume without loss of generality that $\hat\theta_n$ and $\theta_n$ lie in the set
\begin{align}
    \Theta(\delta):= \{\theta \in \Theta:\ |\beta-\beta_0| \leq \delta,\ \alpha_1\geq\alpha_{1,0}/2\},\qquad\delta>0,  \label{eq: Theta small}
\end{align}
where the restriction on $\alpha_1$ is void if $\alpha_{1,0}=0$. Indeed, $\Theta(\delta)$ contains all $\theta\in\Theta$ with $|\theta-\theta_0|\leq\varepsilon$ for sufficiently small $\varepsilon > 0$ (for instance $\varepsilon = \min\{\delta,\alpha_{1,0}/2\}$ if $\alpha_{1,0} >0$ and $\varepsilon=\delta$ if $\alpha_{1,0}=0$), such that, with probability 1, $\hat\theta_n,\theta_n\in\Theta(\delta)$ for all $n$ sufficiently large. 

The reason for introducing $\theta\in\Theta(\delta)$ is that the bounds of Lemma \ref{lem: new process bound} apply for any $\theta\in\Theta(\delta)$ when $\delta$ is chosen sufficiently small. We impose this restriction on $\delta$ throughout. Finally, $\Theta(\delta)$ is convex and compact, which is used in Lemmas \ref{lem: asym norm infeasible}--\ref{lem: uniform convergence of information}.

With a slight abuse of notation we will in the following let $\beta_L=\beta_0-\delta$ and $\beta_U=\beta_0+\delta$ denote the bounds for $\beta$ in $\Theta(\delta)$, where $0<\beta_L<\beta_U<1$.

\bigskip

\begin{lemma} \label{lem: asym norm infeasible}
    It holds in case (a) of Theorem \ref{thm: asym normality} that
    \begin{align*}
        \sqrt{n}(\theta_n - \theta_0) \dto \mathrm{N}(0,\mathbb{V}[z^2_0] \Omega^{-1}/2), \qquad n \to \infty,
    \end{align*}
    and in case (b) that $\sqrt{n}(\theta_n - \theta_0) \dto \hat\lambda$, with $\hat\lambda$ given in Theorem \ref{thm: asym normality}(b), where $\Omega = \E[\zeta_0]$ with $\zeta_t=\zeta_t(\theta_0)$ defined in \eqref{eq: def information}.
\end{lemma}

\noindent\textbf{Proof:} Recalling $ \mathcal{S}_n(\theta)=\sum^n_{t=1}s_t(\theta)$, it holds by Lemma \ref{lem: FTC} that
\begin{align*}
    \mathcal{S}_n(\theta_n) = \mathcal{S}_n(\theta_0) - \sum_{t=1}^n\left(\int_0^1 \zeta_t(\theta_n(u))du\right)(\theta_n-\theta_0),
\end{align*}
with $\theta_n(u) = \theta_0 + u(\theta_n - \theta_0)$, and therefore
\begin{align}
    \Omega_n n^{1/2}(\theta_n-\theta_0) = n^{-1/2}  \mathcal{S}_n(\theta_0)-n^{-1/2}  \mathcal{S}_n(\theta_n), \qquad \Omega_n=\int_0^1 n^{-1}\sum_{t=1}^n \zeta_t(\theta_n(u))du, \label{eq: normalized score in proof}
\end{align}
where we have used linearity to interchange the summation and integration order. We first establish the limits of $\Omega_n$ and $n^{-1/2}\mathcal{S}_n(\theta_0)$, which are the same in the two cases (a) and (b), and then conclude from \eqref{eq: normalized score in proof} in each case.

Note that
\begin{align*}
     \Omega_n = \int_0^1 n^{-1}\sum_{t=1}^n \left(\zeta_t(\theta_n(u))-\Omega\right) du + \Omega.
\end{align*}
Using that $|\theta_n(u)-\theta_0|= u|\theta_n-\theta_0|\leq |\theta_n-\theta_0|$ and the strong consistency of $\theta_n$, there exists for every $\varepsilon > 0$ an $n_0$ such that $|\theta_n(u)-\theta_0| < \varepsilon$ uniformly in $u$ for all $n \geq n_0$. Hence with $\varepsilon >0 $ fixed, it follows
\begin{align*}
   \limsup_{n\to \infty} \int_0^1\left| n^{-1}\sum_{t=1}^n \left(\zeta_t(\theta_n(u))-\Omega\right)\right| du \leq \limsup_{n\to \infty}\sup_{\theta \in B(\varepsilon)} \left|n^{-1}\mathcal{I}_n(\theta)-\Omega\right|
\end{align*}
with probability 1, where $B(\varepsilon)=\{\theta\in\Theta:|\theta-\theta_0|\leq\varepsilon\}$. An application of Lemma \ref{lem: uniform convergence of information} yields that the upper bound tends to zero by taking $\varepsilon \downarrow 0$.
Consequently, we have shown
\begin{align*}
     \Omega_n \asto \Omega, \qquad n \to \infty.
\end{align*}
By the expression for $\zeta_t(\theta_0)=\zeta_t$ in \eqref{eq: def information}, note
\begin{align*}
    \Omega = \E[\zeta_0] = \frac{1}{2}\E\left[\frac{\partial \sigma^2_t}{\sigma^2_t}\left(\frac{\partial \sigma^2_t}{\sigma^2_t}\right)^\top\right],
\end{align*}
which is positive definite by Lemma \ref{lem: positive definite}.
Next, note $\mathcal{S}_n(\theta_0)=\sum^n_{t=1}s_t(\theta_0)$ with
\begin{align*}
    s_t(\theta_0) = -\frac{1}{2}\left(1-z^2_t\right)\frac{\partial \sigma^2_t}{\sigma^2_t},
\end{align*}
where $\partial \sigma^2_t/\sigma^2_t$ is $\mathcal{F}_{t-1}$-measurable and $z_t$ is independent of $\mathcal{F}_{t-1}$. Furthermore, $\E[z^2_t] = 1$ and $\E[z^4_t] < \infty$, and $\partial \sigma^2_t/\sigma^2_t$ has finite second order moments by Lemma \ref{lem: new process bound}, such that $\{s_t(\theta_0)\}$ is a stationary and ergodic martingale difference sequence with
\begin{align*}
    \E[s_t(\theta_0)s_t(\theta_0)^\top]=\frac{1}{4}\mathbb{V}[z^2_0]\E\left[\frac{\partial \sigma^2_t}{\sigma^2_t}\frac{{\partial \sigma^2_t}^\top}{\sigma^2_t}\right]=\frac{1}{2}\mathbb{V}[z^2_0]\Omega.
\end{align*}
Hence by the Cram\'er--Wold device and the central limit theorem for stationary and ergodic martingale difference sequences in \citet{billingsley1961lindeberg},
\begin{align*}
    n^{-1/2}\mathcal{S}_n(\theta_0)\dto G\sim\mathrm{N}(0,\mathbb{V}[z^2_0]\Omega/2), \qquad n \to \infty.
\end{align*}

As $\Omega_n$ is symmetric with $\Omega_n\asto\Omega$ and $\Omega$ is positive definite, $\Omega_n$ is positive definite, and in particular invertible, for $n$ sufficiently large.

\emph{Case (a).} Since $\theta_0$ is an interior point of $\Theta$ and $\theta_n\asto\theta_0$, $\theta_n$ is an interior point of $\Theta$ for $n$ sufficiently large, and as $\theta_n$ maximizes $\mathcal{L}_n$, it holds that $\mathcal{S}_n(\theta_n)=\partial \mathcal{L}_n(\theta_n)/\partial \theta=0$. Therefore, by \eqref{eq: normalized score in proof} and Slutsky's theorem,
\begin{align*}
 n^{1/2}(\theta_n-\theta_0) = \Omega_n^{-1}n^{-1/2}\mathcal{S}_n(\theta_0) \dto  \mathrm{N}(0,\mathbb{V}[z^2_0]\Omega^{-1}/2), \qquad n \to \infty.
\end{align*}

\emph{Case (b).} On the other hand, when $\alpha_{1,0}=0$, $\theta_0$ is a boundary point of $\Theta$ and we are no longer ensured that $\mathcal{S}_n(\theta_n)=0$. However, with $\mathcal{S}^{(i)}_n(\theta_n)$ denoting the $i$th entry of $\mathcal{S}_n(\theta_n)$, we still have that $\mathcal{S}^{(i)}_n(\theta_n)=0$ for $i=1,3,4,5$ and $n$ large enough, as the remaining true parameter values are not on the boundary and $\theta_n$ is consistent. Consequently, \eqref{eq: normalized score in proof} gives
\begin{align}
    n^{1/2}(\theta_n-\theta_0)=\Omega_n^{-1}n^{-1/2}\mathcal{S}_n(\theta_0)-\Omega_n^{-1}e_2\,n^{-1/2}\mathcal{S}^{(2)}_n(\theta_n), \label{eq: expansion boundary}
\end{align}
where $e_2=(0,1,0,0,0)^\top$. Next, we make the following observations, which hold for $n$ sufficiently large:
\begin{itemize}
    \item If $\alpha_{1,n}=0$, then $\mathcal{S}^{(2)}_n(\theta_n)\leq0$. Otherwise, increasing $\alpha_{1,n}$ would increase $\mathcal{L}_n$.
    \item If $\alpha_{1,n}>0$, then $\mathcal{S}^{(2)}_n(\theta_n)=0$. Otherwise, moving $\alpha_{1,n}$ up or down would increase $\mathcal{L}_n$.
\end{itemize}
With these observations in mind, we now show that $\alpha_{1,n}= 0$ iff $e_2^\top\Omega_n^{-1}n^{-1/2}\mathcal{S}_n(\theta_0) \leq 0$.

We have by \eqref{eq: expansion boundary} that
\begin{align}
    n^{1/2}\alpha_{1,n}=e_2^\top\Omega_n^{-1}n^{-1/2}\mathcal{S}_n(\theta_0)-e_2^\top\Omega_n^{-1}e_2\,n^{-1/2}\mathcal{S}^{(2)}_n(\theta_n), \label{eq:alpha1n-express}
\end{align}
where $e_2^\top\Omega_n^{-1}e_2>0$ for $n$ sufficiently large. If $\alpha_{1,n}=0$, the left-hand side vanishes, and by the first observation
\begin{align}
    n^{-1/2}\mathcal{S}^{(2)}_n(\theta_n)=\frac{e_2^\top\Omega_n^{-1}n^{-1/2}\mathcal{S}_n(\theta_0)}{e_2^\top\Omega_n^{-1}e_2}\leq0, \label{eq:sign-of-score}
\end{align}
such that $e_2^\top\Omega_n^{-1}n^{-1/2}\mathcal{S}_n(\theta_0)\leq0$. On the other hand, if $e_2^\top\Omega_n^{-1}n^{-1/2}\mathcal{S}_n(\theta_0)\leq0$, then supposing for contradiction that $\alpha_{1,n} >0$, we obtain a contradiction by \eqref{eq:alpha1n-express} since by the second observation $\mathcal{S}^{(2)}_n(\theta_n)=0$. Hence $\alpha_{1,n} = 0$ in this case, which proves the equivalence. 

Using this equivalence, the observations above, and equation \eqref{eq:sign-of-score}, we have thus shown
\begin{itemize}
    \item if $e_2^\top\Omega_n^{-1}n^{-1/2}\mathcal{S}_n(\theta_0) \leq 0$, then $n^{-1/2}\mathcal{S}^{(2)}_n(\theta_n)=\frac{e_2^\top\Omega_n^{-1}n^{-1/2}\mathcal{S}_n(\theta_0)}{e_2^\top\Omega_n^{-1}e_2}\leq0$,
    \item if $e_2^\top\Omega_n^{-1}n^{-1/2}\mathcal{S}_n(\theta_0) > 0$, then $\mathcal{S}^{(2)}_n(\theta_n)=0$,
\end{itemize}
which we may equivalently represent as
\begin{align*}
    n^{-1/2}\mathcal{S}^{(2)}_n(\theta_n)=\frac{\min\{0,e_2^\top\Omega_n^{-1}n^{-1/2}\mathcal{S}_n(\theta_0)\}}{e_2^\top\Omega_n^{-1}e_2}.
\end{align*}
Collecting terms, it holds for $n$ sufficiently large that \eqref{eq: expansion boundary} becomes
\begin{align*}
    n^{1/2}(\theta_n-\theta_0)=\Omega_n^{-1}n^{-1/2}\mathcal{S}_n(\theta_0)-\Omega_n^{-1}e_2\,\frac{\min\{0,e_2^\top\Omega_n^{-1}n^{-1/2}\mathcal{S}_n(\theta_0)\}}{e_2^\top\Omega_n^{-1}e_2}.
\end{align*}
As the right-hand side is a continuous function of $\Omega_n$ and $n^{-1/2}\mathcal{S}_n(\theta_0)$, it follows by Slutsky's theorem that
\begin{align*}
    n^{1/2}(\theta_n-\theta_0)\dto\Omega^{-1}G-\Omega^{-1}e_2\,\frac{\min\{0,e_2^\top\Omega^{-1}G\}}{e_2^\top\Omega^{-1}e_2}=Z-\frac{\min\{Z_2,0\}}{\Sigma_{22}}\,\Sigma e_2=\hat\lambda,\qquad n\to\infty,
\end{align*}
where $Z=\Omega^{-1}G\sim\mathrm{N}(0,\Sigma)$ with $\Sigma=\mathbb{V}[z^2_0]\Omega^{-1}/2$, $Z_2=e_2^\top Z$ and $\Sigma_{22}=e_2^\top\Sigma e_2$, and where we used that $\Omega^{-1}e_2/(e_2^\top\Omega^{-1}e_2)=\Sigma e_2/\Sigma_{22}$. This completes the proof. \hfill $\square$

\bigskip

\noindent\textbf{Proof of Theorem \ref{thm: asym normality}:} By the conclusion of Lemma \ref{lem: asym norm infeasible}, it suffices to show that $n^{1/2}|\theta_n-\hat\theta_n|\asto0$ as $n\to\infty$. To this end, as we have for sufficiently large $n$ that both $\theta_n,\hat\theta_n\in \Theta(\delta)$, it follows by Lemma \ref{lem: FTC} that the following expansion holds,
\begin{equation}\label{eq:exp proof T4.2}
  \mathcal{S}_n(\theta_n)-\mathcal{S}_n(\hat{\theta}_n)=-\left(\int_0^1\sum_{t=1}^n\zeta_t(\hat\theta_n+u(\theta_n-\hat\theta_n))\,du\right)(\theta_n-\hat\theta_n).
\end{equation}
Next, since both $\theta_n$ and $\hat\theta_n$ are strongly consistent estimators of $\theta_0$, and $\sup_{u\in[0,1]}|\hat\theta_n+u(\theta_n-\hat\theta_n)-\theta_0|\leq\max\{|\hat\theta_n-\theta_0|,|\theta_n-\theta_0|\}$, it also holds that every point $\hat\theta_n+u(\theta_n-\hat\theta_n)$ on the line segment between them satisfies $\hat\theta_n+u(\theta_n-\hat\theta_n)\asto\theta_0$, uniformly in $u\in[0,1]$, as $n\to\infty$. Moreover, since $\Theta(\delta)$ is a convex set, this line segment lies in $\Theta(\delta)$ for sufficiently large $n$. Consequently, it now follows by the same arguments as those applied to $\Omega_n$ in Lemma \ref{lem: asym norm infeasible} that,
\begin{equation*}
  \widetilde{\Omega}_n:=\int_0^1n^{-1}\sum_{t=1}^n\zeta_t(\hat\theta_n+u(\theta_n-\hat\theta_n))\,du\asto \Omega,\qquad n\to\infty.
\end{equation*}
Since $\Omega$ is positive definite, $\widetilde{\Omega}_n$ is invertible for sufficiently large $n$, and $\widetilde{\Omega}_n^{-1}\asto\Omega^{-1}$, as $n\to\infty$. Consequently, we obtain from \eqref{eq:exp proof T4.2} that, for sufficiently large $n$,
\begin{align}
  &n^{1/2}(\theta_n-\hat\theta_n) \notag\\
  &=-\widetilde{\Omega}_n^{-1}\,n^{-1/2}\bigl(\mathcal{S}_n(\theta_n)-\mathcal{S}_n(\hat\theta_n)\bigr)\notag\\
  &=-\widetilde{\Omega}_n^{-1}\Bigl[n^{-1/2}\bigl(\mathcal{S}_n(\theta_n)-\hat{\mathcal{S}}_n(\hat\theta_n)\bigr)+n^{-1/2}\bigl(\hat{\mathcal{S}}_n(\hat\theta_n)-\mathcal{S}_n(\hat\theta_n)\bigr)\Bigr]. \label{eq:exp2 proof T4.2}
\end{align}
Note already that $n^{-1/2}\bigl(\hat{\mathcal{S}}_n(\hat\theta_n)-\mathcal{S}_n(\hat\theta_n)\bigr)\asto 0$, as $n\to\infty$, follows from Lemma \ref{lem: score approximation}. Consequently, as $\widetilde{\Omega}_n^{-1}$ converges, the theorem follows from \eqref{eq:exp2 proof T4.2} provided that $n^{-1/2}\bigl(\mathcal{S}_n(\theta_n)-\hat{\mathcal{S}}_n(\hat\theta_n)\bigr)\asto0$, as $n\to\infty$. We split the argument for this based on the cases \emph{(a)} and \emph{(b)}. \medskip

\emph{Case (a):} Since $\theta_0$ in this case is an interior point of $\Theta$, consistency of $\theta_n$ and $\hat\theta_n$ implies that $\mathcal{S}_n(\theta_n)=\hat{\mathcal{S}}_n(\hat\theta_n)=0$ for sufficiently large $n$. Hence, the result follows.\smallskip

\emph{Case (b):} As the remaining entries of $\theta_0$ are non-boundary points, it follows that $\mathcal{S}^{(i)}_n(\theta_n)=\hat{\mathcal{S}}^{(i)}_n(\hat{\theta}_n)=0$ for $i=1,3,4,5$ and $n$ sufficiently large and, consequently,
\begin{equation}\label{eq:kuhn-tucker T4.2}
  n^{-1/2}\bigl(\mathcal{S}_n(\theta_n)-\hat{\mathcal{S}}_n(\hat\theta_n)\bigr) =e_2\,n^{-1/2}\bigl(\mathcal{S}^{(2)}_n(\theta_n)-\hat{\mathcal{S}}^{(2)}_n(\hat\theta_n)\bigr).
\end{equation}
By the same observations as those we made in the proof of Lemma \ref{lem: asym norm infeasible}, it holds that if $\alpha_{1,n}=0$, then $\mathcal{S}^{(2)}_n(\theta_n)\leq0$, and if $\alpha_{1,n}>0$, then $\mathcal{S}^{(2)}_n(\theta_n)=0$. Similar arguments are true for $\hat{\alpha}_{1,n}$ and $\hat{\mathcal{S}}^{(2)}_n(\hat\theta_n)$. Therefore, one has that $\alpha_{1,n}\,\mathcal{S}^{(2)}_n(\theta_n)=\hat{\alpha}_{1,n}\,\hat{\mathcal{S}}^{(2)}_n(\hat\theta_n)=0$ for sufficiently large $n$, and
\begin{equation}\label{eq:proof T4.2 inequality}
  (\alpha_{1,n}-\hat{\alpha}_{1,n})\bigl(\mathcal{S}^{(2)}_n(\theta_n)-\hat{\mathcal{S}}^{(2)}_n(\hat\theta_n)\bigr)=-\alpha_{1,n}\hat{\mathcal{S}}^{(2)}_n(\hat\theta_n)-\hat{\alpha}_{1,n}\mathcal{S}^{(2)}_n(\theta_n)\geq 0,
\end{equation}
where the inequality holds since $\alpha_{1,n},\hat{\alpha}_{1,n}\geq0$ and $\mathcal{S}^{(2)}_n(\theta_n),\hat{\mathcal{S}}^{(2)}_n(\hat\theta_n)\leq0$. Next, the second entry of \eqref{eq:exp2 proof T4.2} reads,
\begin{align*}
  &n^{1/2}(\alpha_{1,n}-\hat{\alpha}_{1,n}) \\
  &=-e_2^\top\widetilde{\Omega}_n^{-1}e_2\,n^{-1/2}\bigl(\mathcal{S}^{(2)}_n(\theta_n)-\hat{\mathcal{S}}^{(2)}_n(\hat\theta_n)\bigr)-e_2^\top\widetilde{\Omega}_n^{-1}\,n^{-1/2}\bigl(\hat{\mathcal{S}}_n(\hat\theta_n)-\mathcal{S}_n(\hat\theta_n)\bigr).
\end{align*}
Multiplying both sides by $n^{-1/2}(\mathcal{S}^{(2)}_n(\theta_n)-\hat{\mathcal{S}}^{(2)}_n(\hat\theta_n))$, we obtain
\begin{align*}
  0&\leq (\alpha_{1,n}-\hat\alpha_{1,n})(\mathcal{S}^{(2)}_n(\theta_n)-\hat{\mathcal{S}}^{(2)}_n(\hat\theta_n)) \\
  &=-e_2^\top\,\widetilde{\Omega}_n^{-1}\,e_2\,\Bigl[n^{-1/2}(\mathcal{S}^{(2)}_n(\theta_n)-\hat{\mathcal{S}}^{(2)}_n(\hat\theta_n))\Bigr]^2\\
  &\quad -e_2^\top\,\widetilde{\Omega}_n^{-1}\,n^{-1/2}\bigl(\hat{\mathcal{S}}_n(\hat\theta_n)-\mathcal{S}_n(\hat\theta_n)\bigr)\,n^{-1/2}(\mathcal{S}^{(2)}_n(\theta_n)-\hat{\mathcal{S}}^{(2)}_n(\hat\theta_n)),
\end{align*}
where the inequality comes from \eqref{eq:proof T4.2 inequality}. Rearranging this now yields,
\begin{align*}
  &\bigl|n^{-1/2}(\mathcal{S}^{(2)}_n(\theta_n)-\hat{\mathcal{S}}^{(2)}_n(\hat\theta_n))\bigr|\\
  &\leq \frac{\bigl|e_2^\top\,\widetilde{\Omega}_n^{-1}\,n^{-1/2}\bigl(\hat{\mathcal{S}}_n(\hat\theta_n)-\mathcal{S}_n(\hat\theta_n)\bigr)\bigr|}{e_2^\top\,\widetilde{\Omega}_n^{-1}\,e_2}\asto 0,\qquad n\to\infty,
\end{align*}
where we used that $e_2^\top\widetilde{\Omega}_n^{-1}e_2\asto e_2^\top\Omega^{-1}e_2>0$, as $\Omega^{-1}$ is positive definite, and that $n^{-1/2}\bigl(\hat{\mathcal{S}}_n(\hat\theta_n)-\mathcal{S}_n(\hat\theta_n)\bigr)\asto0$ by Lemma \ref{lem: score approximation}. Inserting this limiting relation into \eqref{eq:kuhn-tucker T4.2} completes the proof.\hfill $\square$
\bigskip

\begin{lemma} \label{lem: FTC}
    With $s_t$ defined in \eqref{eq: def of score}, it holds with probability 1 for all $t\in \mathbb{N}$ and all $\theta,\tilde\theta \in \Theta(\delta)$ that
    \begin{align*}
        s_t(\theta)-s_t(\tilde \theta)=-\left(\int_0^1 \zeta_t(\tilde \theta + u(\theta - \tilde \theta))du\right)(\theta-\tilde\theta),
    \end{align*}
    where $\zeta_t(\theta)$ is defined in \eqref{eq: def information}.
\end{lemma}

\noindent\textbf{Proof:} Fix any $\theta, \tilde\theta\in \Theta(\delta)$, and define $h(u)=\tilde\theta + u(\theta-\tilde \theta), \ u \in[0,1]$. Furthermore, let $F$ denote
\begin{align*}
    F(u)=s_t(h(u))=\frac{1}{2}\left(\frac{x^2_t}{\sigma^2_t(h(u))}-1\right)\frac{\partial \sigma^2_t(h(u))}{\sigma^2_t(h(u))}, \qquad u \in [0,1],
\end{align*}
where $\partial \sigma^2_t(\theta)$ is defined in \eqref{eq: partial sig2(theta)} and denotes the first partial derivatives of $\sigma^2_t(\theta)$. The above is well defined since $\Theta(\delta)$ is a convex set such that $h(u)\in\Theta(\delta)$.

Assume initially that $\gamma\neq\tilde\gamma$. Then using that $\partial \sigma^2_t(\theta)$ is continuously partially differentiable, and hence differentiable, at all points $\theta\in\Theta$ with $\gamma\notin D$, where $D$ is the countable set in \eqref{eq: countable set}, it follows that $F$ is differentiable on $(0,1)\setminus E_t$, where
\begin{align*}
    E_t = \left\{u \in [0,1] : \tilde \gamma + u(\gamma - \tilde \gamma)=x_{t-i} \text{ for some } i \geq 1 \right\}
\end{align*}
is countable. Here we note that the $\alpha_1$-coordinate of $h(u)$ is either zero for all $u\in[0,1]$, in which case the $\alpha_1$-coordinate of $h'(u)=\theta-\tilde\theta$ is zero, or positive for all $u\in(0,1)$. In both cases, the chain rule applies with the (right) partial derivatives of $s_t$. Therefore, by the chain rule and using that $h'(u)=\theta-\tilde \theta$, it holds for any $u \in (0,1)\setminus E_t$ that $F'(u)= \partial s_t(h(u))/\partial \theta^\top(\theta-\tilde\theta)= -\zeta_t(h(u))(\theta-\tilde \theta)$. By Lemma \ref{lem: new process bound}, it holds that $|\zeta_t(h(u))|\leq X_t$, where $X_t<\infty$ almost surely. This bound and the fact that $E_t$ is countable allow for an application of Theorem $8.5.2$ in \citet{dieudonne1960foundations} to conclude
\begin{align}
    |F(u)-F(\overline u)|\leq X_t|\theta-\tilde\theta||u - \overline u|, \qquad u, \overline{u}\in[0,1], \label{eq: Lipschitz in u}
\end{align}
i.e.\ that $F$ is Lipschitz continuous on $[0,1]$.

Next, in the case $\gamma=\tilde\gamma$, $h(u)=\tilde\theta+u(\theta-\tilde\theta)$ is constant in the $\gamma$-coordinate and hence $F$ as defined above is differentiable on $(0,1)$, and \eqref{eq: Lipschitz in u} holds by the same argument.

Now an application of the second fundamental theorem of calculus for Lipschitz functions (second equality), which holds as Lipschitz functions are absolutely continuous (see, e.g., Theorem 3.35 in \citealt{folland1999real}), yields that
\begin{align*}
    s_t(\theta)-s_t(\tilde \theta)=F(1)-F(0)=\int_0^1F'(u)du = -\left(\int_0^1 \zeta_t(\tilde \theta+u(\theta-\tilde \theta))du\right)(\theta-\tilde \theta),
\end{align*}
where we emphasize that it suffices that $F'$ is defined Lebesgue almost everywhere on $[0,1]$. \hfill $\square$
\bigskip

\begin{lemma} \label{lem: uniform convergence of information}
   It holds with probability 1 that
\begin{align*}
\lim_{\varepsilon\downarrow0}\ \limsup_{n\to\infty}\ \sup_{\theta\in B(\varepsilon)}
\big|n^{-1}\mathcal{I}_n(\theta)-\Omega\big|=0,
\end{align*}
where $B(\varepsilon)=\{\theta\in\Theta:|\theta-\theta_0|\leq\varepsilon\}$ for $\varepsilon>0$.
\end{lemma}
\noindent\textbf{Proof:} As the statement concerns $\varepsilon\downarrow0$, we may assume that $\varepsilon$ is sufficiently small such that $B(\varepsilon)\subset\Theta(\delta)$, see below \eqref{eq: Theta small}. Note that $n^{-1}\mathcal{I}_n(\theta)-\Omega=\big(n^{-1}\sum_{t=1}^n\zeta_t-\Omega\big)
+n^{-1}\sum_{t=1}^n\big(\zeta_t(\theta)-\zeta_t\big)$. Since $|\zeta_t(\theta)|\le X_t$ uniformly
in $\theta\in\Theta(\delta)$ with $\E[X_0]<\infty$ by Lemma \ref{lem: new process bound}, ergodicity of $\{\zeta_t\}$ gives
$n^{-1}\sum_t\zeta_t\asto\E[\zeta_0]=\Omega$ showing that the first term converges to zero with probability 1.

Considering the second term, note that all entries of $\zeta_t(\theta)$ except the $(\gamma,\gamma)$-entry are continuous in $\theta\in\Theta(\delta)$. Furthermore, for the $(\gamma,\gamma)$-entry, observe that for any $\theta \in \Theta(\delta)$
\begin{align*}
    \zeta_t^{(\gamma,\gamma)}(\theta) &= -\frac{1}{2}\left(\frac{x^2_t}{\sigma^2_t(\theta)}-1\right)\frac{\partial^2\sigma^2_t(\theta)/\partial\gamma^2}{\sigma^2_t(\theta)} + \frac{1}{2}\left(2\frac{x^2_t}{\sigma^2_t(\theta)} -1\right)\left(\frac{\partial \sigma^2_t(\theta)/\partial \gamma}{\sigma^2_t(\theta)}\right)^2\\
    &= J_{1,t}(\theta)+J_{2,t}(\theta),
\end{align*}
where $J_{2,t}(\theta)$ is continuous in $\theta$. Let $\zeta^{c}_t(\theta)$ denote the matrix obtained from $\zeta_t(\theta)$ by replacing the $(\gamma,\gamma)$-entry with $J_{2,t}(\theta)$, such that $\zeta^c_t(\theta)$ is continuous in $\theta\in\Theta(\delta)$. Moreover, as $x_t^2/\sigma^2_t(\theta)\leq z_t^2Y_{1,t}$ by Lemma \ref{lem: new process bound}, $|\partial^2\sigma^2_t(\theta)/\partial\gamma^2|\leq c$ by \eqref{eq: second order deriv sig2(theta) wrt gamma} and $\sigma^2_t(\theta)\geq\omega_L$, it holds that $|\zeta^c_t(\theta)|\leq|\zeta_t(\theta)|+|J_{1,t}(\theta)|\leq X_t+c(1+z_t^2)Y_{1,t}=:X^c_t$, where $\E[X^c_0]<\infty$ by Lemma \ref{lem: new process bound}. Then
\begin{align*}
    \sup_{\theta\in B(\varepsilon)}\Big|n^{-1}\sum_{t=1}^n\big(\zeta^c_t(\theta)-\zeta^c_t\big)\Big|
&\le2\sup_{\theta\in\Theta(\delta)}\Big|n^{-1}\sum_{t=1}^n\zeta^c_t(\theta)-\E[\zeta^c_0(\theta)]\Big|\\
&\quad+\sup_{\theta\in B(\varepsilon)}\big|\E[\zeta^c_0(\theta)]-\E[\zeta^c_0(\theta_0)]\big| .
\end{align*}
The first term vanishes a.s.\ by the uniform law of large numbers of Theorem 2.2.1 in \citet{straumann2005estimation}, which applies since $\Theta(\delta)$ is compact, and $\theta\mapsto\zeta^c_t(\theta)$ is continuous and dominated by $X^c_t$ with $\E[X^c_0]<\infty$. Furthermore, the second term vanishes as $\varepsilon\downarrow0$ by continuity and the dominated convergence theorem.

It remains to show that
\begin{align*}
\lim_{\varepsilon\downarrow0}\limsup_{n\to\infty}\sup_{\theta\in B(\varepsilon)}n^{-1}\Big|\sum_{t=1}^n\big(J_{1,t}(\theta)-J_{1,t}(\theta_0)\big)\Big|=0.
\end{align*}
We decompose as follows,
\begin{align}
    2\big(J_{1,t}(\theta)-J_{1,t}(\theta_0)\big) &=\left(\frac{1}{\sigma^2_t}-\frac{1}{\sigma^2_t(\theta)}\right) 
    z^2_t\frac{\sigma^2_t}{\sigma^2_t(\theta)}
   \frac{\partial^2\sigma^2_t(\theta)}{\partial\gamma^2}+\left(1-z^2_t\right)\frac{\partial^2\sigma^2_t(\theta)}{\partial\gamma^2}\left(\frac{1}{\sigma^2_t(\theta)}-\frac{1}{\sigma^2_t}\right) \notag\\
   &+\left(1-z^2_t\right)\left(\frac{\partial^2\sigma^2_t(\theta)/\partial\gamma^2-\partial^2\sigma^2_t(\theta_0)/\partial\gamma^2}{\sigma^2_t}\right) \notag \\
   &=T_{1,t}(\theta)+T_{2,t}(\theta)+T_{3,t}(\theta),\label{eq: diff in J}
\end{align}
where $T_{i,t}$ is implicitly defined for $i=1,2,3$. 

Considering initially $T_{1,t}$ and $T_{2,t}$, note that since $\Theta(\delta)$ is convex and
$\theta\mapsto\sigma^2_t(\theta)$ is continuously differentiable,
$\sigma^2_t(\theta)-\sigma^2_t=\int_0^1\partial\sigma^2_t(\theta(u))^\top(\theta-\theta_0)\,du$ with
$\theta(u)=\theta_0+u(\theta-\theta_0)\in\Theta(\delta)$, it follows that
\begin{align*}
\left|\frac{1}{\sigma^2_t(\theta)}-\frac{1}{\sigma^2_t}\right|&\leq c\frac{|\sigma^2_t(\theta)-\sigma^2_t|}{\sigma^2_t}
\\
&\leq c|\theta-\theta_0|\int_0^1\frac{|\partial\sigma^2_t(\theta(u))|}{\sigma^2_t}du
\le c|\theta-\theta_0|\,Y_{2,t},
\end{align*}
where we have used in the first inequality that $1/\sigma^2_t(\theta)\leq c$, and Lemma \ref{lem: new process bound} in the last inequality.

Using this upper bound, $\sigma^2_t/\sigma^2_t(\theta)\leq Y_{1,t}$ with $Y_{1,t}\geq1$ by Lemma \ref{lem: new process bound}, and the fact that $|\partial^2\sigma_t^2(\theta)/\partial\gamma^2|\leq c$ from \eqref{eq: second order deriv sig2(theta) wrt gamma}, we obtain
\begin{align*}
    \sup_{\theta\in B(\varepsilon)}|T_{1,t}(\theta) + T_{2,t}(\theta)|\leq \varepsilon cY_{1,t}Y_{2,t}(1+z_t^2),
\end{align*}
whose average converges a.s.\ by the ergodic theorem to $\varepsilon c\,\E[Y_{1,0}Y_{2,0}(1+z_0^2)]$, which is finite by independence of $z_t$
and $(Y_{1,t},Y_{2,t})$ and Lemma \ref{lem: new process bound}, and tends to zero as $\varepsilon\downarrow0$.

Considering the final term of \eqref{eq: diff in J}, as $\sigma^2_t\geq\omega_0$,
\begin{align*}
    \left|n^{-1}\sum_{t=1}^nT_{3,t}(\theta)\right|\leq c n^{-1}\sum^n_{t=1}\left(1+z^2_t\right)|\partial^2\sigma^2_t(\theta)/\partial\gamma^2 - \partial^2\sigma^2_t(\theta_0)/\partial\gamma^2|.
\end{align*}
Recall from \eqref{eq: second order deriv sig2(theta) wrt gamma} that for any $\theta \in \Theta(\delta)$
\begin{align*}
    &\partial^2\sigma^2_t(\theta)/\partial\gamma^2 - \partial^2\sigma^2_t(\theta_0)/\partial\gamma^2 \\&= \sum_{i=1}^\infty \beta^{i-1}2(\alpha_1\I(x_{t-i}> \gamma)+\alpha_2\I(x_{t-i}\leq \gamma))\\&\quad- \sum_{i=1}^\infty \beta_0^{i-1}2(\alpha_{1,0}\I(x_{t-i}> \gamma_0)+\alpha_{2,0}\I(x_{t-i}\leq \gamma_0)).
\end{align*}
We focus here on the terms involving the indicators $\I(x_{t-i} > \gamma), \I(x_{t-i} > \gamma_0)$. The argument for the terms involving $\I(x_{t-i} \leq \gamma), \I(x_{t-i} \leq \gamma_0)$ is similar and yields the same upper bound as provided in \eqref{eq: upper bound on infinite sums}.
To this end, observe that for any $t,i$, we have for $\gamma \geq \gamma_0$ that
\begin{align*}
    &\beta^{i-1}\alpha_{1}\I(x_{t-i}> \gamma) - \beta_0^{i-1}\alpha_{1,0}\I(x_{t-i}> \gamma_0)\\
    &= \beta_0^{i-1}(\alpha_{1}-\alpha_{1,0})\I(x_{t-i}> \gamma) + (\beta^{i-1}-\beta_0^{i-1})\alpha_{1}\I(x_{t-i}> \gamma) - \beta_0^{i-1}\alpha_{1,0}\I(\gamma_0 < x_{t-i} \leq \gamma).
\end{align*}
In the case $\gamma_0 > \gamma$, the indicator in the last term is replaced by $-\I(\gamma < x_{t-i} \leq \gamma_0)$.
Hence
\begin{align}
    &\left|\sum_{i=1}^\infty \beta^{i-1}\alpha_{1}\I(x_{t-i}> \gamma) - \sum_{i=1}^\infty \beta_0^{i-1}\alpha_{1,0}\I(x_{t-i}> \gamma_0)\right| \notag \\
    &\leq c(|\theta-\theta_0| + \sum_{i=1}^\infty \beta_0^{i-1}\I(\gamma_0 \wedge \gamma < x_{t-i} \leq \gamma_0 \vee \gamma)), \label{eq: upper bound on infinite sums}
\end{align}
where we have used that $\beta^{i-1}-\beta_0^{i-1}=(\beta-\beta_0)(i-1)(\beta^{\ast}_{i})^{i-2}$ for some $\beta^{\ast}_{i} \leq \beta_U < 1$ for all $i$. 
We have thus established that
\begin{align*}
    \left|n^{-1}\sum_{t=1}^nT_{3,t}(\theta)\right|&\leq cn^{-1}\sum^n_{t=1}\left(1+z^2_t\right)\left(|\theta-\theta_0| + \sum_{i=1}^\infty \beta_0^{i-1}\I(\gamma_0 \wedge \gamma < x_{t-i} \leq \gamma_0 \vee \gamma)\right).
\end{align*}
Next, it holds for any $\gamma$ satisfying that $|\gamma-\gamma_0|\leq\varepsilon$ that
\begin{align*}
    \I(\gamma_0 \wedge \gamma < x_{t-i} \leq \gamma_0 \vee \gamma)\leq \I(|\gamma_0-x_{t-i}|\leq\varepsilon),
\end{align*}
which implies that
\begin{align*}
    &\sup_{\theta \in B(\varepsilon)}\left|n^{-1}\sum_{t=1}^nT_{3,t}(\theta)\right| \leq \varepsilon c n^{-1}\sum_{t=1}^n(1+z^2_t)+cn^{-1}\sum_{t=1}^n(1+z^2_t)\sum_{i=1}^\infty \beta_0^{i-1}\I(|\gamma_0-x_{t-i}|\leq\varepsilon)\\
    &\asto \varepsilon c\E[(1+z^2_0)] + c\frac{\E[(1+z^2_0)]\P(|\gamma_0-x_{-1}|\leq\varepsilon)}{1-\beta_0}, \qquad n \to \infty
\end{align*}
using the ergodic theorem and that $z_t$ is independent of $\{x_{t-i} : i \geq 1\}$. Finally letting $\varepsilon \downarrow 0$, and using $\P(|\gamma_0-x_{-1}|\leq\varepsilon)\to\P(x_{-1}=\gamma_0)=0$, proves the lemma. \hfill$\square$
\bigskip

\begin{lemma} \label{lem: score approximation}
    With probability 1,
    \begin{equation*}
        \sum_{t=1}^\infty\sup_{\theta\in\Theta(\delta)}|s_t(\theta)-\hat s_t(\theta)|<\infty.
    \end{equation*}
    In particular, $n^{-1/2}\sup_{\theta\in\Theta(\delta)}|\mathcal{S}_n(\theta)-\hat{\mathcal{S}}_n(\theta)|\asto0$ as $n\to\infty$.
\end{lemma}
\noindent\textbf{Proof:} We show that
\begin{equation}
    \sup_{\theta\in\Theta(\delta)}|s_t(\theta)-\hat s_t(\theta)|\leq(1+V)^2\,t\beta_U^{t-1}W_t,\qquad t\geq1, \label{eq: score approx bound}
\end{equation}
where $V$ and $W_t$ are random variables, defined in Step 2 below, which do not depend on $\theta$ or $n$, and which satisfy $V<\infty$ almost surely and $\E[W_t]=\E[W_1]<\infty$, see Step 3. This implies the lemma: by monotone convergence, $\E[\sum_{t=1}^\infty t\beta_U^{t-1}W_t]=\E[W_1](1-\beta_U)^{-2}<\infty$, such that $\sum_{t=1}^\infty t\beta_U^{t-1}W_t<\infty$ almost surely, and the first statement follows from \eqref{eq: score approx bound}, as $V<\infty$ almost surely. The second statement follows from the first, as $\sup_{\theta\in\Theta(\delta)}|\mathcal{S}_n(\theta)-\hat{\mathcal{S}}_n(\theta)|\leq\sum_{t=1}^\infty\sup_{\theta\in\Theta(\delta)}|s_t(\theta)-\hat s_t(\theta)|$, which does not depend on $n$.

We prove \eqref{eq: score approx bound} in three steps. In Step 1, we write $s_t(\theta)-\hat s_t(\theta)$ as a sum of two terms, each of which is a product of factors. In Step 2, we bound these factors, which gives \eqref{eq: score approx bound}. In Step 3, we show that $V$ and $W_t$ have the properties stated above.

\emph{Step 1.} By \eqref{eq: def of score} and \eqref{eq: def shat}, the score contributions of the infeasible and the feasible log-likelihood are
\begin{equation*}
    s_t(\theta)=\frac12\Big(\frac{x_t^2}{\sigma^4_t(\theta)}-\frac{1}{\sigma^2_t(\theta)}\Big)\partial\sigma^2_t(\theta),\qquad\hat s_t(\theta)=\frac12\Big(\frac{x_t^2}{\hat\sigma^4_t(\theta)}-\frac{1}{\hat\sigma^2_t(\theta)}\Big)\frac{\partial\hat\sigma^2_t(\theta)}{\partial\theta}.
\end{equation*}
Adding and subtracting $\big(x_t^2/\hat\sigma^4_t(\theta)-1/\hat\sigma^2_t(\theta)\big)\partial\sigma^2_t(\theta)$ gives
\begin{align}
    2\big(s_t(\theta)-\hat s_t(\theta)\big)&=\Big(1-\frac{\sigma^2_t(\theta)}{\hat\sigma^2_t(\theta)}\Big)\Big(\frac{x_t^2}{\sigma^2_t(\theta)}+\frac{x_t^2}{\hat\sigma^2_t(\theta)}-1\Big)\frac{\partial\sigma^2_t(\theta)}{\sigma^2_t(\theta)}\notag\\
    &\quad+\frac{\partial\sigma^2_t(\theta)-\partial\hat\sigma^2_t(\theta)/\partial\theta}{\hat\sigma^2_t(\theta)}\Big(\frac{x_t^2}{\hat\sigma^2_t(\theta)}-1\Big). \label{eq: score difference}
\end{align}

\emph{Step 2.} We bound the factors in \eqref{eq: score difference}, uniformly in $\theta\in\Theta(\delta)$. For $t\geq1$, $\sigma^2_t(\theta)$ and $\hat\sigma^2_t(\theta)$ satisfy the same recursion, see \eqref{eq: def sig2hat(theta)} and \eqref{eq: sig2(theta)}, with initial values $\sigma^2_0(\theta)$ and $\hat\sigma^2_0$, respectively. Hence $\sigma^2_t(\theta)-\hat\sigma^2_t(\theta)=\beta^t(\sigma^2_0(\theta)-\hat\sigma^2_0)$, and, as $\hat\sigma^2_0$ does not depend on $\theta$,
\begin{equation*}
    \partial\sigma^2_t(\theta)-\frac{\partial\hat\sigma^2_t(\theta)}{\partial\theta}=\beta^t\partial\sigma^2_0(\theta)+t\beta^{t-1}\big(\sigma^2_0(\theta)-\hat\sigma^2_0\big)e_5,\qquad e_5=(0,0,0,0,1)^\top.
\end{equation*}
Moreover, $\hat\sigma^2_t(\theta)\geq\omega_L$ for $t\geq1$ by \eqref{eq: def sig2hat(theta)}, $\beta^t\leq\beta_U^t\leq t\beta_U^{t-1}$, $|\sigma^2_0(\theta)-\hat\sigma^2_0|\leq\sup_{\theta\in\Theta}\sigma^2_0(\theta)+\hat\sigma^2_0$, and $|\partial\sigma^2_0(\theta)|\leq\sigma^2_0Y_{2,0}$ by Lemma \ref{lem: new process bound}(i). With
\begin{equation*}
    V=\frac{1}{\omega_L}\Big(\sup_{\theta\in\Theta}\sigma^2_0(\theta)+\hat\sigma^2_0+\sigma^2_0Y_{2,0}\Big),
\end{equation*}
the factors in \eqref{eq: score difference} are bounded as follows.
\begin{itemize}
    \item[(a)] The first factor of the first term satisfies
    \begin{equation*}
        \Big|1-\frac{\sigma^2_t(\theta)}{\hat\sigma^2_t(\theta)}\Big|=\frac{\beta^t|\sigma^2_0(\theta)-\hat\sigma^2_0|}{\hat\sigma^2_t(\theta)}\leq V\beta_U^t.
    \end{equation*}
    In particular, $\sigma^2_t(\theta)/\hat\sigma^2_t(\theta)\leq1+V$.
    \item[(b)] The first factor of the second term satisfies
    \begin{equation*}
        \frac{|\partial\sigma^2_t(\theta)-\partial\hat\sigma^2_t(\theta)/\partial\theta|}{\hat\sigma^2_t(\theta)}\leq\frac{\beta^t|\partial\sigma^2_0(\theta)|+t\beta^{t-1}|\sigma^2_0(\theta)-\hat\sigma^2_0|}{\omega_L}\leq Vt\beta_U^{t-1}.
    \end{equation*}
    \item[(c)] For the remaining factors, let $u_t(\theta)=x_t^2/\sigma^2_t(\theta)$ and $v_t(\theta)=|\partial\sigma^2_t(\theta)|/\sigma^2_t(\theta)$. By (a), $x_t^2/\hat\sigma^2_t(\theta)=(\sigma^2_t(\theta)/\hat\sigma^2_t(\theta))\,u_t(\theta)\leq(1+V)u_t(\theta)$, such that
    \begin{equation*}
        \Big|\frac{x_t^2}{\sigma^2_t(\theta)}+\frac{x_t^2}{\hat\sigma^2_t(\theta)}-1\Big|\leq(2+V)\big(1+u_t(\theta)\big),\qquad\Big|\frac{x_t^2}{\hat\sigma^2_t(\theta)}-1\Big|\leq(2+V)\big(1+u_t(\theta)\big),
    \end{equation*}
    while the last factor of the first term has norm $v_t(\theta)$.
\end{itemize}
Inserting (a)--(c) into \eqref{eq: score difference}, and using that $\beta_U^t\leq t\beta_U^{t-1}$ and $V(2+V)\leq(1+V)^2$, gives
\begin{align*}
    2|s_t(\theta)-\hat s_t(\theta)|&\leq V(2+V)\,t\beta_U^{t-1}\big(1+u_t(\theta)\big)\big(v_t(\theta)+1\big)\\
&\leq(1+V)^2\,t\beta_U^{t-1}\big(1+u_t(\theta)\big)\big(1+v_t(\theta)\big).
\end{align*}
Finally, by Lemma \ref{lem: new process bound}(i), $u_t(\theta)=z_t^2\sigma^2_t/\sigma^2_t(\theta)\leq z_t^2Y_{1,t}$ and $v_t(\theta)=(|\partial\sigma^2_t(\theta)|/\sigma^2_t)(\sigma^2_t/\sigma^2_t(\theta))\leq Y_{2,t}Y_{1,t}$. As $Y_{1,t},Y_{2,t}\geq1$, it follows that $1+u_t(\theta)\leq(1+z_t^2)Y_{1,t}$ and $1+v_t(\theta)\leq2Y_{1,t}Y_{2,t}$, and hence
\begin{equation*}
    |s_t(\theta)-\hat s_t(\theta)|\leq(1+V)^2\,t\beta_U^{t-1}W_t,\qquad W_t=(1+z_t^2)Y_{1,t}^2Y_{2,t}.
\end{equation*}
As $V$ and $W_t$ depend neither on $\theta$ nor on $n$, this proves \eqref{eq: score approx bound}.

\emph{Step 3.} Next, $V<\infty$ almost surely, as $\sup_{\theta\in\Theta}\sigma^2_0(\theta)<\infty$ almost surely by the proof of Theorem \ref{thm: consistency}, and $Y_{2,0}<\infty$ almost surely, as $\E[Y_{2,0}^2]<\infty$ by Lemma \ref{lem: new process bound}(ii). Finally, by Lemma \ref{lem: new process bound}(i), $z_t$ is independent of $(Y_{1,t},Y_{2,t})$, and $\{(z_t,Y_{1,t},Y_{2,t})\}$ is stationary. Hence, $\E[W_t]=\E[1+z_t^2]\,\E[Y_{1,t}^2Y_{2,t}]=2\,\E[Y_{1,1}^2Y_{2,1}]=\E[W_1]$, which is finite by Lemma \ref{lem: new process bound}(ii). \hfill $\square$
\bigskip

\begin{lemma} \label{lem: positive definite}
    With $\Omega$ as in Theorem \ref{thm: asym normality}, it holds that $\Omega$ is a positive definite matrix.
\end{lemma}
\noindent\textbf{Proof:} Recall that
\begin{align*}
    \Omega = \frac{1}{2}\E\left[\frac{\partial \sigma^2_t(\theta_0)}{\sigma^2_t}\left(\frac{\partial \sigma^2_t(\theta_0)}{\sigma^2_t}\right)^\top\right],
\end{align*}
which is finite by Lemma \ref{lem: new process bound}. As $\Omega$ is an outer product, it is positive semi-definite. Suppose that $v^\top \Omega v = 0$ for some $v=(v_\omega,v_{\alpha_1},v_{\alpha_2},v_\gamma,v_{\beta})^\top\in \mathbb{R}^5$. Then $v^\top \partial \sigma^2_t(\theta_0)=0$ with probability 1, for all $t$ by stationarity. By \eqref{eq: partial sig2(theta)}, $\partial \sigma^2_t(\theta_0)=w_{t-1} + \beta_0\, \partial\sigma^2_{t-1}(\theta_0)$, where
\begin{align*}
    w_{t-1}=\begin{pmatrix}
        1 \\
        \I(x_{t-1}\geq \gamma_0)(x_{t-1}-\gamma_0)^2 \\
        \I(x_{t-1}< \gamma_0)(x_{t-1}-\gamma_0)^2 \\
        -2a_0(x_{t-1})(x_{t-1}-\gamma_0) \\
        \sigma_{t-1}^2
    \end{pmatrix},
\end{align*}
and hence $v^\top w_{t-1}=v^\top\partial\sigma^2_t(\theta_0)-\beta_0v^\top\partial\sigma^2_{t-1}(\theta_0)=0$ with probability 1. Conditionally on $\mathcal{F}_{t-2}$, $\sigma_{t-1}$ is fixed and $x_{t-1}=\sigma_{t-1}z_{t-1}$ has a density which is strictly positive on $\mathbb{R}$, and $v^\top w_{t-1}$ is a continuous function of $x_{t-1}$. Hence, almost surely, for all $x\geq\gamma_0$,
\begin{align*}
    0 = v_{\omega}+v_\beta \sigma_{t-1}^2 + v_{\alpha_1}(x-\gamma_0)^2 - 2v_\gamma \alpha_{1,0}(x-\gamma_0),
\end{align*}
and for all $x<\gamma_0$,
\begin{align*}
    0 = v_{\omega}+v_\beta \sigma_{t-1}^2 + v_{\alpha_2}(x-\gamma_0)^2 - 2v_\gamma \alpha_{2,0}(x-\gamma_0).
\end{align*}
As these are polynomials in $x-\gamma_0$ which vanish on intervals, their coefficients are zero. Therefore $v_{\alpha_1}=v_{\alpha_2}=v_\gamma \alpha_{1,0}=v_\gamma \alpha_{2,0}=v_\omega+v_{\beta}\sigma^2_{t-1}=0$ with probability 1. As $\alpha_{2,0}>0$, $v_\gamma=0$, also when $\alpha_{1,0}=0$. Finally, $v_\omega + v_\beta \sigma^2_{t-1} = 0$ almost surely implies $v_{\omega}=v_{\beta}=0$, as $\sigma^2_{t-1}$ is non-degenerate (see the proof of Lemma \ref{lem: cons 2}). This proves the claim. \hfill $\square$
\bigskip

\begin{lemma}\label{lem: new process bound}
    Let $\partial\sigma_t^2(\theta)$ be defined as in \eqref{eq: partial sig2(theta)}. There exists $\delta_0>0$ such that, with $\Theta(\delta_0)$ given by \eqref{eq: Theta small}, the following hold.
    \begin{itemize}
        \item[(i)] For any $\theta\in\Theta(\delta_0)$,
        \begin{align*}
            \frac{\sigma_t^2}{\sigma_t^2(\theta)}\leq Y_{1,t},\qquad \frac{|\partial\sigma_t^2(\theta)|}{\sigma_t^2}\leq Y_{2,t},
        \end{align*}
        where ${Y_{i,t}=f_i(z_{t-1},z_{t-2},\dots)\geq1}$, $i=1,2$, for measurable functions $f_i$ that do not depend on $t$ or $\theta$. In particular, $\{(Y_{1,t},Y_{2,t})\}$ is stationary and ergodic, and $(Y_{1,t},Y_{2,t})$ is independent of $z_t$.
        \item[(ii)] It holds that $\E[Y_{1,t}^aY_{2,t}^b]<\infty$ for $a,b\in\{0,1,2\}$.
        \item[(iii)] With $\zeta_t(\theta)$ defined in \eqref{eq: def information}, there exists a random variable $X_t$ with $\E[X_t]<\infty$, such that $|\zeta_t(\theta)|\leq X_t$ for any $\theta\in\Theta(\delta_0)$.
    \end{itemize}
\end{lemma}

\noindent\textbf{Proof:} Throughout, let $\theta\in\Theta(\delta_0)$, where $\delta_0>0$ is chosen in Step 4, such that $\beta_L=\beta_0-\delta_0\leq\beta\leq\beta_0+\delta_0=\beta_U$, and let $\theta_0(\beta)=(\omega_0,\alpha_{1,0},\alpha_{2,0},\gamma_0,\beta)^\top$, which differs from $\theta_0$ only in $\beta$. Moreover, $c\geq1$ denotes a generic constant independent of $\theta$ whose value may change from line to line. The proof consists of the following six steps, which together give the lemma. The steps are proved below.
\begin{itemize}
    \item \emph{Step 1:} With $\sigma_t^2=\sigma_t^2(\theta_0)$, we write
\begin{equation}\label{eq: factorization of ratio}
    \frac{\sigma_t^2}{\sigma_t^2(\theta)}=\frac{\sigma_t^2}{\sigma_t^2(\theta_0(\beta))}\cdot\frac{\sigma_t^2(\theta_0(\beta))}{\sigma_t^2(\theta)}.
\end{equation}
The second factor in \eqref{eq: factorization of ratio} is bounded by a constant $c$.
    \item \emph{Step 2:} The first factor in \eqref{eq: factorization of ratio}, which involves only $\beta$, is bounded by sums of products of iid variables $\widetilde Y_k=g(z_k)$, where $g$ is a measurable function with $0<g\leq\beta_U/\beta_L$, which does not depend on $k$ or $\theta$. Together with Step 1, this gives the first bound in (i),
    \begin{equation*}
        \frac{\sigma_t^2}{\sigma_t^2(\theta)}\leq Y_{1,t}:=c\Big(1+\sum_{i=1}^\infty\prod_{k=t-i}^{t-1}\widetilde Y_k\Big).
    \end{equation*}
    \item \emph{Step 3:} The first and second order partial derivatives $\partial^r\sigma_t^2(\theta)$, $r=1,2$, of $\sigma_t^2(\theta)$ are bounded similarly,
    \begin{equation*}
        \frac{|\partial^r\sigma_t^2(\theta)|}{\sigma_t^2}\leq Y_{r+1,t}:=c\Big(1+\frac{1}{\beta_L}\sum_{i=1}^\infty i^rz_{t-i}^2\prod_{k=t-i+1}^{t-1}\widetilde Y_k\Big),\qquad r=1,2.
    \end{equation*}
    As $c\geq1$ and $i\leq i^2$, it holds that $Y_{1,t}\geq1$ and $1\leq Y_{2,t}\leq Y_{3,t}$. Moreover, each $Y_{j,t}$, $j=1,2,3$, is a measurable function of $z_{t-1},z_{t-2},\dots$, which does not depend on $t$ or $\theta$. Hence, as $\{z_t\}$ is iid, $\{(Y_{1,t},Y_{2,t},Y_{3,t})\}$ is stationary and ergodic, and $(Y_{1,t},Y_{2,t},Y_{3,t})$ is independent of $z_t$. Together with Step 2, this proves (i).
    \item \emph{Step 4:} We choose $\delta_0>0$ such that $\E[\widetilde Y_k^8]<1$, which gives $\E[Y_{1,t}^8]<\infty$.
    \item \emph{Step 5:} We show that $\E[Y_{1,t}^3Y_{3,t}^2]<\infty$. As $Y_{1,t}^aY_{2,t}^b\leq Y_{1,t}^3Y_{3,t}^2$ for $a,b\in\{0,1,2\}$ by the inequalities in Step 3, this proves (ii).
    \item \emph{Step 6:} We show that $|\zeta_t(\theta)|\leq X_t$ with $X_t=3(1+z_t^2)Y_{1,t}^3Y_{3,t}^2$. As $z_t$ is independent of $(Y_{1,t},Y_{3,t})$ and $\E[z_t^2]=1$, it holds that $\E[X_t]=6\,\E[Y_{1,t}^3Y_{3,t}^2]$, which is finite by Step 5. This proves (iii).
\end{itemize}
It remains to show Steps 1--6.

\emph{Step 1.} We show that $\sigma_t^2(\theta_0(\beta))\leq c\,\sigma_t^2(\theta)$. By \eqref{eq: sig2(theta)},
\begin{equation*}
    \sigma_t^2(\theta)=\sum_{i=1}^\infty\beta^{i-1}\big(\omega+a_\theta(x_{t-i})(x_{t-i}-\gamma)^2\big),
\end{equation*}
and $\sigma_t^2(\theta_0(\beta))$ is given by the same series with $\omega_0$, $a_0$ and $\gamma_0$ in place of $\omega$, $a_\theta$ and $\gamma$. Hence it suffices to show that, for all $x\in\mathbb{R}$,
\begin{equation*}
    \omega_0+a_0(x)(x-\gamma_0)^2\leq c\big(\omega+a_\theta(x)(x-\gamma)^2\big).
\end{equation*}
Here $\omega\geq\omega_L$, and $(x-\gamma_0)^2\leq2(x-\gamma)^2+8\gamma_{\max}^2$ with $\gamma_{\max}=\max\{|\gamma_L|,|\gamma_U|\}$. In case (a), $\alpha_1\geq\alpha_{1,0}/2$ by \eqref{eq: Theta small}, such that $a_\theta(x)\geq\min\{\alpha_{1,0}/2,\alpha_L\}>0$, and the inequality follows. In case (b), $a_0(x)=\alpha_{2,0}\I(x<\gamma_0)$ and $a_\theta(x)\geq\alpha_L\I(x<\gamma)$, and the inequality follows, as $(x-\gamma_0)^2\leq4\gamma_{\max}^2$ if $\gamma\leq x<\gamma_0$.

\emph{Step 2.} We bound the first factor in \eqref{eq: factorization of ratio}. If $\beta\geq\beta_0$, then $\sigma_t^2\leq\sigma_t^2(\theta_0(\beta))$, as $\sigma_t^2(\theta)$ is increasing in $\beta$ by \eqref{eq: sig2(theta)}. If $\beta<\beta_0$, we make the following observation, which is due to \citet{jensen2004asymptotic},
\begin{align*}
    \sigma_t^2-\sigma_t^2(\theta_0(\beta))&=\beta_0\sigma_{t-1}^2-\beta\sigma_{t-1}^2(\theta_0(\beta))\\
    &=(\beta_0-\beta)\sigma_{t-1}^2(\theta_0(\beta))+\beta_0\big(\sigma_{t-1}^2-\sigma_{t-1}^2(\theta_0(\beta))\big)\\
    &=(\beta_0-\beta)\sum_{i=1}^\infty\beta_0^{i-1}\sigma_{t-i}^2(\theta_0(\beta)),
\end{align*}
where the first equality holds since $\theta_0(\beta)$ differs from $\theta_0$ only in $\beta$, and the last equality follows by recursion, using that $\beta_0<1$. Using that $(\beta_0-\beta)\beta_0^{i-1}\leq\beta_U^i$, and writing $\sigma_{t-i}^2(\theta_0(\beta))/\sigma_t^2(\theta_0(\beta))$ as a telescoping product, it follows that
\begin{equation*}
    \frac{\sigma_t^2}{\sigma_t^2(\theta_0(\beta))}\leq1+\sum_{i=1}^\infty\prod_{k=t-i}^{t-1}\frac{\beta_U\,\sigma_k^2(\theta_0(\beta))}{\sigma_{k+1}^2(\theta_0(\beta))}.
\end{equation*}
To bound the factors, let first $\gamma_0\geq0$. On the event $(x_k<0)$, $a_0(x_k)=\alpha_{2,0}\geq\alpha_L$ and $(x_k-\gamma_0)^2\geq x_k^2=z_k^2\sigma_k^2\geq z_k^2\sigma_k^2(\theta_0(\beta))$, where the last inequality holds as $\beta<\beta_0$. Hence, by the recursion for $\sigma_{k+1}^2(\theta_0(\beta))$,
\begin{equation*}
    \sigma_{k+1}^2(\theta_0(\beta))\geq\alpha_L\I(x_k<0)x_k^2+\beta_L\sigma_k^2(\theta_0(\beta))\geq\big(\alpha_L\I(z_k<0)z_k^2+\beta_L\big)\sigma_k^2(\theta_0(\beta)),
\end{equation*}
such that each factor is bounded by
\begin{equation*}
    \widetilde Y_k=\frac{\beta_U}{\alpha_L\I(z_k<0)z_k^2+\beta_L}.
\end{equation*}
If $\gamma_0<0$, the same argument on the event $(x_k<2\gamma_0)$, where $(x_k-\gamma_0)^2\geq x_k^2/4$, gives this bound with $\alpha_L\I(z_k<0)$ replaced by $\alpha_L\I(z_k<\kappa)/4$, where $\kappa=2\gamma_0/\omega_0^{1/2}$, since $\sigma_k^2\geq\omega_0$ implies that $\I(x_k<2\gamma_0)\geq\I(z_k<\kappa)$. Note that only the left branch of the NIC is used, with slope $\alpha_{2,0}\geq\alpha_L$, as $\alpha_{1,0}$ may be zero. In both cases, $\widetilde Y_k\leq\beta_U/\beta_L$ is a function of $z_k$ only. Combining the cases $\beta\geq\beta_0$ and $\beta<\beta_0$ with Step 1 and \eqref{eq: factorization of ratio}, we have shown that
\begin{equation}\label{eq: bound ratio1}
    \frac{\sigma_t^2}{\sigma_t^2(\theta)}\leq c\Big(1+\sum_{i=1}^\infty\prod_{k=t-i}^{t-1}\widetilde Y_{k}\Big)=Y_{1,t}.
\end{equation}

\emph{Step 3.} We treat the first and second order partial derivatives simultaneously, writing $\partial^r\sigma_t^2(\theta)$ for $r=1,2$. From the expressions \eqref{eq: partial sig2(theta)} and \eqref{eq: 2nd partial deriv}, each component of $\partial^r\sigma_t^2(\theta)$ is bounded in absolute value by $c\sum_{i=1}^\infty i^{r-1}\beta^{i-1}(1+x_{t-i}^2+\sigma_{t-i}^2(\theta))$, where we used that $(x-\gamma)^2+|x-\gamma|\leq c(1+x^2)$. Moreover, by \eqref{eq: sig2(theta)}, $\sigma_{t-i}^2(\theta)\leq c\sum_{j=1}^\infty\beta^{j-1}(1+x_{t-i-j}^2)$. Inserting this and using that $\sigma_t^2\geq\omega_0$, it is straightforward to show that
\begin{equation}\label{eq: derivative bound}
    \frac{|\partial^r\sigma_t^2(\theta)|}{\sigma_t^2}\leq c\Big(1+\sum_{i=1}^\infty i^r\beta_U^{i-1}\frac{x_{t-i}^2}{\sigma_t^2}\Big),\qquad r=1,2.
\end{equation}
Note that we compare with the true volatility $\sigma_t^2$, since the terms $x_{t-i}^2$ cannot be bounded by $\sigma_{t-i+1}^2(\theta)$ when $\alpha_1$ may be zero. By the argument in Step 2 with $\beta=\beta_0$, $\beta_U\sigma_k^2/\sigma_{k+1}^2\leq\widetilde Y_k$, and moreover $\sigma_{k+1}^2\geq\beta_0\sigma_k^2\geq\beta_L\sigma_k^2$. Hence, writing $\sigma_{t-i}^2/\sigma_t^2$ as a telescoping product,
\begin{equation*}
    \beta_U^{i-1}\frac{x_{t-i}^2}{\sigma_t^2}=\frac{z_{t-i}^2}{\beta_U}\prod_{k=t-i}^{t-1}\frac{\beta_U\sigma_k^2}{\sigma_{k+1}^2}\leq\frac{z_{t-i}^2}{\beta_L}\prod_{k=t-i+1}^{t-1}\widetilde Y_k,
\end{equation*}
and inserting this into \eqref{eq: derivative bound} yields
\begin{equation}\label{eq: def Y23}
    \frac{|\partial^r\sigma_t^2(\theta)|}{\sigma_t^2}\leq c\Big(1+\frac{1}{\beta_L}\sum_{i=1}^\infty i^rz_{t-i}^2\prod_{k=t-i+1}^{t-1}\widetilde Y_k\Big)=Y_{r+1,t},\qquad r=1,2.
\end{equation}

\emph{Step 4.} We choose $\delta_0$ such that $\rho=\lVert\widetilde Y_k\rVert_8<1$, where $\rho$ does not depend on $k$, as $\{z_k\}$ is iid. Let $\gamma_0\geq0$; the case $\gamma_0<0$ is similar. Since $z_0$ has a density function which is bounded away from zero on compact subsets of $\mathbb{R}$, it holds for some $\varepsilon>0$ that $\P(z_0<-\varepsilon)>0$, and splitting the expectation according to $z_0<-\varepsilon$ and $z_0\geq-\varepsilon$ yields
\begin{equation*}
    \E[\widetilde Y_k^8]\leq\bigg(\frac{\beta_0+\delta_0}{\alpha_L\varepsilon^2+\beta_0-\delta_0}\bigg)^8\P(z_0<-\varepsilon)+\bigg(\frac{\beta_0+\delta_0}{\beta_0-\delta_0}\bigg)^8\P(z_0\geq-\varepsilon).
\end{equation*}
As $\delta_0\downarrow0$, the upper bound converges to $(\beta_0/(\alpha_L\varepsilon^2+\beta_0))^8\,\P(z_0<-\varepsilon)+\P(z_0\geq-\varepsilon)<1$, and hence $\rho<1$ for $\delta_0>0$ sufficiently small. As $\{\widetilde Y_k\}$ is iid, Minkowski's inequality then gives $\lVert Y_{1,t}\rVert_8\leq c(1+\sum_{i=1}^\infty\rho^i)<\infty$.

\emph{Step 5.} We show that $\E[Y_{1,t}^3Y_{3,t}^2]<\infty$. By \eqref{eq: def Y23} and Minkowski's inequality,
\begin{equation*}
    \E\big[Y_{1,t}^3Y_{3,t}^2\big]^{1/2}\leq c\,\E\big[Y_{1,t}^3\big]^{1/2}+c\sum_{i=1}^\infty i^2\,\E\Big[Y_{1,t}^3z_{t-i}^4\prod_{k=t-i+1}^{t-1}\widetilde Y_k^2\Big]^{1/2},
\end{equation*}
where $\E[Y_{1,t}^3]<\infty$ by Step 4. In the $i$th term, $z_{t-i}$ is independent of the product, but not of $Y_{1,t}$, which depends on $z_{t-i}$ only through $\widetilde Y_{t-i}\leq\beta_U/\beta_L$. Let $Y^{(i)}_{1,t}$ be defined as $Y_{1,t}$, but with $\widetilde Y_{t-i}$ replaced by $\beta_U/\beta_L$. Then $Y_{1,t}\leq Y^{(i)}_{1,t}$, $Y^{(i)}_{1,t}$ and the product are jointly independent of $z_{t-i}$, and, by Minkowski's inequality as in Step 4, $\lVert Y^{(i)}_{1,t}\rVert_8\leq c(1+(\beta_U/\beta_L)\sum_{j=1}^\infty\rho^{j-1})$, which does not depend on $i$. Hence, by independence, the Cauchy--Schwarz inequality and $\lVert\prod_{k=t-i+1}^{t-1}\widetilde Y_k\rVert_4\leq\rho^{i-1}$, the $i$th expectation is bounded by $\E[z_0^4]\lVert Y^{(i)}_{1,t}\rVert_8^3\rho^{2(i-1)}\leq c\rho^{2(i-1)}$, and the sum is finite, as $\sum_{i=1}^\infty i^2\rho^{i-1}<\infty$ and $\E[z_0^4]<\infty$.

\emph{Step 6.} The statement regarding $\zeta_t(\theta)$ follows, as $x_t^2/\sigma_t^2(\theta)=z_t^2\sigma_t^2/\sigma_t^2(\theta)\leq z_t^2Y_{1,t}$, and as the bounds above give $|\partial\sigma_t^2(\theta)|/\sigma_t^2(\theta)\leq Y_{1,t}Y_{2,t}$ and $|\partial^2\sigma_t^2(\theta)|/\sigma_t^2(\theta)\leq Y_{1,t}Y_{3,t}$. Hence, using that $Y_{1,t}\geq1$ and $1\leq Y_{2,t}\leq Y_{3,t}$,
\begin{align*}
    |\zeta_t(\theta)|\leq(z_t^2Y_{1,t}+1)Y_{1,t}Y_{3,t}+(2z_t^2Y_{1,t}+1)Y_{1,t}^2Y_{2,t}^2\leq3(1+z_t^2)Y_{1,t}^3Y_{3,t}^2=X_t.
\end{align*}
\hfill $\square$

\end{document}